\documentclass[12pt]{article}

\usepackage{epsf,epsfig,epstopdf}
\usepackage{graphics}
\usepackage{cite}
\usepackage{amsmath,amssymb}
\usepackage{url}
\usepackage{a4wide}

\newcommand{\dirac}[1]{\displaystyle{\not} #1}

\numberwithin{equation}{section}
  \newcommand{\ccaption}[2]{
    \begin{center}
    \parbox{0.95\textwidth}{
      \caption[#1]{\small{#2}}
      }
    \end{center}
    }

\begin{document}
\thispagestyle{empty}
\allowdisplaybreaks

\begin{flushright}
{\small
IPPP/11/12\\
ITP-UU-11/07\\
SPIN-11/05\\
BARI-TH/640-11}
\end{flushright}

\vspace{\baselineskip}

\begin{center}

\vspace{0.2\baselineskip} \textbf{\Large\boldmath
Off-shell effects for $t$-channel and $s$-channel\\[5pt]
single-top production at NLO in QCD}
\\
\vspace{3\baselineskip}
{\sc P.~Falgari$^a$, F.~Giannuzzi$^b$, P.~Mellor$^c$, A.~Signer$^c$}\\
\vspace{0.7cm}
{\sl 
$^a$Institute for Theoretical Physics and Spinoza Institute,\\ 
Utrecht University, 3508 TD Utrecht, The Netherlands\\ 
\vspace{0.3cm}

$^b$Dipartimento di Fisica, Universit\`a degli Studi di Bari,\\
I-70126 Bari, Italy\\
I.N.F.N., Sezione di Bari, I-70126 Bari, Italy\\ 
\vspace{0.3cm}

$^c$IPPP, Department of Physics, University of Durham, \\
Durham DH1 3LE, England}

\vspace*{1.2cm}

\textbf{Abstract}\\

\vspace{1\baselineskip}

\parbox{0.9\textwidth}{ In this work we present a calculation of both
  $t$-channel and $s$-channel single-top production at next-to-leading
  order in QCD for the Tevatron and for the LHC at a centre-of-mass energy
  of $7$ TeV. All the cross sections and kinematical distributions
  presented include leading non-factorizable corrections arising from
  interferences of the production and decay subprocesses, extending
  previous results beyond the narrow-width approximation. The new
  off-shell effects are found to be generally small, but can be
  sizeable close to kinematical end-points and for specific
  distributions.}

\end{center}

\newpage
\setcounter{page}{1}


\section{Introduction}

The production of a single top quark in hadronic collisions was
observed, for the first time, two years ago by both the D0 and CDF
collaborations at the Fermilab Tevatron \cite{Aaltonen:2009jj,
  Abazov:2009ii}. Contrary to the top-pair production mechanism, which
proceeds via strong interactions in the Standard Model (SM),
single-top production is mediated by electroweak couplings, and thus
represents an important window into the charged-current interactions
of the top quark.  Though experimentally challenging, due to the
presence of large backgrounds from $Wj$ and $t \bar{t}$ production,
this process will play an important role in the physics programme of
the LHC, where top quarks will be produced at much higher rates than
at the Tevatron.  Measurements of the cross section allow for a direct
determination of the CKM matrix element, $V_{tb}$, whose precise value
provides a test of the unitarity of the weak-flavour mixing matrix
\cite {Alwall:2006bx}. Moreover, the angular correlations of the
products of the top-quark decay encode information on the spin
structure of the $W t b$ vertex and on the production dynamics of the
top quark \cite{Mahlon:1999gz,Motylinski:2009kt}, and can thus be used
to set limits on the strength of possible anomalous couplings and
new-physics effects~\cite{Tait:2000sh,Cao:2007ea}.  Being sensitive to
the bottom-quark content of the proton, the single-top production
cross section also encodes information on the bottom-quark parton
density, which so far has been theoretically computed from the gluon
PDF, rather than extracted from experimental data. Finally, beside
being an important signal in itself, single-top production also
represents a background to a number of new-physics production
channels, including some relevant for Higgs-boson
searches~\cite{Dittmar:1996ss}. For these reasons, an accurate
theoretical understanding of single-top production is clearly
desirable.

In the SM, a single top quark can be produced via exchange of a virtual,
$t$-channel $W$ boson ($q b \rightarrow q' t$ or $\bar{q} b
\rightarrow \bar{q}' t$), a virtual, $s$-channel $W$ boson ($q \bar{q}'
\rightarrow t \bar{b}$) or in association with a real $W$ ($b g
\rightarrow W^- t$). At the Tevatron, the largest contribution to the
cross section comes from the $t$-channel process, followed by
$s$-channel production, whereas associated $Wt$ production is
negligible. At the LHC, the $t$-channel production process still
dominates the cross section, but the associated production mechanism
gives the second largest contribution, while the $s$-channel cross
section is very small (see, e.g., Ref.~\cite{Bernreuther:2008ju}). Note
that the distinction of $t$-channel and $s$-channel production is
somewhat artificial since, starting at next-to-leading order (NLO) in
QCD, both processes can give rise to the same physical final
state. While, from a theoretical point of view, the two channels can
always be disentangled on a diagram-by-diagram basis, depending on the
sign of the invariant mass of the intermediate $W$ boson,
experimentally they are defined by two distinct sets of selection
criteria and kinematical cuts, chosen in such a way to suppress one of
the two processes in favour of the other. However, as it will be shown
in more detail in Section \ref{sec:results}, for some observables the
contribution of the ``wrong" channel can still be sizeable, even after
cuts. For this reason, a detailed phenomenological analysis
generally requires the inclusion of both $t$-channel and $s$-channel
production.  The associated production of a top quark with a $W$ boson
produces very different final-state topologies and, experimentally, its
signal can be much more easily disentangled from the other two
channels. Hence, in the following we will ignore this production
mechanism and focus our attention on $t$-channel and $s$-channel
single-top production only.

Both $t$-channel and $s$-channel single-top production have been
studied extensively in the literature. The NLO corrections to both
channels, as well as $Wt$ associated production, have been known for a
while \cite{Bordes:1994ki, Stelzer:1997ns,Smith:1996ij, Giele:1995kr},
and a fully differential calculation was presented in
Refs.~\cite{Harris:2002md,Sullivan:2004ie}.  Electroweak corrections,
both in the SM and in the MSSM, have also been computed and
appeared in more recent works \cite{Beccaria:2008av, Macorini:2010bp},
while resummation of soft-gluon corrections was studied in, among others, 
Refs.~\cite{Kidonakis:2007ej,Kidonakis:2010tc} using the
Mellin-space formalism, and Refs.~\cite{Zhu:2010mr,Wang:2010ue}
using renormalization-group techniques in the context of
soft-collinear effective theory. All these works considered the case
of a stable top quark, ignoring the subsequent decay of the top to
lighter, longer-lived particles.

The decay of the top quark can be included in the, so-called,
narrow-width approximation (NWA), where the top is produced on shell
and then allowed to decay. This approach allows for the inclusion of
realistic cuts on the decay products of the top quark and preserves
spin correlations between initial and final states. Also, the
complexity of the calculation is only sightly increased compared to
the stable-quark approximation. In the NWA framework, NLO corrections
correspond to the, so-called, \emph{factorizable} corrections,
i.e. (on-shell) corrections to either the production or decay
subprocess. NLO QCD analyses in this framework for the semileptonic
top decay were published in Refs.~\cite{Campbell:2004ch,
  Cao:2004ky,Cao:2005pq, Campbell:2005bb} and, more recently, with an
extended and detailed phenomenological analysis in
Refs.~\cite{Heim:2009ku, Schwienhorst:2010je}.

None of the works mentioned so far have included
\emph{non-factorizable} corrections, i.e. contributions from virtual
and real gluons connecting production and decay subprocesses.  Such
off-shell effects were studied recently in Ref.~\cite{Falgari:2010sf},
for $t$-channel single-top production, where the fully differential
cross section was computed for a \emph{resonant} ($p_t^2-m_t^2 \sim
m_t \Gamma_t$), rather than on-shell ($p_t^2-m_t^2=0$), top quark.  A
similar calculation for $s$-channel production was presented in
Ref.~\cite{Pittau:1996rp} and non-factorizable corrections to other
processes have been considered as well~\cite{Beenakker:1997ir,
  Denner:1997ia}.  The effective-theory approach adopted in
Ref.~\cite{Falgari:2010sf} will be reviewed briefly in Section
\ref{sec:method}. Here it is sufficient to say that the effect of
non-factorizable corrections was found to be generally small for
inclusive-enough observables, consistent with previous calculations
\cite{Fadin:1993dz, Melnikov:1993np}, though they can be locally
sizeable near kinematic thresholds and for specific distributions.

Here we extend the results of Ref.~\cite{Falgari:2010sf} to the
$s$-channel production mechanism, getting an overall view of the
process, and present a more detailed analysis of the size of these
off-shell effects.  We also discuss how the $t$-channel and $s$-channel
mechanisms can be experimentally identified. All the results presented
in Section \ref{sec:results} include non-factorizable corrections and,
in most cases, the corresponding prediction in the narrow-width
approximation is given for comparison. In our calculation we adopt the
$5$-flavour scheme, setting the bottom-quark mass to zero. The
relation between the $5$-flavour and $4$-flavour scheme has been
extensively studied in Refs.~\cite{Campbell:2009ss,
  Campbell:2009gj}. Furthermore, we generally take the CKM matrix to
be the unit matrix. However, in Section \ref{sec:CKM} we will briefly
discuss how CKM-suppressed initial-state configurations can
significantly modify the shape of the top-rapidity distribution, as
first observed by the authors of Ref.~\cite{AguilarSaavedra:2010wf}
for the tree-level processes.  Finally, in the present calculation we
do not include the matching of the NLO result and parton shower Monte
Carlo.  This was implemented in the programs
MC@NLO~\cite{Frixione:2005vw, Frixione:2008yi} and
POWHEG~\cite{Alioli:2009je, Re:2010bp}.

The paper is organized as follows: In Section \ref{sec:method} we will
review the effective-theory formalism, introduced in
Ref.~\cite{Falgari:2010sf}, for the calculation of the non-factorizable
corrections to $t$-channel single-top production, and illustrate the
(trivial) extensions necessary to include the $s$-channel process.  In
Section \ref{sec:results} we will first give the precise selection
rules and kinematical cuts we used to define $t$-channel-like and
$s$-channel-like signatures, and then present cross sections and
relevant kinematical distributions for the Tevatron (in Section
\ref{sec:Tevatron}) and for the $7$ TeV LHC (in Section
\ref{sec:LHC}).  This section also contains a discussion of the scale
dependence of the NLO results and, as previously mentioned, a brief discussion 
of the importance of CKM off-diagonal partonic channels.  Finally, in Section
\ref{sec:conclusion} we draw our conclusions.

\section{Calculation} \label{sec:method}

In Ref.~\cite{Falgari:2010sf} a formalism was presented that allows for the
systematic inclusion of non-factorizable and background contributions
to the production of a massive, unstable particle. The method discussed
there is based on a previously developed framework
\cite{Beneke:2004km} in which the heavy-resonance production is
described in an effective-theory (ET) language.  Here we limit
ourselves to a brief review of this method and refer the reader to
Refs.~\cite{Falgari:2010sf,Beneke:2004km} for more detail.

\subsection{The ET approach to unstable-particle production}

The key point of the approach adopted here is the relaxation of the
assumption made in the narrow-width approximation of an exactly
on-shell massive particle, $p_X^2=m_X^2$, where $m_X$ denotes the mass
of the particle, for the moment generically referred to as
$X$. Instead we consider a \emph{resonant}, unstable particle with a
non-vanishing virtuality, $p_X^2-m_X^2 \neq 0$, which is assumed to be
much smaller than the unstable-particle mass, $p_X^2-m_X^2 \ll m_X^2$.
The hierarchy between the virtuality and the mass of the unstable
particle provides a small parameter, $D_X/m_X^2 \equiv
(p_X^2-m_X^2)/m_X^2$, from which a systematic expansion of
the full matrix element is performed.  In this respect, the
effective-theory approach to unstable-particle production can be seen
as a systematisation and extension of the pole
approximation~\cite{Stuart:1991xk, Aeppli:1993rs}.  To make this
hierarchy explicit, we introduce a generic small parameter, $\delta \ll
1$, and count $D_X/m_X^2 \sim \delta$.

In the effective theory, only low-virtuality modes with $q^2 \lesssim
m_X^2 \delta^2$ are still dynamical, and are described by a set of
effective fields.  These include, in particular, a \emph {resonant}
field, $\Phi_X$, to describe the heavy, unstable particle;
\emph{collinear} fields, $\psi_c$, describing massless, energetic
particles; and \emph{soft} fields, $A_s$, corresponding to low-scale
gluonic fluctuations.  The effects related to the non-vanishing width
of particle $X$ are resummed into the leading, massive-particle
kinetic term of the effective Lagrangian,
\begin{equation}
\label{bilinear}
2 \hat{M}_X\phi^\dagger_X 
\left( i v \cdot \partial -\frac{\Omega_X}{2}\right) \phi_X \,,  
\end{equation}
where the coefficient, $\Omega_X$, can be related to $\bar{s}=m_X^2-i
m_X \Gamma_X$, the gauge-invariant complex pole of the full
heavy-particle propagator~\cite{Beneke:2004km}
\footnote{In the pole scheme that we are using in this paper, one has
  the simple relation $\Omega_X=- i \Gamma_X$.}.  \emph{Hard} modes
with virtuality of order $m_X^2$ are not part of the effective
Lagrangian. Their effect is instead included in the effective-theory
calculation through hard matching coefficients, i.e. effective
couplings extracted from fixed-order, on-shell matrix elements of the
full theory.  Note that in this context, ``on-shell" refers to the
complex pole of the heavy-particle propagator, i.e. $p_X^2= \bar{s}
\equiv m_X^2-i m_X \Gamma_X$, meaning that the couplings of the
effective Lagrangian are generally complex. This is a feature shared
with the Complex Mass Scheme, in which the gauge-invariant resummation
of finite-width effects is obtained through a complex renormalization
of masses and couplings of the SM Lagrangian~\cite{Denner:2005fg}.
  
\begin{figure}[t]
\begin{center}
\includegraphics[width=0.6 \linewidth]{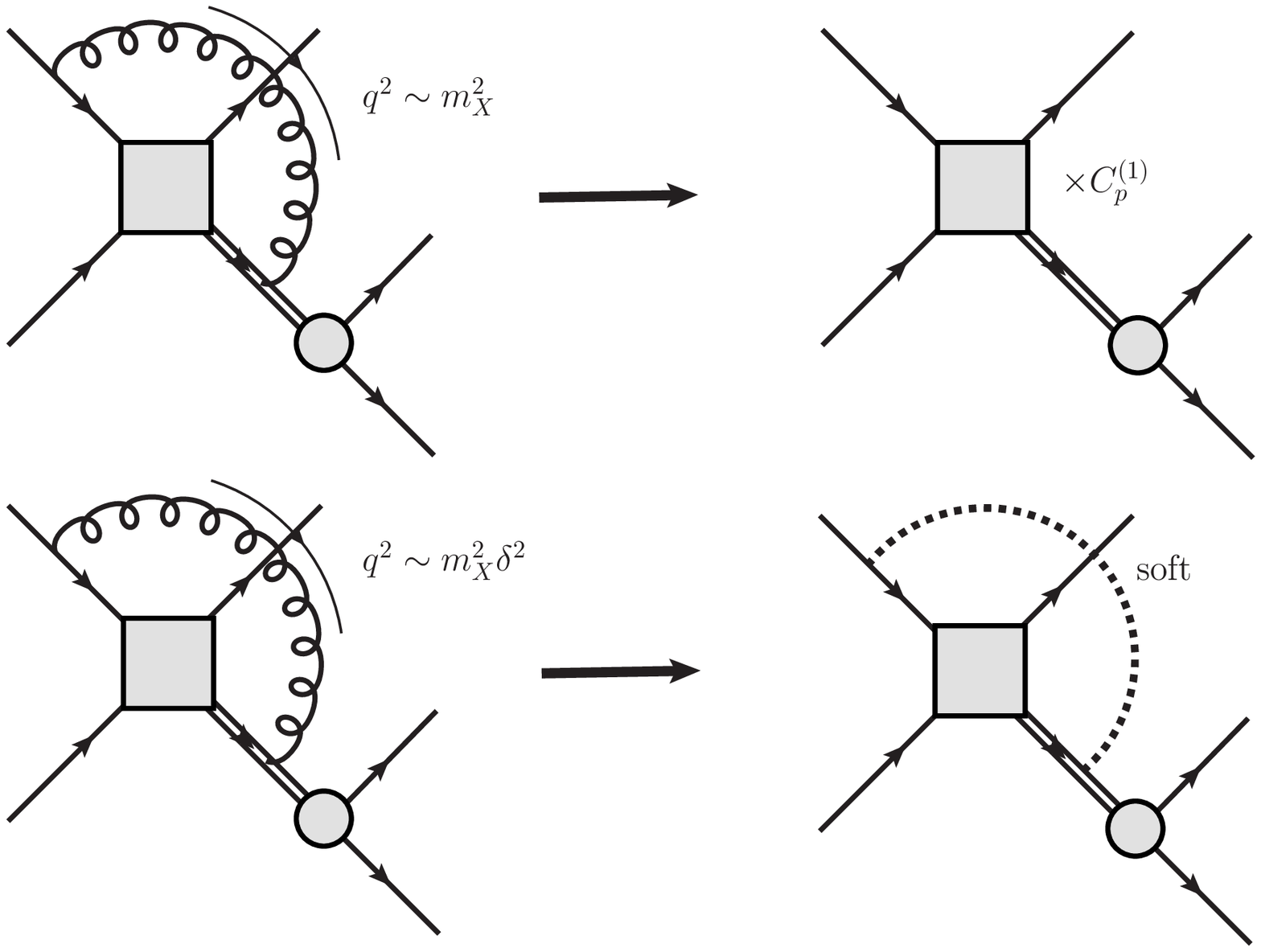}
\end{center}
\ccaption{}{Correspondence between the expansion by regions and the
  effective-theory calculation: hard loops (top left) coincide with
  corrections to the matching coefficient of the production vertex
  (top right), whereas soft loops (bottom left) reproduce the effect
  of loop diagrams in the effective theory (bottom
  right). \label{fig:regions}}
\end{figure}

From a practical point of view, the hard matching coefficients and the
effective-theory matrix elements required can be easily computed using
the method of regions (see Refs.~\cite{Beneke:1997zp,Smirnov:2002pj})
to expand, in powers of $\delta$, the loop integrals of the full theory. In
the \emph{hard region}, the integrand is expanded under the assumption
that the loop momentum, $q$, scales as $q \sim m_X$. Hard corrections
coincide with what, in the language of the double-pole approximation,
are usually defined as factorizable corrections, i.e. corrections that
can unambiguously be assigned to the production or decay of the
unstable particle.  On the other hand, the loop corrections in the
effective theory encode non-factorizable effects, and can be obtained
from the \emph{soft part} of full loop integrals, computed by
expanding the integrand according to the assumption that the loop
momentum scales as $q \sim m_X \delta$. The correspondence between
a strict effective-theory calculation and the expansion by regions is
schematically given in Figure~\ref{fig:regions}.  Note that the
separation of hard and soft corrections is gauge invariant and can,
in principle, be pursued to an arbitrarily-high number of
loops~\cite{Chapovsky:2001zt}.

For the total cross section, real radiative corrections can be treated
as virtual ones, using the optical theorem to relate phase-space
integrals to loop integrals and expanding with the method of
regions. In this context, the effective-theory formalism has been
previously applied to linear-collider
phenomenology~\cite{Beneke:2007zg,Actis:2008rb}.  For an arbitrary
observable, the inclusion of real corrections in the effective-theory
formalism poses some problems. First of all, the method of regions
relies on dimensional regularisation to take care of possible extra
singularities introduced by the expansion in $\delta$, whereas real
corrections in high-energy physics calculations are typically computed
numerically in $D=4$ dimensions. Furthermore, in the presence of an
extra gluon with momentum $q$ in the final state, it is, a priori, not
clear what the correct expansion parameter is, given that both $p_X^2$
and $(p_X-q)^2$ can become resonant. Finally, an arbitrary observable
introduces, in general, new scales that can hamper the
effective-theory expansion in $\delta$.

For the aforementioned reasons, in Ref.~\cite{Falgari:2010sf} we
deviated from a strict ET calculation and used the full matrix element
to compute real corrections.  This can, in principle, pose a problem
since the different treatment of virtual and real corrections could
lead to uncancelled infrared singularities in the cross section. The
cancellation of infrared singularities is usually made manifest by
adding and subtracting from the NLO cross section a suitably chosen
term that correctly reproduces the soft and collinear singularities of
the real matrix element,
 \begin{eqnarray} \label{eq:subt}
 d \sigma^\text{NLO} &=& d \sigma_V+\int d \Phi\, d \sigma_R\nonumber\\
 &=& \left(d \sigma_V+\int d \Phi\, d \sigma_{\text{subt}}\right)
  +\int d \Phi (d \sigma_R-d \sigma_{\text{subt}}) \, ,
 \end{eqnarray}
 where $d \sigma_V$ and $d \sigma_R$ denote virtual and real
 corrections respectively, and $d \Phi$ represents the integration
 over the phase space of the additional final-state particle. The
 first term in the second line of Eq. (\ref{eq:subt}) is integrated
 analytically in $D=4 -2 \epsilon$ dimensions, ideally leading to an
 explicit cancellation of the poles in $\epsilon$, whereas the second
 term, which is free of singularities by construction, can be computed
 numerically in $D=4$ dimensions. Since, in our case, the
 virtual-correction term, $d \sigma^V$, is expanded in $\delta$, an
 exact cancellation of the singularities requires that $d
 \sigma_{\text{subt}}$ in the first term of (\ref{eq:subt}) is also
 expanded consistently, i.e.
 \begin{equation}\label{eq:subt_exp}
  d \sigma^\text{NLO} \sim  \left(d \sigma^{\text{exp}}_V
  +\int d \Phi\, d \sigma^{\text{exp}}_{\text{subt}}\right)
  +\int d \Phi\, (d \sigma_R-d \sigma_{\text{subt}}) \, .
 \end{equation}
Given that the kinematical configurations described by $d
\sigma_{\text{subt}}$ correspond to a gluon being soft or two partons
being collinear, in this case the expansion parameter can always be
clearly identified. Note that Eq. (\ref{eq:subt_exp}) is formally
gauge invariant as long as $d \sigma_R$ contains the full set of
relevant Feynman diagrams.

\subsection{$t$-channel and $s$-channel single-top production}
\label{sec:s_top}

In Ref.~\cite{Falgari:2010sf} the effective theory formalism just
described was applied to $t$-channel single-top production. In this
work we extend that calculation to include the $s$-channel production
mechanism. At the Born level we are thus interested in the three
processes:
\begin{eqnarray} \label{eq:processes}
t\text{-channel} && 
q(p_1)  b(p_2) \rightarrow q'(p_3) b(p_4) W^+(p_W) \, , \nonumber\\ 
&& \bar{q}'(p_1) b(p_2) \rightarrow \bar{q}(p_3) b(p_4) W^+(p_W) \, ,
\nonumber\\ 
s\text{-channel} && 
q(p_1) \bar{q}'(p_2) \rightarrow \bar{b}(p_3) b(p_4) W^+(p_W) \, ,
\end{eqnarray}
where $(q,q')\in \{(u,d), \, (c,s)\}$. The leptonic decay, $W^+(p_W)
\rightarrow e^+(p_5) \nu_e(p_6)$, is included in the narrow-width
approximation and, in the following, will be understood. In this
article we will not give explicit results for the production of an
anti-top, $\bar{t}$, but our results could, of course, be trivially
extended to this case.  Along the same lines as the calculation
presented in Ref.~\cite{Falgari:2010sf}, we systematically expand the
amplitudes for these processes in the small parameters $\alpha_s$,
$\alpha_{ew}$ and $\Delta_t \equiv (p_W+p_b)^2-m_t^2+i m_t \Gamma_t$.
Throughout the paper we will denote these parameters collectively as
$\delta$ and assume the relative scaling
\begin{equation}
\delta \sim \alpha_s^2 \sim \alpha_{ew} \sim \frac{\Delta_t}{m_t^2} \, .
\end{equation}  

At tree-level, the amplitudes for the three processes
(\ref{eq:processes}) contain contributions from \emph{resonant
  diagrams}, containing an intermediate top-quark propagator, and
non-resonant ones. The latter are further grouped into
\emph{electroweak non-resonant diagrams} and \emph{QCD background
  diagrams}, in which two electroweak vertices are replaced by strong
couplings.  Accordingly, the amplitude for $q b \rightarrow q' b W^+$
can be parameterised as~\cite{Falgari:2010sf}
\begin{equation}
{\cal A}^{{\rm tree}} = \delta_{31} \delta_{42} \Big(  
g_{ew}^3\, A^{(3,0)}_{(-1)} + g_{ew}^3\, A^{(3,0)}_{(0)} +
\ldots \Big) +
T^a_{31} T^a_{42}\,g_{ew} g_s^2\, A^{(1,2)} \, ,
\label{treeA}
\end{equation}
where the powers of the strong and electroweak couplings associated
with the amplitude have been made explicit and appear as upper
indices, whereas a lower index denotes a power of the resonant
propagator $\Delta_t$, i.e.  $A^{(i,j)}_{(k)} \sim (\Delta_t)^k$ and has
a prefactor $g_{ew}^i g_s^j$. 
A similar representation holds for the processes $\bar{q}' b \rightarrow
\bar{q} b W^+$ and $q \bar{q}' \rightarrow \bar{b} b W^+$, whose
amplitudes can be obtained from the first $t$-channel process by
crossing $p_1\rightarrow -p_3$ and $p_2\rightarrow -p_3$
respectively.  The term $g_{ew}^3 A^{(3,0)}_{(-1)}$, which scales as
$g_{ew}^3 (\Delta_t)^{-1} \sim \delta^{1/2}$, arises from the leading
contribution of the resonant diagram, whereas $g_{ew}^3
A_{(0)}^{(3,0)} \sim \delta^{3/2}$ accounts for subleading
contributions from the same diagram, where the resonant propagator is
cancelled by higher-order terms in $\delta$, or from the electroweak
background diagrams. The leading-order term of the matrix-element
squared thus scales as $\delta$. The contribution of the QCD
background diagrams is given by $g_{ew} g_s^2 A^{(1,2)} 
\sim \delta$, i.e. they are, in principle, suppressed by only 
$\delta^{1/2} \sim \alpha_s \sim 10\%$ compared to the leading 
resonant contribution. However, the interference between these diagrams
and the resonant diagram does not contribute to the squared matrix element
due to the different colour structure of the QCD and purely EW
contributions. Thus, at Born level, the first corrections to the 
matrix-element squared are suppressed by a factor $\delta$ and 
not $\delta^{1/2}$.

While terms of ${\cal O}(\delta^{3/2})$ in the cross section vanish at
the Born level, terms with this parametric scaling arise from a subset
of one-loop QCD corrections and real-gluon corrections. These
contributions are what, in the following, we define as our NLO
approximation. The one-loop diagrams are given by virtual corrections
to the resonant tree-level diagram.  Note that only upon expansion in
$\delta$ is this subset of corrections gauge invariant.  The leading
term in the expansion in $\Delta_t/m_t^2$, given by the sum of leading
soft and hard contributions, was computed in
Ref.~\cite{Falgari:2010sf} for the $t$-channel production
processes. Again, the analogous result for the $s$-channel process can
be obtained via the replacement $p_2 \leftrightarrow -p_3$.

While Eq.~(\ref{eq:subt_exp}) requires the use of the full, unexpanded
matrix element for the process $q b \rightarrow q' b W^+ g$ for the
computation of real corrections, here we include only a subset of
diagrams corresponding to gluonic corrections to the resonant
electroweak diagram. This clearly violates gauge invariance. However,
the omitted diagrams, missing a resonant top propagator, are
numerically suppressed by an extra power of $\delta$. This is also the
case for the gauge-violating effects, as long as the condition 
$\Delta_t^2/m_t^2 \ll 1$ is satisfied. We would like to stress that the 
approximation made here is not a limitation of the method, but a choice 
made out of convenience, which could be easily relaxed. The real-correction
amplitudes can be found in Ref.~\cite{Falgari:2010sf}.

Alongside the aforementioned real diagrams, corrections from
gluon-initiated partonic processes also contribute to the NLO cross
section. More precisely, one has to include the processes
\begin{table}[h!]
\begin{center}
\begin{tabular}{ccccc}
$g b \rightarrow q \bar{q}' b W^+ \, ,$ & \hspace{1 cm}
& $q g \rightarrow q' \bar{b} b W^+ \, ,$ 
& \hspace{1 cm}&
$\bar{q}' g \rightarrow \bar{q} \bar{b} b W^+ \, .$ 
\end{tabular}
\label{tab:gluon_in} 
\end{center}
\end{table}\\
The required matrix elements can be obtained by suitably crossing the
results for the process $q b \rightarrow q' b W^+ g$, given in
Ref.~\cite{Falgari:2010sf}.  While the first process can be
unambiguously assigned to $t$-channel top production, the processes
with a bottom and anti-bottom in the final state can arise from
diagrams with a $W$ boson exchanged in either the $t$- or
$s$-channel. More precisely, the amplitude for the process $q g
\rightarrow q' \bar{b} b W^+$ can be written as
\begin{equation}
{\cal A}^{{\rm tree}}_{qg} = 
g_s\, g_{ew}^3\, \Big(
T^{a_2}_{47} \delta_{31}\,A^{47}_{qg} +
T^{a_2}_{31} \delta_{47}\,A^{31}_{qg} \Big) \, ,
\label{treeAqg}
\end{equation}
where the term proportional to $A^{47}_{qg}$ is usually assigned to
$t$-channel single top production, and the term proportional to
$A^{31}_{qg}$ to $s$-channel production. As anticipated in the
introduction, while this separation of the two channels on a
diagram-by-diagram basis is possible from a theoretical point of
view, in general both of them will contribute to an experimentally
defined observable. This is discussed in detail and quantified in the
next section.

\section{Results}\label{sec:results}

As mentioned earlier, all necessary tree-level, virtual and real
amplitudes for the $t$-channel processes were computed in
Ref.~\cite{Falgari:2010sf}. The necessary amplitudes for $s$-channel single-top
production can thus be obtained from there by crossing. These results
were implemented in two independent Monte Carlo codes, one adopting
the Catani-Seymour dipole subtraction scheme~\cite{Catani:1996vz}, and
the other the FKS subtraction method~\cite{Frixione:1995ms}.

In this section we will present results for the two hadronic processes
\begin{eqnarray}\label{eq:proc1} 
&& N_1 N_2 \rightarrow J_{b} J_{l} e^+\dirac{E}_T+X \, , \\
\label{eq:proc2}
&&  N_1 N_2 \rightarrow J_{b} J_{\bar{b}} e^+\dirac{E}_T+X \, , 
\end{eqnarray}
where $N_1 N_2=p\bar{p}$ for Tevatron and $N_1 N_2=pp$ for LHC. $J_b$
($J_{\bar{b}}$) represents a jet generated from a bottom quark
(antiquark), $J_l$ is a jet generated from a light parton and
$\dirac{E}_T$ denotes missing transverse energy.  Clearly, from an
experimental point of view, jets generated from a bottom quark or
antiquark are not distinguishable. This assumption can easily be
relaxed, however, this is unessential to the discussion of
non-factorizable corrections, which is the main point of this work.
In Eqs.~(\ref{eq:proc1}) and (\ref{eq:proc2}), $X$ represents an arbitrary
number of additional jets which are not generated from $b$ or $\bar{b}$
quarks.

Process (\ref{eq:proc1}) clearly represents a $t$-channel-like
signature, while (\ref{eq:proc2}) represents an $s$-channel-like signal. 
In fact, at LO in QCD, only $t$-channel diagrams contribute to the cross 
section for (\ref{eq:proc1}), and (\ref{eq:proc2}) is determined by
$s$-channel diagrams only. However, as mentioned at the end of Section~
\ref{sec:s_top}, at NLO in QCD both production channels can contribute
to both experimental signatures due to the mixing arising from the
$qg$ processes in Eq. (\ref{treeAqg}). While the contribution of
$s$-channel configurations to the first process is negligible, the
$t$-channel contribution to $N_1 N_2 \rightarrow J_{b} J_{\bar{b}}
e^+\dirac{E}_T+X$ can be numerically important, even after applying
suitable cuts to suppress it.  This is discussed more quantitatively
in Sections~\ref{sec:Tevatron} and \ref{sec:LHC}.

The input parameters for our numerical analysis are given in 
Table~\ref{tab:input}. The NLO top-decay width is used for both Born and
next-to-leading order cross sections. Within a strict effective-theory
approach, in the leading part of the bilinear operator,
Eq.~(\ref{bilinear}), $\Omega_X$ corresponds to the leading-order
width within the pole mass scheme. NLO corrections to the width
(i.e. the matching coefficient $\Omega_X$) would be taken into account
perturbatively. However, we decided to resum NLO corrections to
$\Omega_X$ as well. The difference between these two approaches is
beyond ${\cal O}(\delta^{3/2})$ and our approach avoids large
differences due to using different input parameters, which could
obscure the effect of genuine NLO corrections.  For the same reason,
we use the MSTW2008 NLO PDF set~\cite{Martin:2009iq} and the
corresponding value of strong coupling, $\alpha_s$, everywhere. Jets are
constructed using a standard $k_\perp$ cluster algorithm with a
resolution parameter $D_{\text{res}}=0.7$, but any other jet
definition could equally well be used. Unless otherwise specified,
the renormalization and factorization scales are, by default, set to
$\mu_R=\mu_F=m_t/2$.
     
\begin{table}
  \begin{center}
  \begin{tabular}{lr}

  \hline
  \hline
  $m_t=172$~GeV & $\alpha_{ew}=0.03394$ \\[2pt]
  $M_W=80.4$~GeV & $\Gamma_W=2.14$~GeV \\[2pt]
  $M_Z=91.2$~GeV & $\Gamma_t^{{\rm NLO}}=1.32813$~GeV  \\
  \hline
  \hline

  \end{tabular}
  \end{center}
  \ccaption{}{Input parameters used for calculating the cross sections
    and distributions shown in Sections~\ref{sec:Tevatron},
    \ref{sec:LHC} and \ref{sec:comparison}. \label{tab:input}}
\end{table}

\subsection{Single-top production at the Tevatron} \label{sec:Tevatron}

We start by presenting results for single-top production for
proton-antiproton collisions at a centre-of-mass energy of $1.96$
TeV. The kinematical cuts and vetoes applied to the two processes,
(\ref{eq:proc1}) and (\ref{eq:proc2}), are presented in
Table~\ref{tab:Tev}.  In both cases we apply a (loose) constraint on
the invariant mass of the $J_b e^+ \nu_e$ system, defined by
\begin{equation}
m_{\text{inv}}(t)=\sqrt{(p(J_b)+p(e)+p(\nu))^2} \, ,
\label{eq:inv_mass}
\end{equation}
where the three-momentum of the invisible neutrino can be exactly
reconstructed by imposing an on-shell condition for the $e^+ \nu_e$
system, i.e. $(p(e)+p(\nu))^2=M_W^2$. The cut on $m_\text{inv}$
ensures that $\delta < 1$ and, thus, that the effective-theory counting
is satisfied. Further standard cuts are applied on the transverse
momenta of the $b$-tagged jet and the charged lepton, on the momentum
of the hardest light jet in process (\ref{eq:proc1}) or the
$\bar{b}$-tagged jet in process (\ref{eq:proc2}), and on the
transverse missing energy, $\dirac{E}_T$. Finally, for the process $p
\bar{p} \rightarrow J_{b} J_{l} e^+\dirac{E}_T+X$ we impose a veto on
extra $b$-tagged jets. This suppresses the contributions from the
$s$-channel production diagrams, except for the kinematic
configurations in which the $\bar{b}$ jet is very forward, and thus
undetected. A similar veto is imposed on extra light jets in the
process $p \bar{p} \rightarrow J_{b} J_{\bar{b}} e^+\dirac{E}_T+X$ to
suppress contributions from the $t$-channel diagrams.  We would like
to point out that our calculation is fully differential, and the cuts
can, therefore, be easily varied at will. The cuts used here represent a
minimal, but still realistic, set-up that allows the discussion of
non-factorizable corrections.

\begin{table}
  \begin{center}
  \begin{tabular}{l|r}

  \hline
  \hline
  $p \bar{p} \rightarrow J_{b} J_{l} e^+\dirac{E}_T+X$ & $p \bar{p}
  \rightarrow J_{b} J_{\bar{b}} e^+\dirac{E}_T+X$ \\[2pt] 
  \hline
  \hline
  $p_T(J_b) > 20$~GeV & $p_T(J_b) > 20$~GeV \\[2pt]
   $p_T(\text{hardest} \, J_l) > 20$~GeV & $p_T(J_{\bar{b}}) > 20$~GeV \\[2pt]
      $p_T(\text{extra} \, J_{\bar{b}}) < 20$~GeV & $p_T(\text{extra}
  \, J_l) < 15$~GeV \\[2pt] 
   $\dirac{E_T} +p_T(e)> 30$~GeV & $\dirac{E_T} +p_T(e)> 30$~GeV \\[2pt]
   $120 < m_{{\rm inv}} < 200$~GeV  & $120 < m_{{\rm inv}} < 200$~GeV \\
  \hline
  \hline

  \end{tabular}
  \end{center}
  \ccaption{}{Kinematical cuts and vetoes used for Tevatron
    results. \label{tab:Tev}} 
\end{table} 

Results for the total cross sections, for both the effective-theory
calculation (ET) and the spin-correlated narrow-width approximation
(NWA), are presented in Table~\ref{tab:totalTev}. As previously
mentioned, the NLO top-quark width and PDFs are used for both the LO
and NLO cross sections.
\begin{table}[t!]
  \begin{center}
  \begin{tabular}{l|c|c|c}

  \hline
  \hline
  $p \bar{p} \rightarrow J_{b} J_{l} e^+\dirac{E}_T+X$ & \hspace{1 cm}
  & {\bf ET} & {\bf NWA}  \\[2pt] 
  \hline \phantom{$\left(\frac{1}{2}\right)^X$}
   & {\bf LO}[fb] & $86.89(1)^{+ 0.45}_{-4.01} $ & $88.11(1)$ \\[2pt]
  \hline \phantom{$\left(\frac{1}{2}\right)^X$}
   &  {\bf NLO}[fb] & $53.62(5)^{+7.76}_{-15.24}$ & $54.43(1)$ \\[2pt]
  \hline
  \hline
  $p \bar{p} \rightarrow J_{b} J_{\bar{b}} e^+\dirac{E}_T+X$ & 
  \hspace{1 cm} & \hspace{0.3 cm} {\bf ET} \hspace{0.3 cm}& {\bf NWA} \\[2pt]  
 \hline \phantom{$\left(\frac{1}{2}\right)^X$}
   & {\bf LO}[fb] & $34.68(1)^{+3.53}_{-2.97}$ & $35.16(1)$ \\[2pt]
  \hline \phantom{$\left(\frac{1}{2}\right)^X$}
   &  {\bf NLO}[fb] & $27.42(2)^{+1.95}_{-1.00}$ & $27.79(1)$ \\[2pt]
  \hline
  \hline
  \end{tabular}
  \end{center}
  \ccaption{}{LO and NLO cross sections for the processes
    (\ref{eq:proc1}) and (\ref{eq:proc2}), computed using the parameters 
    given in Table~\ref{tab:input} and imposing the kinematical cuts and 
    vetoes given in Table~\ref{tab:Tev}. The errors come
    from scale uncertainty only.  All numbers are in
    femtobarns. \label{tab:totalTev}}
\end{table}
\begin{figure}[t!]
\begin{center}
\includegraphics[width=0.49 \linewidth]{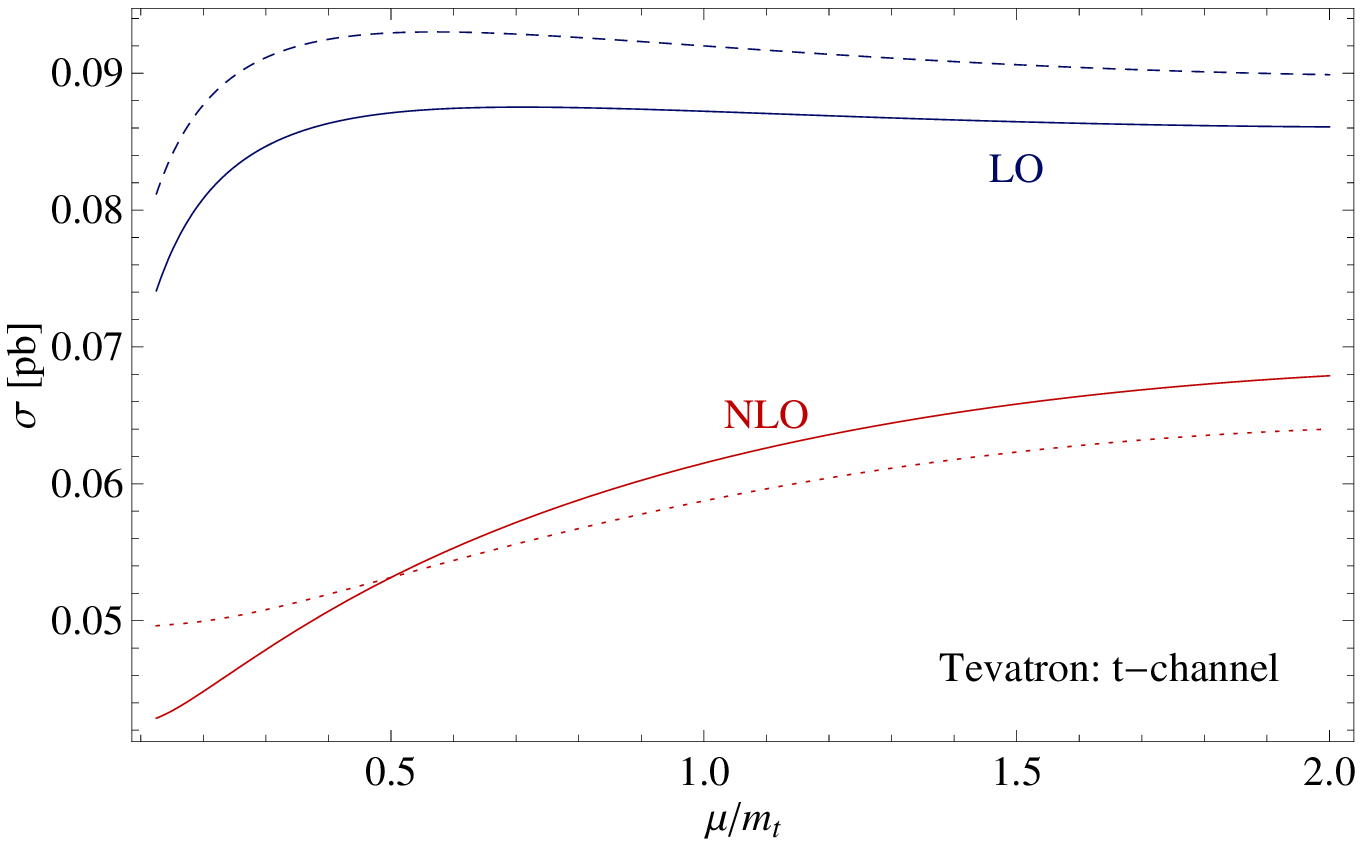}
\includegraphics[width=0.49 \linewidth]{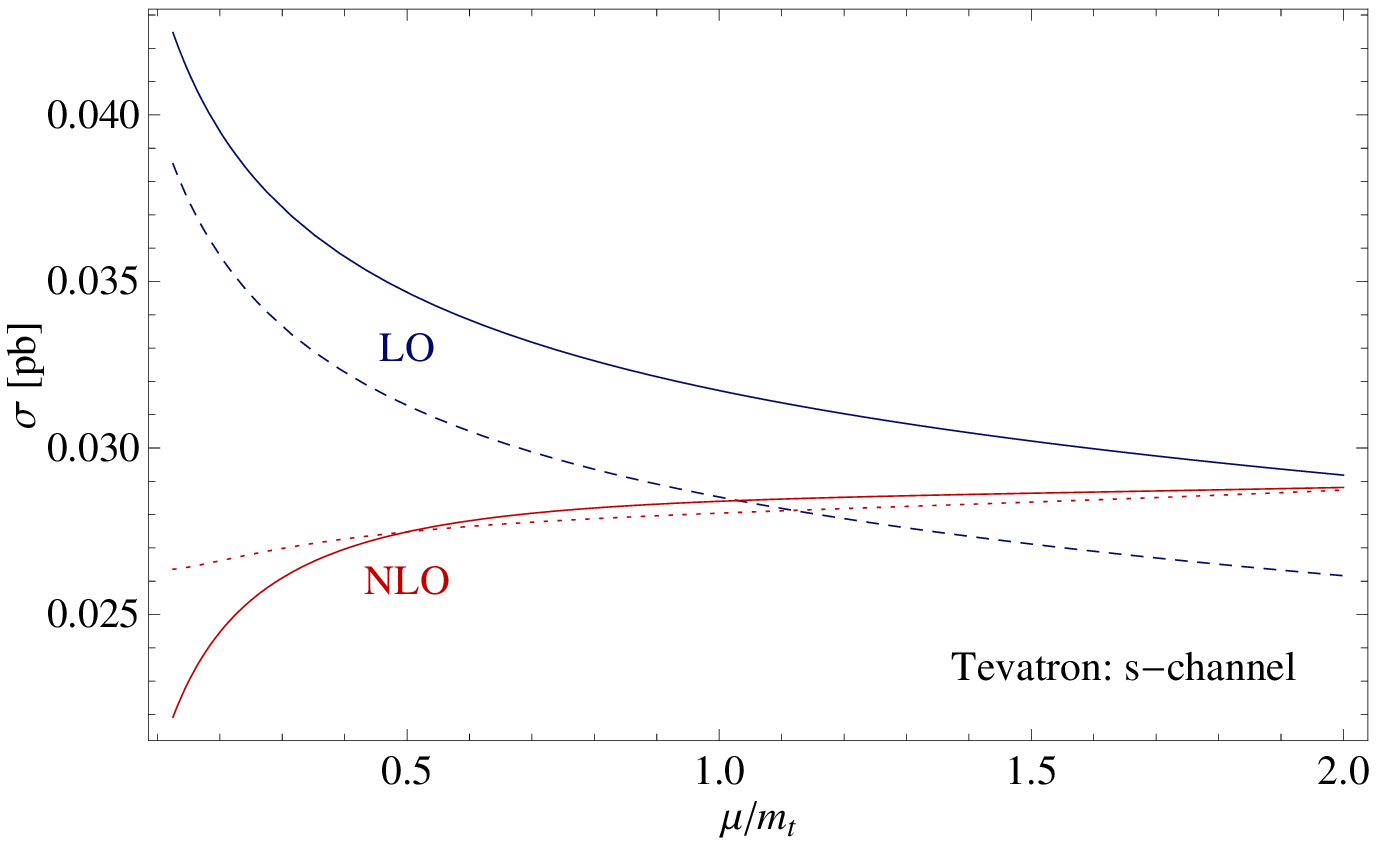}
\end{center}
\ccaption{}{Scale dependence of the total cross section for 
  $p \bar{p} \rightarrow J_{b}J_{l} e^+\dirac{E}_T$ (left) and 
  $p \bar{p} \rightarrow J_{b} J_{\bar{b}} e^+\dirac{E}_T$
  (right) at the Tevatron. The plot shows the LO cross section with LO
  (dashed blue) and NLO (solid blue) PDFs, and the NLO cross section
  with simultaneous variation of factorization and renormalization
  scale (solid red) and for fixed factorization scale (dashed
  red). \label{fig:scaleTevatron}}
\end{figure}
The total NLO corrections are large, and amount to $\sim-38 \%$ and
$\sim-21\%$ for $p \bar{p} \rightarrow J_{b} J_{l} e^+\dirac{E}_T$ and
$p \bar{p} \rightarrow J_{b} J_{\bar{b}} e^+\dirac{E}_T$ respectively.
As expected for the total cross section, both at LO and NLO the
difference between the ET calculation, which includes non-factorizable
corrections, and the NWA result is small, amounting to $\sim 1.5 \%$,
which is compatible with the na\"ive expectation of the accuracy of
the NWA, i.e.  $\Gamma_t/m_t \sim 1\%$.  For the process $p \bar{p}
\rightarrow J_{b} J_{l} e^+\dirac{E}_T+X$ about $96\%$ of the NLO
cross section is given by $t$-channel diagrams, whereas the
$s$-channel contribution amounts to $4\%$.  For $p \bar{p} \rightarrow
J_{b} J_{\bar{b}} e^+\dirac{E}_T+X$ the $s$-channel diagrams
contribute about $87\%$ of the NLO cross section. In this case the
contamination from $t$-channel configurations is sizeable, even after
applying cuts specifically designed to suppress it.

The errors on the values presented in Table~\ref{tab:totalTev} are
obtained from scale variation only (the errors in parenthesis are
statistical Monte Carlo errors), i.e. no PDF uncertainty is
considered.  Both renormalization and factorization scale are varied
in the interval $m_t/4 \leq \mu_R=\mu_F \leq m_t$.  The scale
dependence from LO to NLO results is only moderately reduced for the
$p \bar{p} \rightarrow J_{b} J_{\bar{b}} e^+\dirac{E}_T$ process,
whereas it is increased for $p \bar{p} \rightarrow J_{b} J_{l}
e^+\dirac{E}_T$. This is partly due to the renormalization-scale
dependence of the NLO result, which is absent at LO since the resonant
Born cross section does not depend on the strong coupling constant,
$\alpha_s$. This is better seen in Figure~\ref{fig:scaleTevatron},
where several cross sections are shown as a function of the scale
$\mu$. The dashed red curve, obtained varying the factorization scale
in the NLO result while keeping the renormalization scale fixed, is
clearly flatter than the solid red curve, where the two scales were
varied simultaneously.  The plot also shows that the unusually small
scale dependence of the LO results does not stem from using the NLO
PDFs, since this affects the normalization of the result but not the
shape.  Furthermore, the scale uncertainty of the results depends
quite strongly on the choice of the central value for $\mu$, and our
default choice lies close to the region where the scale dependence of
the NLO result is stronger.

We would like to point out that, in a strict effective-theory
approach, soft and hard contributions are typically renormalized at
different scales, $\mu_s \ll \mu_h$, and then evolved to a common
factorization/renormalization scale, $\mu$, using suitable
renormalization-group equations. This evolution effectively resums
large logarithms of the ratio $ \ln \mu_s/\mu_h \sim \ln \Gamma_t/m_t
\sim \ln \delta$, generally improving the scale dependence of the
cross sections. In the approach used in this work, where real
corrections are computed in the full theory, the application of the
aforementioned procedure is problematic since the separation of real
hard and soft contributions is not transparent.  For this reason, in
the present work we evaluate hard and soft contributions at the same
scale $\mu = \mu_h$.  Given that $\alpha_s(\mu_s)$ is substantially
larger than $\alpha_s(\mu_h)$, this might lead to an underestimation of
the importance of non-factorizable corrections and it may, therefore, be
worthwhile studying the effects of large-logarithm resummation for
generic observables.

\begin{figure}[t!]
\begin{center}
\includegraphics[width=0.7 \linewidth]{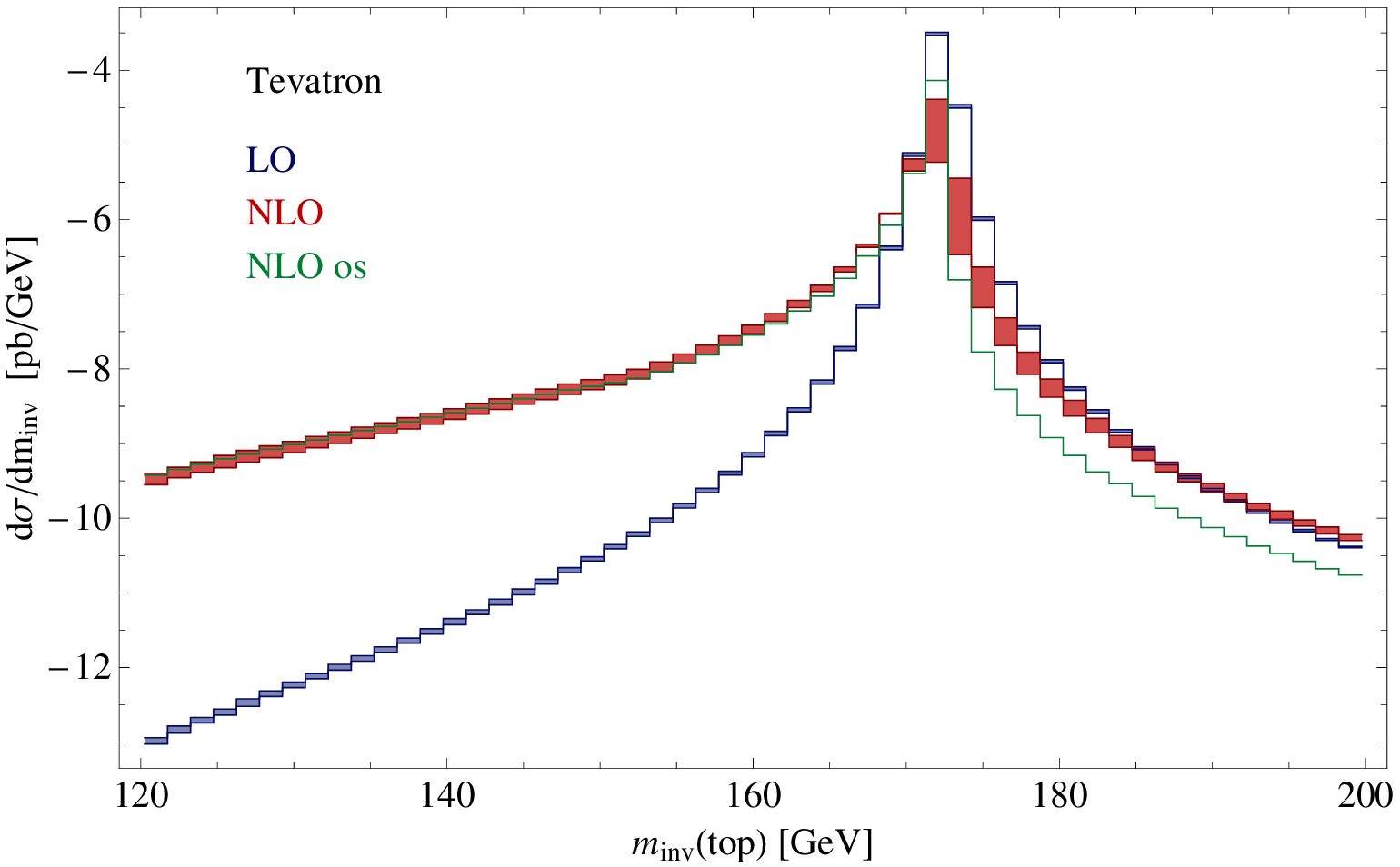}
\includegraphics[width=0.7 \linewidth]{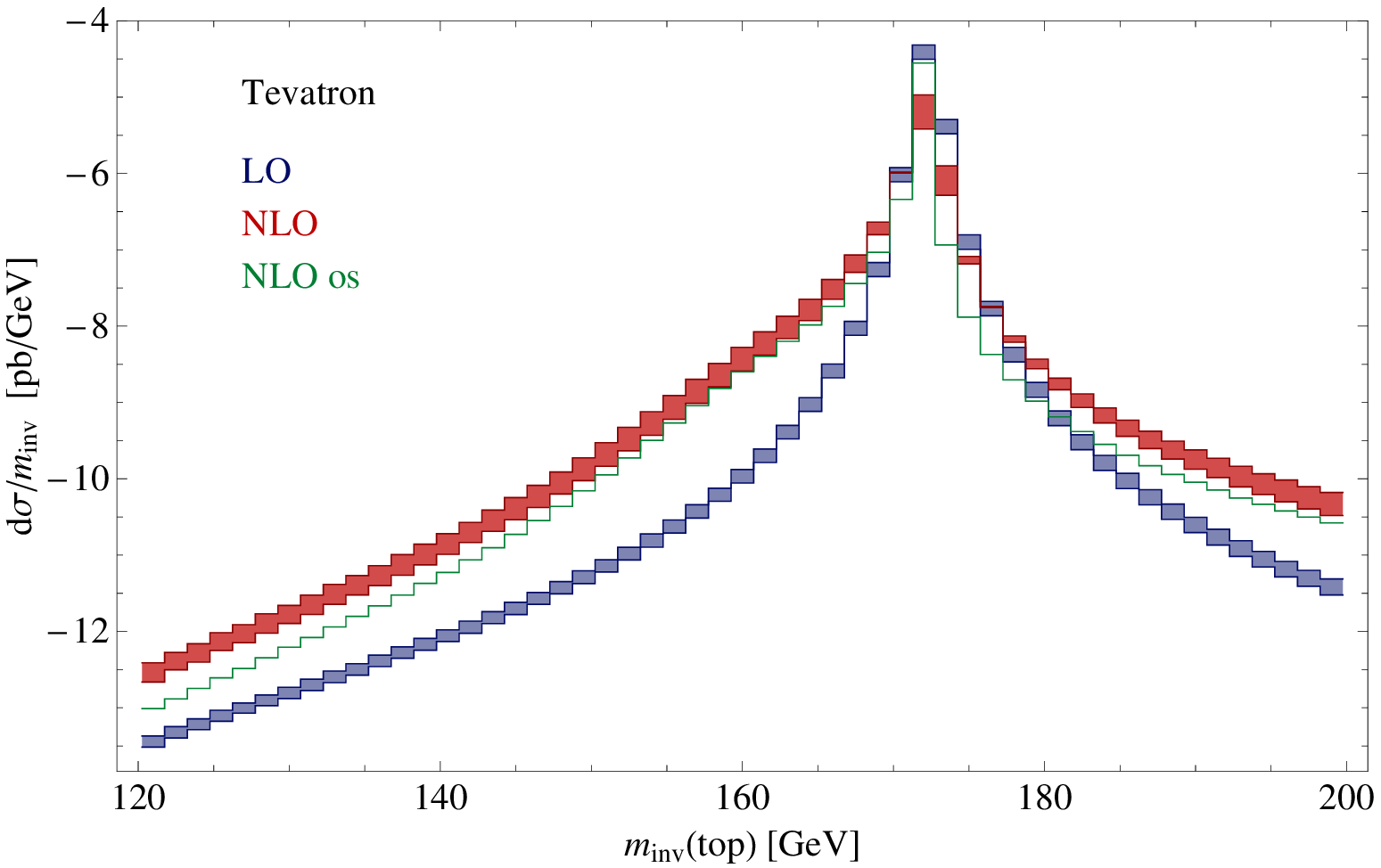} \ccaption{}{Top
  invariant-mass distributions for the process $p \bar{p} \rightarrow
  J_{b} J_{l} e^+\dirac{E}_T+X$ (upper plot) and $p \bar{p}
  \rightarrow J_{b} J_{\bar{b}} e^+\dirac{E}_T+X$ (lower plot) at the
  Tevatron. The blue band represents the LO ET result, the red band
  the NLO ET result, and the green curve the NLO spin-correlated NWA
  prediction. For the ET results the band width is obtained by varying
  the factorization and renormalization scales in the interval $m_t/4
  \leq \mu_R = \mu_F \leq m_t$.
\label{fig:inv_mass_Tev}}
\end{center}
\end{figure}

Next we consider the top invariant-mass (defined in
Eq.~(\ref{eq:inv_mass})) distributions which are plotted in
Figure~\ref {fig:inv_mass_Tev}.  The two plots show results for the LO
and NLO effective-theory predictions (blue and red bands respectively)
and for the NLO NWA result (green curve). The LO prediction in
the NWA is a delta function centered at the top mass, and in the plot
would show as a spike at $m_{\text{inv}}=m_t$. The blue and red bands
are obtained by varying the scales in the usual interval, $m_t/4 \leq
\mu_R=\mu_F \leq m_t$, in the LO and NLO EFT results.  The total NLO
corrections are very large and negative in the peak region,
$m_{\text{inv}} \sim m_t$, whereas they are still large, but positive,
in the tail region where the cross section is numerically very small
due to the rapid Breit-Wigner fall-off.  The difference between the
red and green curves, which gives the non-factorizable and off-shell
effects, is relatively small far below and above the peak of the
distribution.  However, non-factorizable corrections are large and
negative at the peak (especially for the $s$-channel-like process),
and large and positive immediately above it.  The change in sign of
non-factorizable contributions around $m_{\text{inv}}$ explains the
small difference between the off-shell and on-shell predictions found
in observables which are inclusive with respect to the top-quark
invariant mass, for example the total cross section, discussed above.
As was the case for the total cross section, close to the interesting
region of the peak the scale dependence of the NLO ET result is only
mildly reduced compared to the LO result for the process $p \bar{p}
\rightarrow J_{b} J_{\bar{b}} e^+\dirac{E}_T$ (the seemingly larger
band shown in the central bin of the plot is an effect of the
logarithmic scale), and is actually increased for the $t$-channel-like
signal.

\begin{figure}[t!]
\begin{center}
\includegraphics[angle=-90,width=0.49 \linewidth]{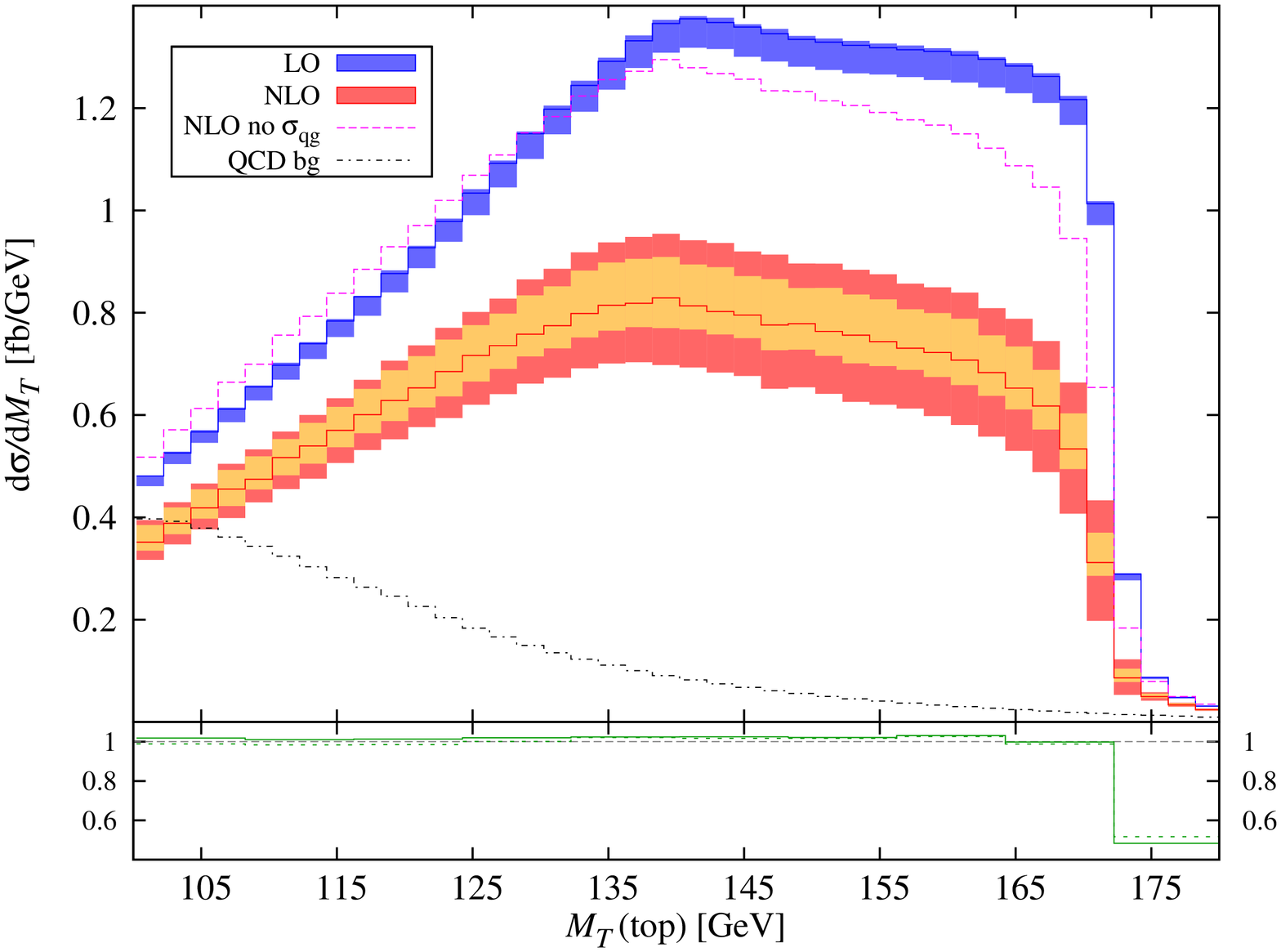}
\includegraphics[angle=-90,width=0.49 \linewidth]{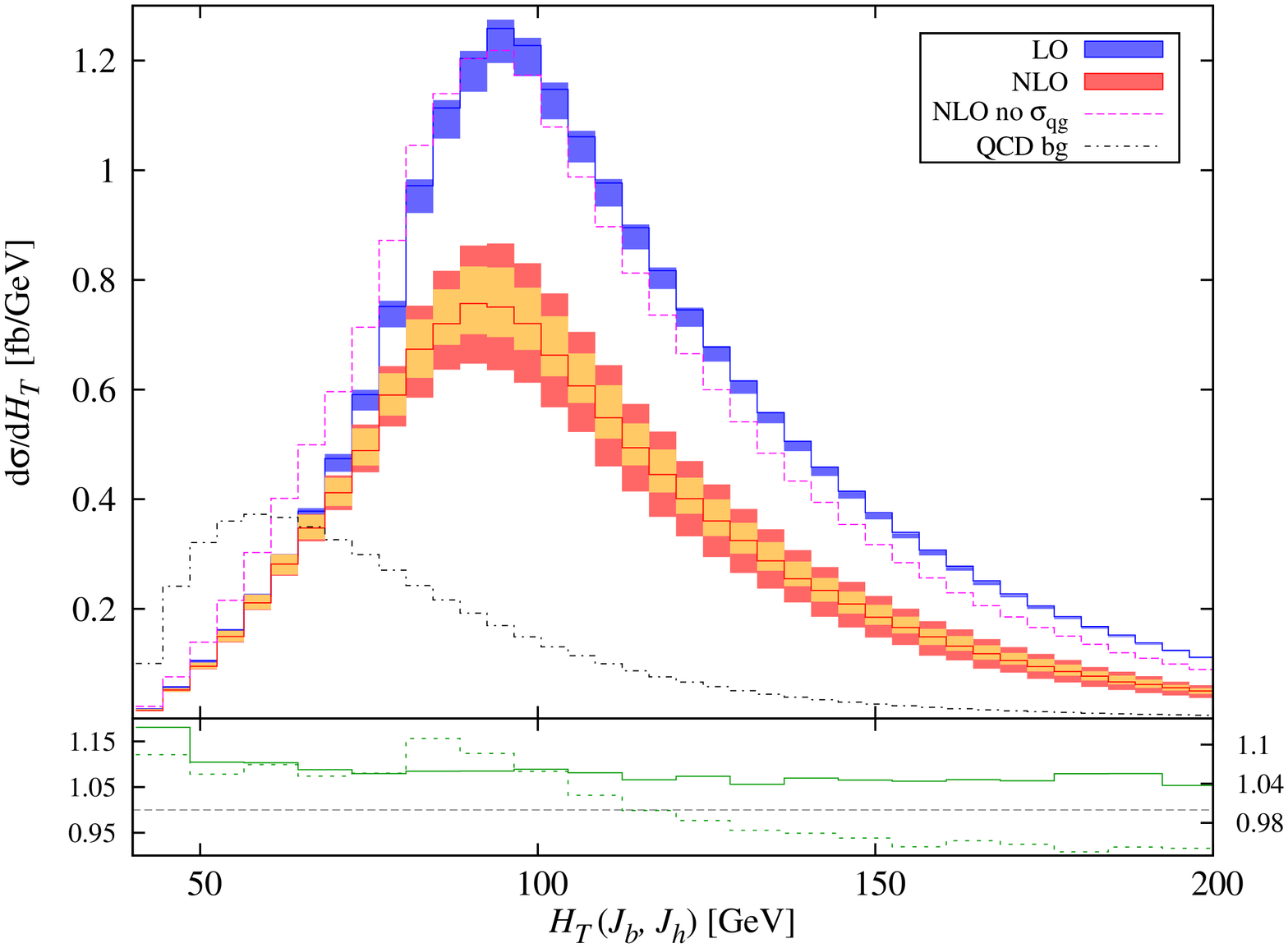}
\includegraphics[angle=-90,width=0.49 \linewidth]{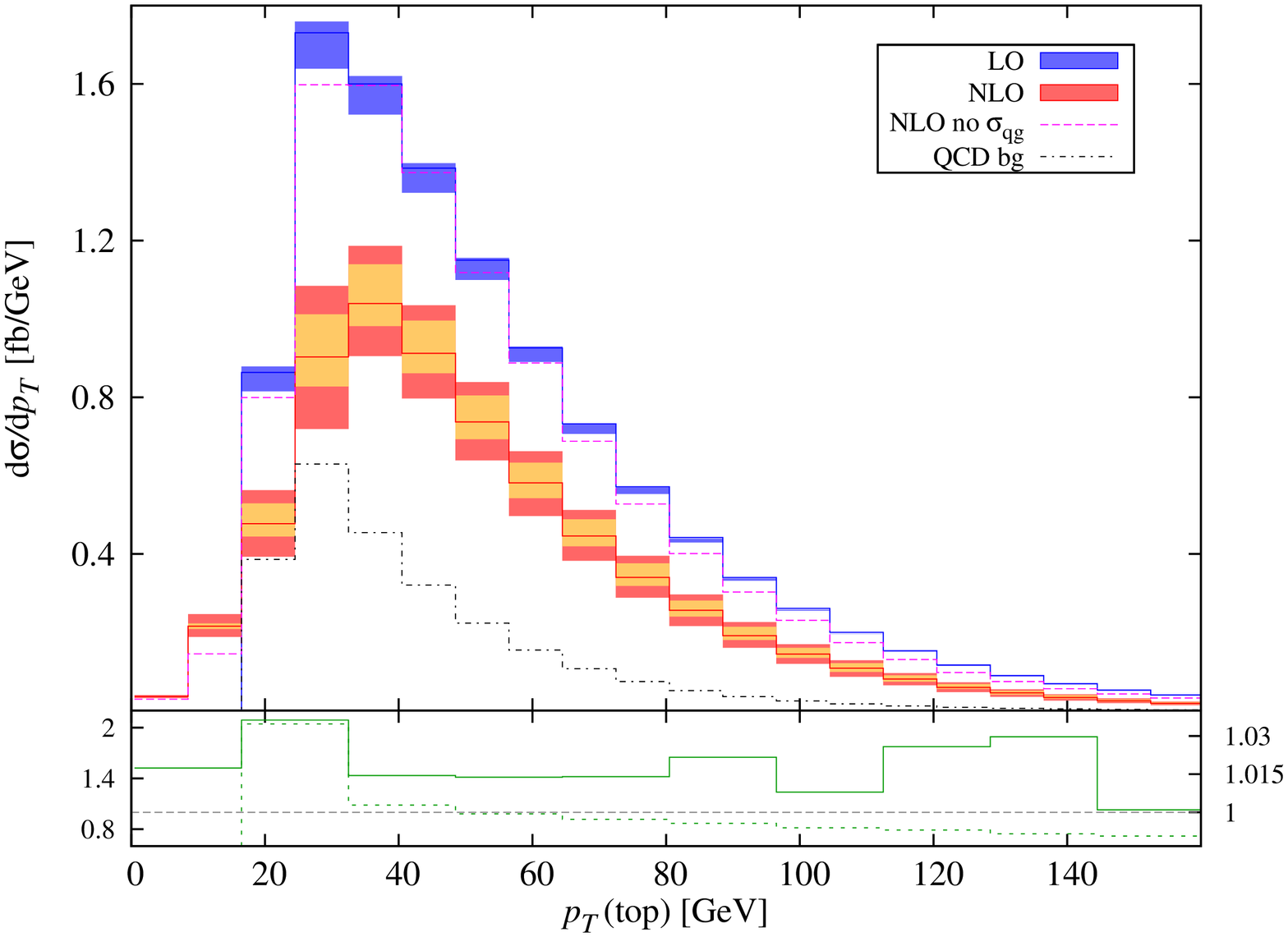}
\includegraphics[angle=-90,width=0.49 \linewidth]{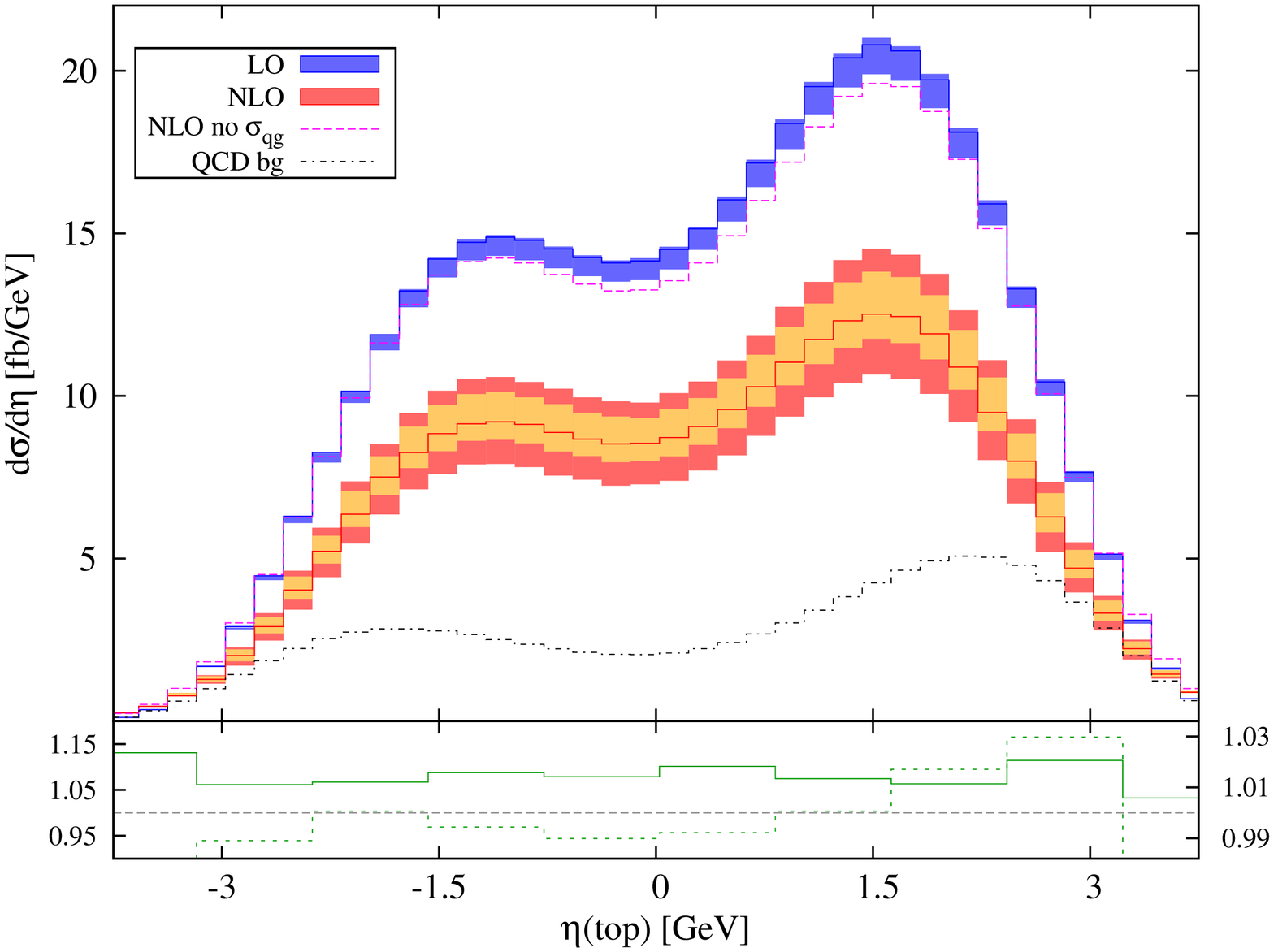}
\includegraphics[angle=-90,width=0.49 \linewidth]{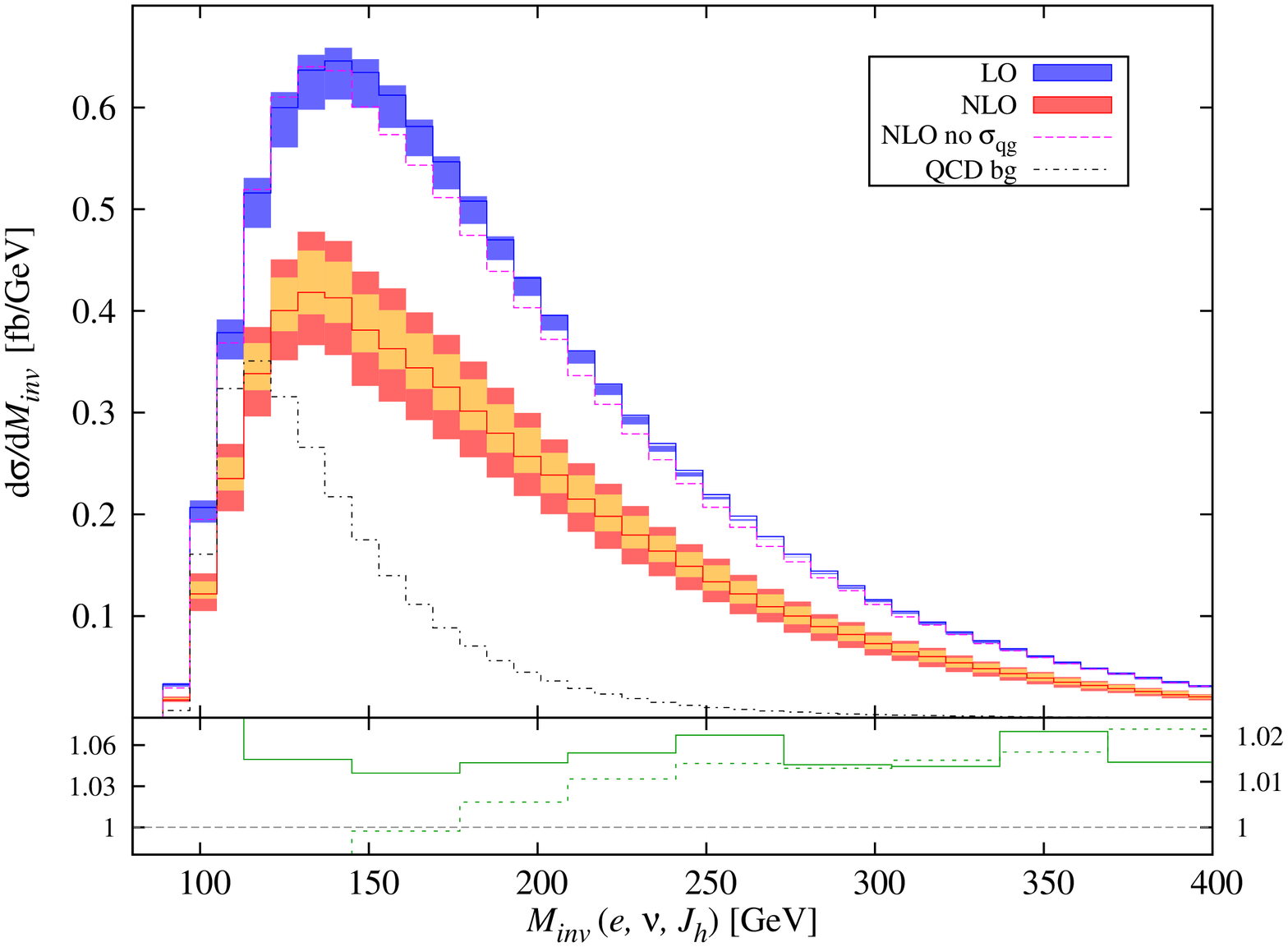}
\includegraphics[angle=-90,width=0.49 \linewidth]{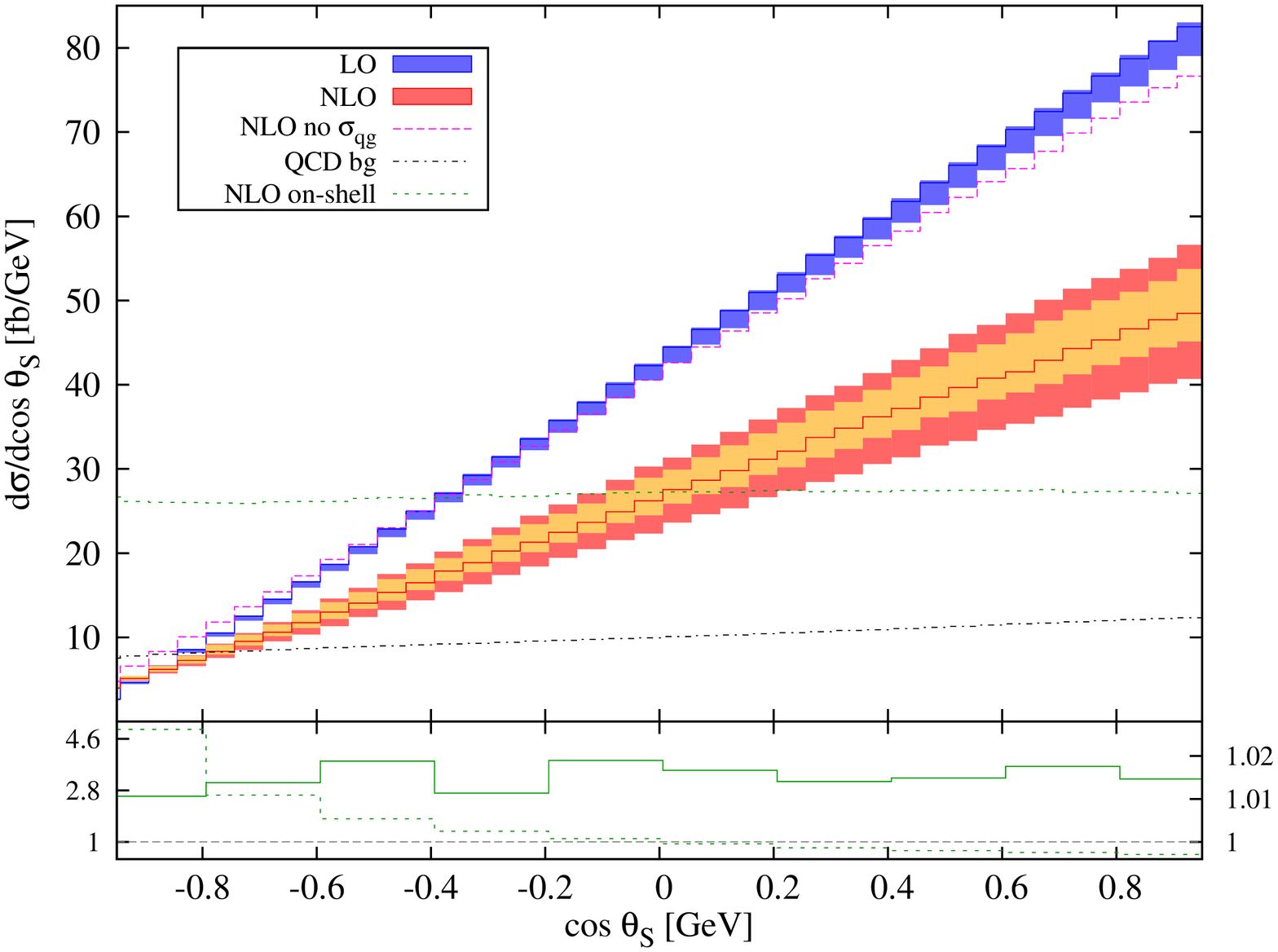}
\ccaption{}{Kinematical distributions for $p \bar{p} \rightarrow J_{b}
  J_{l} e^+\dirac{E}_T+X$ at the Tevatron. Top: top-quark transverse
  mass (left) and hadronic transverse energy (right).  Centre:
  top-quark transverse momentum (left) and top-quark pseudorapidity
  (right).  Bottom: $e^+ \nu J_h$ invariant mass (left) and
  $\cos\theta_S$ (right).  See the text for a precise definition of
  the observables and further explanations.
\label{fig:tchan_dis_Tev}}
\end{center}
\vspace{0.5 cm}
\end{figure}

\begin{figure}[t!]
\begin{center}
\includegraphics[angle=-90,width=0.49 \linewidth]{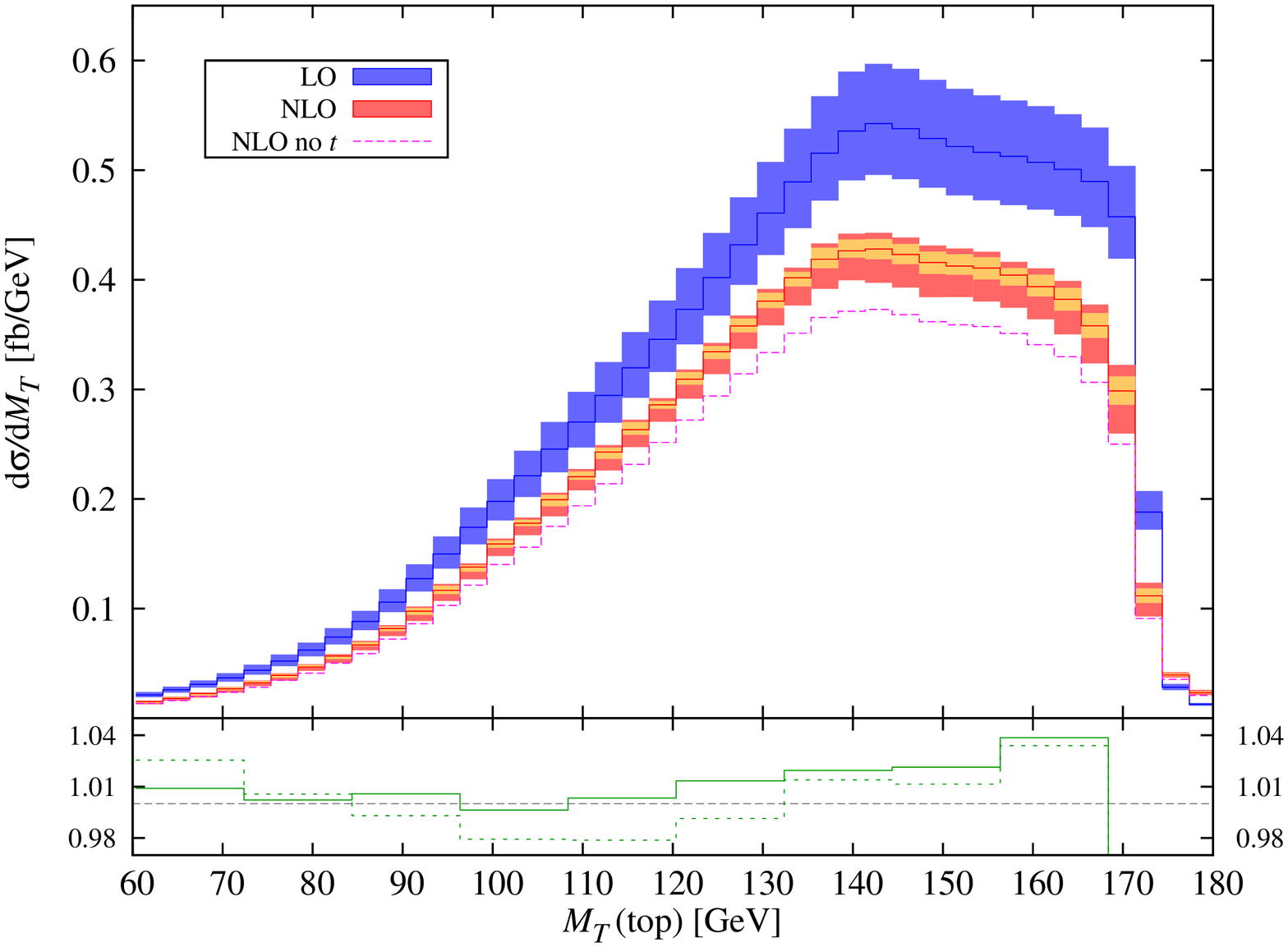}
\includegraphics[angle=-90,width=0.49 \linewidth]{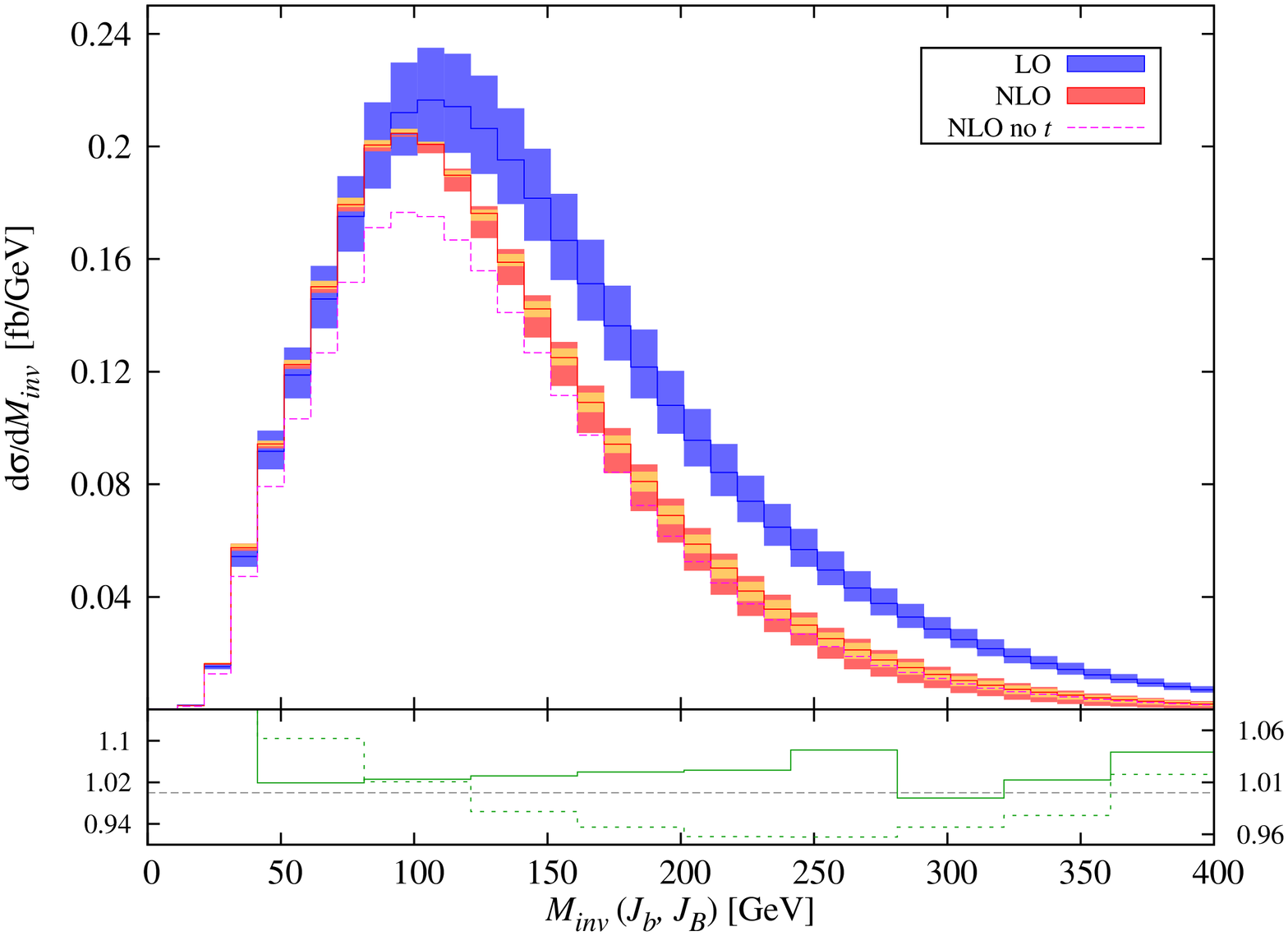}
\includegraphics[angle=-90,width=0.49 \linewidth]{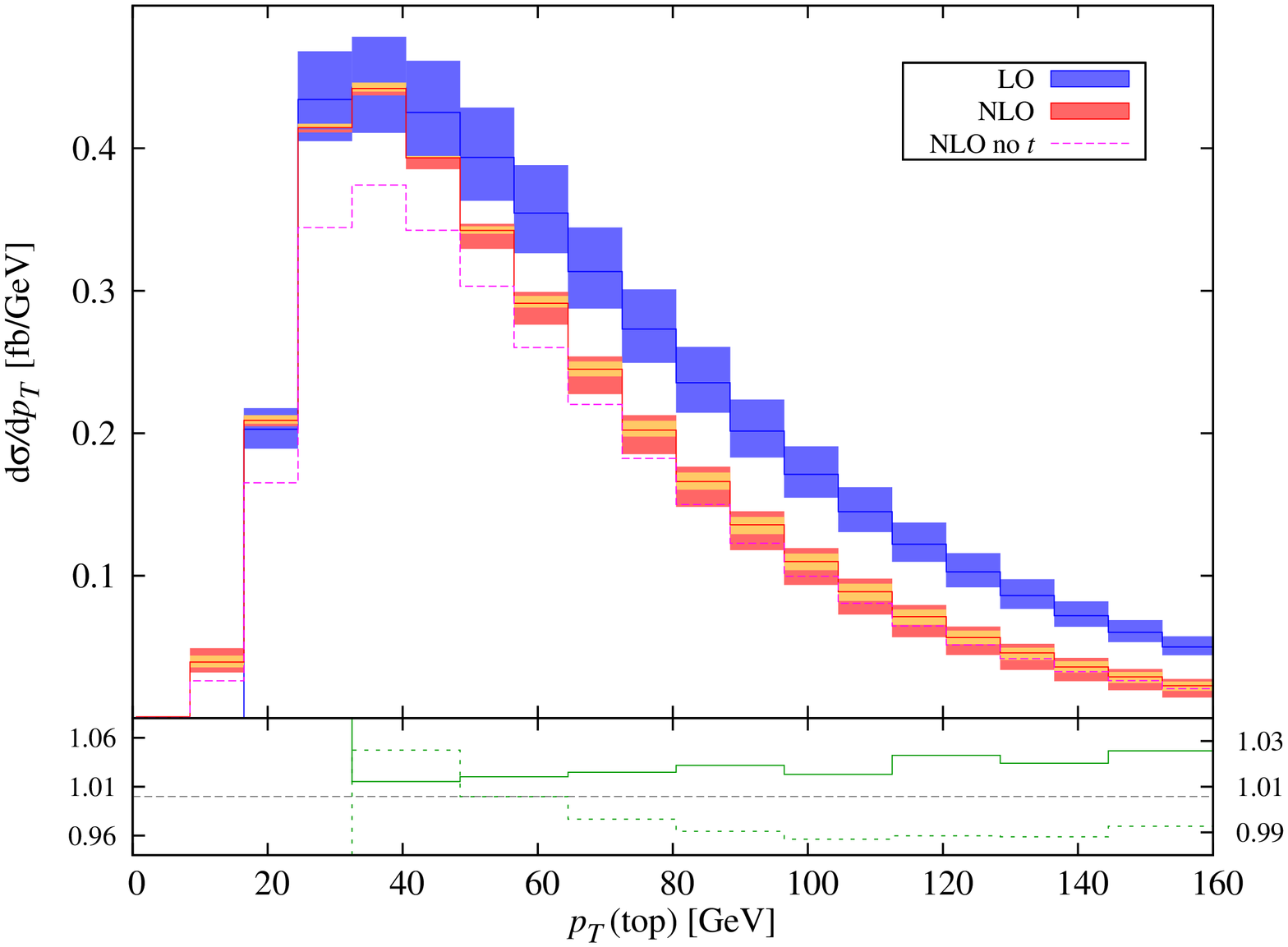}
\includegraphics[angle=-90,width=0.49 \linewidth]{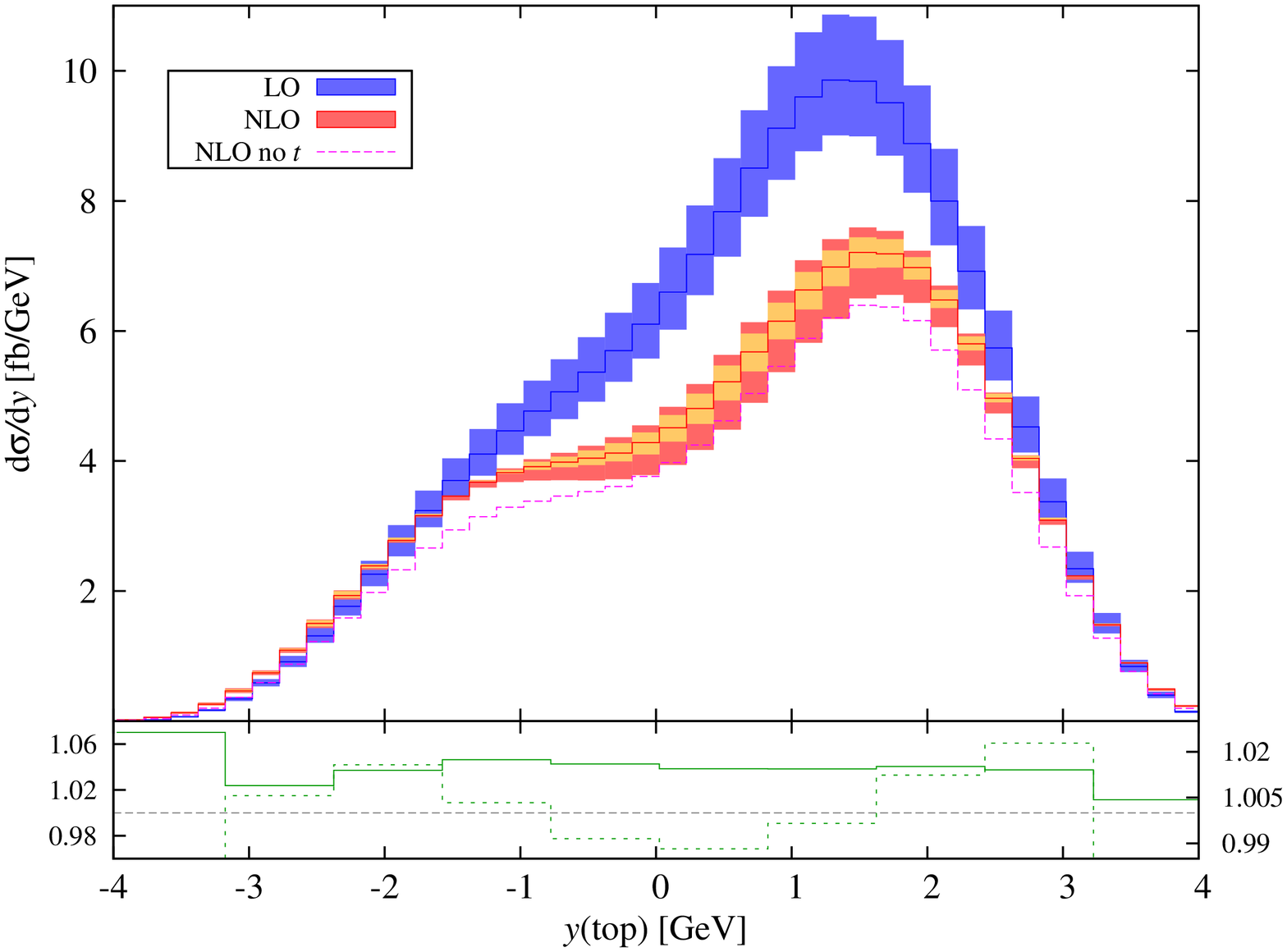}
\includegraphics[angle=-90,width=0.49 \linewidth]{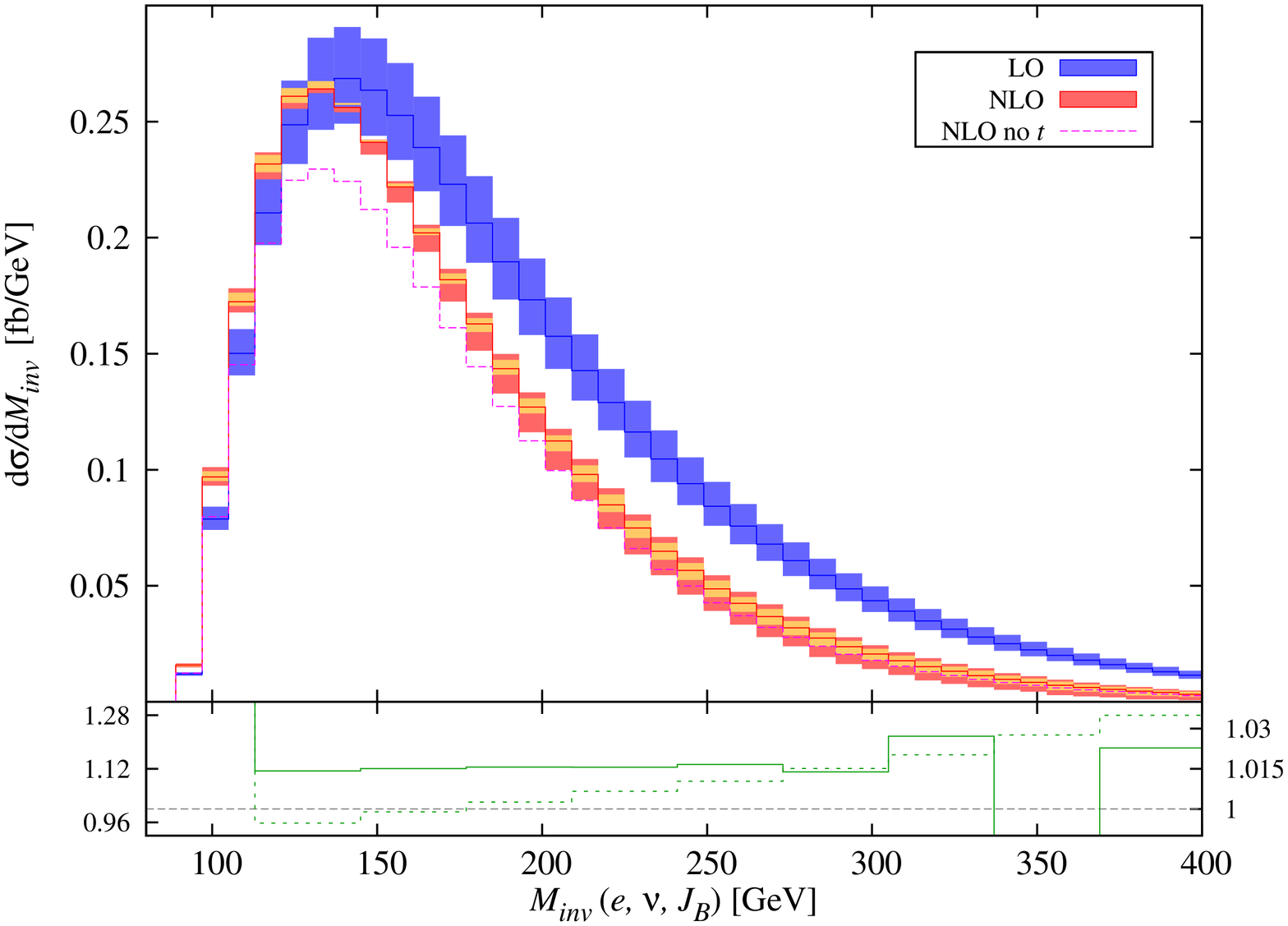}
\includegraphics[angle=-90,width=0.49 \linewidth]{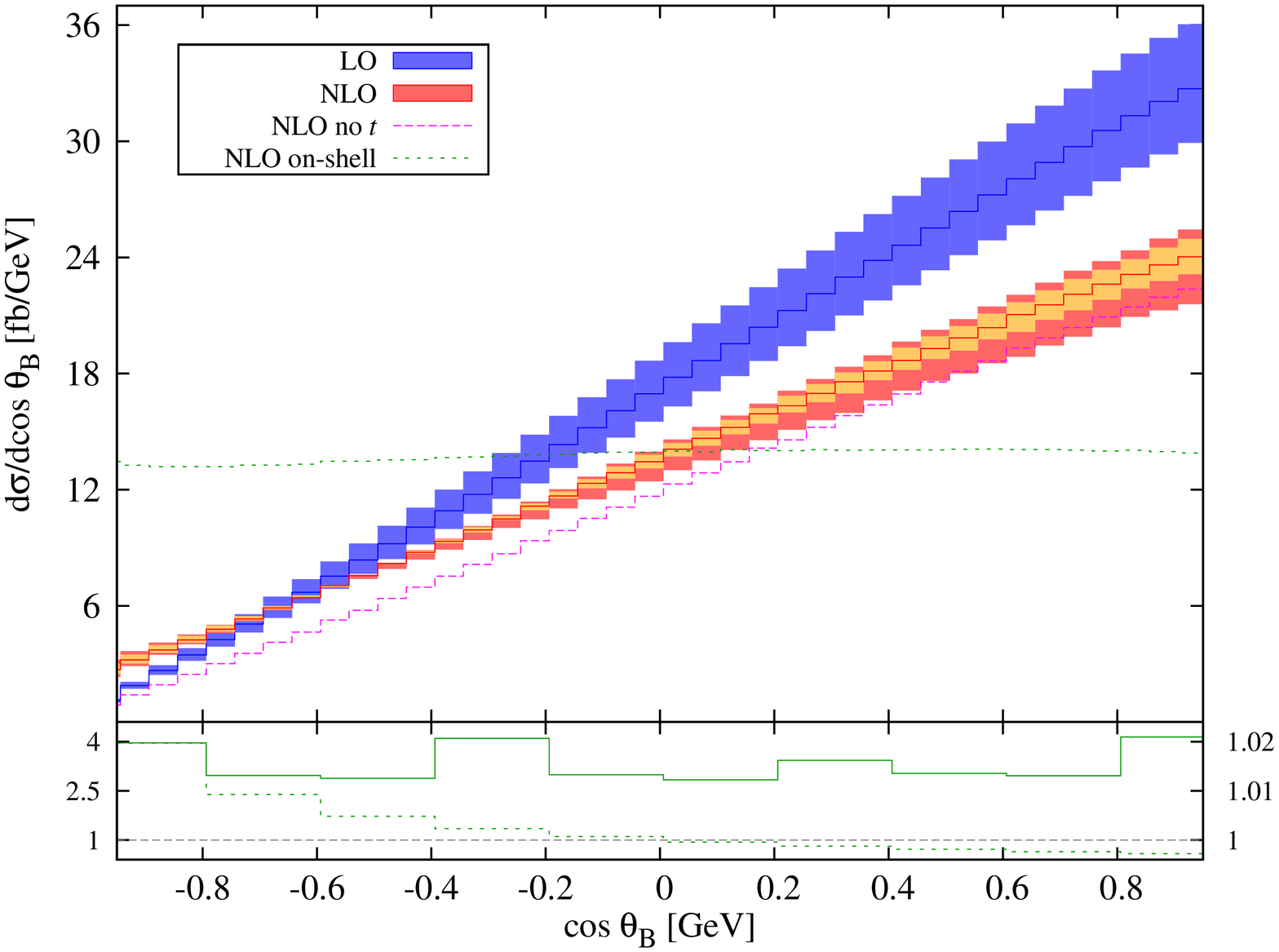}
\ccaption{}{Kinematical distributions for $p \bar{p} \rightarrow J_{b}
  J_{\bar{b}} e^+\dirac{E}_T+X$ at the Tevatron. Top: top-quark
  transverse mass (left) and $J_b J_{\bar{b}}$ invariant mass (right).
  Centre: top-quark transverse momentum (left) and top-quark rapidity
  (right).  Bottom: $e^+ \nu_e J_{\bar{b}}$ invariant mass (left) and
  $\cos\theta_B$ (right).  See the text for a precise definition of
  the observables and further explanations.
\label{fig:schan_dis_Tev}}
\vspace{0.5 cm}
\end{center}
\end{figure}

In Figures~\ref{fig:tchan_dis_Tev} and \ref{fig:schan_dis_Tev} we
present a selection of relevant kinematical distributions. For the
process $p \bar{p} \rightarrow J_{b} J_{l} e^+\dirac{E}_T+X$ (Figure~
\ref{fig:tchan_dis_Tev}) we plot the following observables:
\begin{eqnarray}
&& M_T(t) \equiv \sqrt{ |\vec{p}_\perp(J_b)|^2
+|\vec{p}_\perp(e)|^2+\dirac{E}_T^2
-\left(\vec{p}_\perp(J_b)+\vec{p}_\perp(e)
+\vec{p}_\perp(\nu)\right)^2} \, , \nonumber\\
&& H_T(J_b,J_h) \equiv |\vec{p}_\perp(J_b)|+|\vec{p}_\perp(J_h)| \, ,
\nonumber\\ 
&& p_T(t) \equiv 
|\vec{p}_\perp(e)+\vec{p}_\perp(\nu)+\vec{p}_\perp(J_b)| \, , \nonumber\\
&& \eta(t) \equiv \frac{1}{2} 
\ln \left( \frac{|\vec{p}(t)|+|\vec{p}_{\parallel}(t)|}
{|\vec{p}(t)|-|\vec{p}_{\parallel}(t)|}\right) \, , \nonumber\\
&& M_\text{inv}(e,\nu,J_h) \equiv 
\sqrt{(p(e)+p(\nu)+p(J_h))^2} \, , \nonumber\\
&& \cos\theta_S \equiv 
\frac{\vec{p}(e) \cdot \vec{p}(J_h)}{|\vec{p}(e)| |\vec{p}(J_h)|}
\Bigg \vert_{\mbox{top  r.f.}} \, ,
\end{eqnarray}
where $\vec{p}(t) = \vec{p}(J_b) + \vec{p}(e) + \vec{p}(\nu)$, the
lower indices $\perp$ and $\parallel$ denote the momentum components
perpendicular and parallel to the beam axis, $\vec{p}(\nu)$ and
$\vec{p}_\perp(\nu)$ represent the missing total and missing
transverse momenta respectively, and $J_h$ is the hardest,
non-$b$-tagged jet. $M_T(t)$ represents the top-quark transverse mass,
$H_T(J_b,J_h)$ the hadronic transverse energy, $p_T(t)$ the transverse
momentum of the reconstructed top quark, $\eta(t)$ the pseudorapidity
of the top quark, $M_\text{inv}(e,\nu,J_h)$ the invariant mass of the
$e^+ \nu_e J_h$ system, and $\cos\theta_S$ the angle between the
charged lepton and the hardest light jet in the rest frame of the top
quark. In all histograms the dark blue solid line represents the LO ET
result for the central value of the scales, whereas the NLO result is
given by the red solid line. The light blue and red bands are obtained
by varying the factorization and renormalization scales in the
interval $m_t/4 \leq \mu_R=\mu_F \leq m_t$, while the orange band
comes from varying the factorization scale only in the NLO results,
keeping the renormalization scale fixed. The dashed magenta curve is
the (unphysical) NLO result obtained by omitting the processes with a
quark and gluon in the partonic initial state and both a $b$ and
$\bar{b}$ quark in the final state. This artificially removes any
processes which could lead to a $\bar{b}$ jet in the final state, even
if that jet would not be resolved, and acts as a perfect $\bar{b}$-jet
veto. The black dot-dashed curve represents the (tree-level) QCD
background.

The total NLO corrections are generally large and negative for the
cuts used here, reaching up to $50\%$ of the LO result depending on
the distribution and on the bin chosen, though the shape of the
distributions is usually only mildly modified. The bulk of the
correction arises from the gluon-initiated partonic channels, as
follows from comparing the solid red and dashed magenta lines.
Non-factorizable corrections are typically small. This is shown, in
detail, in the bottom plots in Figure~\ref{fig:tchan_dis_Tev}. Here
the ratio of the NWA result and the resonant result is given for two
different implementations of the narrow-width approximation, one with
exact spin correlations (solid green curve, right scale) and one
without (dashed green curve, left scale). Off-shell effects amount to
about $2-3 \%$, except close to the edges of particular distributions
where they can be much larger (see, e.g., the plot for $M_T(t)$). It
is interesting to note that spin-correlation effects can be much more
important than non-factorizable corrections. This is the case for the
$p_T(t)$ distribution and for $\cos\theta_S$, whereas for other
observables the difference between the two implementations of the NWA
is small. To further illustrate the size of the spin-correlation
effects in the $\cos\theta_S$ distribution, the NLO
narrow-width-approximation result without spin-correlation is shown in
Figure~\ref{fig:tchan_dis_Tev} (dashed green curve, bottom right
plot). Comparing this curve to the ET NLO curve (red) shows a dramatic
change in the shape of the distribution.

Non-factorizable effects are much smaller than one would na\"{i}vely
expect from the result found for the invariant-mass distribution,
where off-shell corrections at the peak are sizeable. This is partly
due to an averaging effect which stems from the fact that, for a fixed
value of a given observable, ${O}$, the top invariant mass,
$m_\text{inv}$, can, in general, vary over a large range of values.
Consider, as an example, the top transverse-mass distribution,
$M_T(t)$. For a fixed-value of $M_T(t)$, the top-quark invariant mass
varies in the interval $M_T(t)<m_{\text{inv}}<200$~GeV, where the
upper cut-off follows from the cuts we imposed. Because of the change
in sign of the off-shell effects around the peak at
$m_{\text{inv}}=m_t$, for $M_T(t)$ smaller than the top mass, large
cancellations between positive and negative corrections take place,
leading to small non-factorizable corrections for $M_T(t) \lesssim
160$~GeV. On the other hand, if $M_T(t) \gtrsim m_t$ the cancellations
are less effective, explaining the relatively large off-shell effects
close to the distribution edge. A similar reasoning can also be
applied to other observables.

In the plots in Figure~\ref{fig:tchan_dis_Tev} we also show the
contribution of QCD background diagrams, represented by the black
dot-dashed curve. In the effective-theory counting these terms are of
${\cal O}(\delta^2)$, i.e. subleading compared to the ${\cal
  O}(\delta^{3/2})$ non-factorizable corrections, and are thus
expected to be numerically negligible. While this is true for some
observables, like, for example, $M_T$ and $H_T$, where the shape of
the distribution for signal and background are qualitatively very
different and peak in different kinematical ranges. For others, like
$p_T(t)$ and $\eta(t)$, the shape of the QCD background is similar to
the signal, and its size comparable to the contribution of NLO
corrections. For $\cos\theta_S$ the QCD background contributes a
constant shift of the distribution. The apparent breakdown in the
effective-theory counting could possibly be explained by the fact that
the relation $\alpha_s^2\sim \alpha_{ew} \sim (p_t^2-m_t^2)/m_t^2$ is
only approximately satisfied since, for the invariant-mass cuts
adopted here, $|(p_t^2-m_t^2)/m_t^2| \sim \delta$ can be as big as
$0.5$.

A set of interesting observables for the process $p \bar{p}
\rightarrow J_{b} J_{\bar{b}} e^+\dirac{E}_T+X$ are plotted in 
Figure~\ref{fig:schan_dis_Tev}. Besides $M_T(t)$ and $p_T(t)$, for the 
$s$-channel-like signal we present histograms for:
\begin{eqnarray}
&& M_{\text{inv}}(J_b,J_{\bar{b}}) \equiv 
\sqrt{(p(J_b)+p(J_{\bar{b}}))^2} \, , \nonumber\\
&& y(t) \equiv 
\frac{1}{2} \ln  \left( \frac{E_t+|\vec{p}_{\parallel}(t)|}
           {E_t-|\vec{p}_{\parallel}(t)|}\right) \, ,\nonumber\\
&& M_\text{inv}(e,\nu,J_{\bar{b}}) \equiv 
\sqrt{(p(e)+p(\nu)+p(J_{\bar{b}}))^2} \, ,\nonumber\\
&& \cos\theta_B \equiv \frac{\vec{p}(e) \cdot \vec{p}(\bar{p})}
{|\vec{p}(e)| |\vec{p}(\bar{p})|} \Bigg \vert_{\mbox{top  r.f.}} \, ,
\end{eqnarray}
where $E_t$ denotes the energy of the reconstructed top quark and
$\vec{p}(\bar{p})$ is the three-momentum of the incoming
antiproton. $M_{\text{inv}}(J_b,J_{\bar{b}})$ is the invariant mass of
the two $b$-tagged jets, $y(t)$ is the rapidity of the top quark,
$M_\text{inv}(e,\nu,J_{\bar{b}})$ represents the mis-reconstructed
mass of the top quark (i.e. the invariant mass obtained using the tagged
bottom quark which does not come from the decay of the top quark) and
$\cos\theta_B$, the angle between the final-state lepton and the
antiproton beam in the top-quark rest frame. As before, the solid blue
and red curves represent the LO and NLO resonant result, while the
three bands are obtained by scale variations, as in the case of the
$t$-channel-like process. In this case, the dashed magenta line gives
the NLO prediction without the contribution of $t$-channel-like diagrams,
while, as before, the lower plots show the ratio of the two
implementations of the NWA and the effective-theory result at NLO.

For $p \bar{p} \rightarrow J_{b} J_{\bar{b}} e^+\dirac{E}_T+X$, the total NLO
corrections are smaller than for the $t$-channel-like process, and about
$10-20 \%$ of the LO result for most distributions in the peak region.
Contrary to $p \bar{p} \rightarrow J_{b} J_{l} e^+\dirac{E}_T+X$, 
the scale dependence of the NLO result is reduced compared to the Born 
prediction, though only mildly. As pointed out for the total cross section, 
the contribution of $t$-channel diagrams is quite important, even after cuts, 
and comparable to the effect of NLO corrections for some distributions
(e.g. $M_{\text{inv}}(J_b,J_{\bar{b}})$, $p_T(t)$ and
$M_\text{inv}(e,\nu,J_{\bar{b}})$). The non-factorizable corrections
are again generally small, and usually in the $1-3\%$ range. Once more
we point out that the two implementations of the NWA, with and without
exact spin correlations, are significantly different for observables
involving angles, like $\cos\theta_B$, whereas spin-correlation
effects are small for most of other distributions.

\subsection{Single-top production at the $7$ TeV LHC} \label{sec:LHC}

\begin{table}
  \begin{center}
  \begin{tabular}{l|r}

  \hline
  \hline
  $p p \rightarrow J_{b} J_{l} e^+\dirac{E}_T+X$ & 
  $p p \rightarrow J_{b} J_{\bar{b}} e^+\dirac{E}_T+X$ \\[2pt]
  \hline
  \hline
  $p_T(J_b) > 20$~GeV & $p_T(J_b) > 20$~GeV \\[2pt]
   $p_T(\text{hardest} \, J_l) > 20$~GeV & 
   $p_T(J_{\bar{b}}) > 30$~GeV \\[2pt]
      $p_T(\text{extra} \, J_{\bar{b}}) < 15$~GeV & 
   $p_T(\text{extra} \, J_l) < 15$~GeV \\[2pt]
   $\dirac{E_T} +p_T(e)> 60$~GeV & $\dirac{E_T} +p_T(e)> 60$~GeV \\[2pt]
   $120 < m_{{\rm inv}} < 200$~GeV  & $120 < m_{{\rm inv}} < 200$~GeV \\
  \hline
  \hline

  \end{tabular}
  \end{center}
  \ccaption{}{Kinematical cuts and vetoes used for LHC
    results. \label{tab:LHC}}
\end{table} 
In this section we presents results for single-top production at the
LHC at a centre-of-mass energy of $7$ TeV. Table~\ref{tab:LHC} shows
the kinematical cuts and vetoes applied to the two processes, $p p
\rightarrow J_{b} J_{l} e^+\dirac{E}_T+X$ and $p p \rightarrow J_{b}
J_{\bar{b}} e^+\dirac{E}_T+X$, in this case. The constraints are very
similar to the ones used for Tevatron, except for harder cuts on the
transverse missing energy and transverse lepton momentum.

\begin{table}[t!]
  \begin{center}
  \begin{tabular}{l|c|c|c}

  \hline
  \hline
  $p p \rightarrow J_{b} J_{l} e^+\dirac{E}_T+X$ & \hspace{1 cm} &
      {\bf ET} & {\bf NWA}  \\[2pt] 
  \hline \phantom{$\left(\frac{1}{2}\right)^X$}
   & {\bf LO}[pb] & $3.460(1)^{+0.278}_{-0.403} $ & $3.505(1)$ \\[2pt]
  \hline \phantom{$\left(\frac{1}{2}\right)^X$}
   &  {\bf NLO}[pb] & $1.609(6)^{+0.303}_{-0.240}$ & $1.642(1)$ \\[2pt]
  \hline
  \hline
  $p p \rightarrow J_{b} J_{\bar{b}} e^+\dirac{E}_T+X$ & \hspace{1 cm}
  & \hspace{0.3 cm} {\bf ET} \hspace{0.3 cm}& {\bf NWA} \\[2pt]   
 \hline \phantom{$\left(\frac{1}{2}\right)^X$}
   & {\bf LO}[pb] & $0.1654(1)^{+0.0001}_{-0.0010}$ & $0.1677(1)$ \\[2pt]
  \hline \phantom{$\left(\frac{1}{2}\right)^X$}
   &  {\bf NLO}[pb] & $0.1618(4)^{+0.0021}_{-0.0005}$ & $0.1635(1)$ \\[2pt]
  \hline
  \hline
  \end{tabular}
  \end{center}
  \ccaption{}{LO and NLO cross sections for the processes
    (\ref{eq:proc1}) and (\ref{eq:proc2}), computed using the parameters 
    given in Table~\ref{tab:input} and imposing the kinematical cuts and 
    vetoes given in Table~\ref{tab:LHC}. The errors come
    from scale uncertainty only.  All numbers are in
    picobarns. \label{tab:totalLHC}}
\end{table}
\begin{figure}[t!]
\begin{center}
\includegraphics[width=0.49 \linewidth]{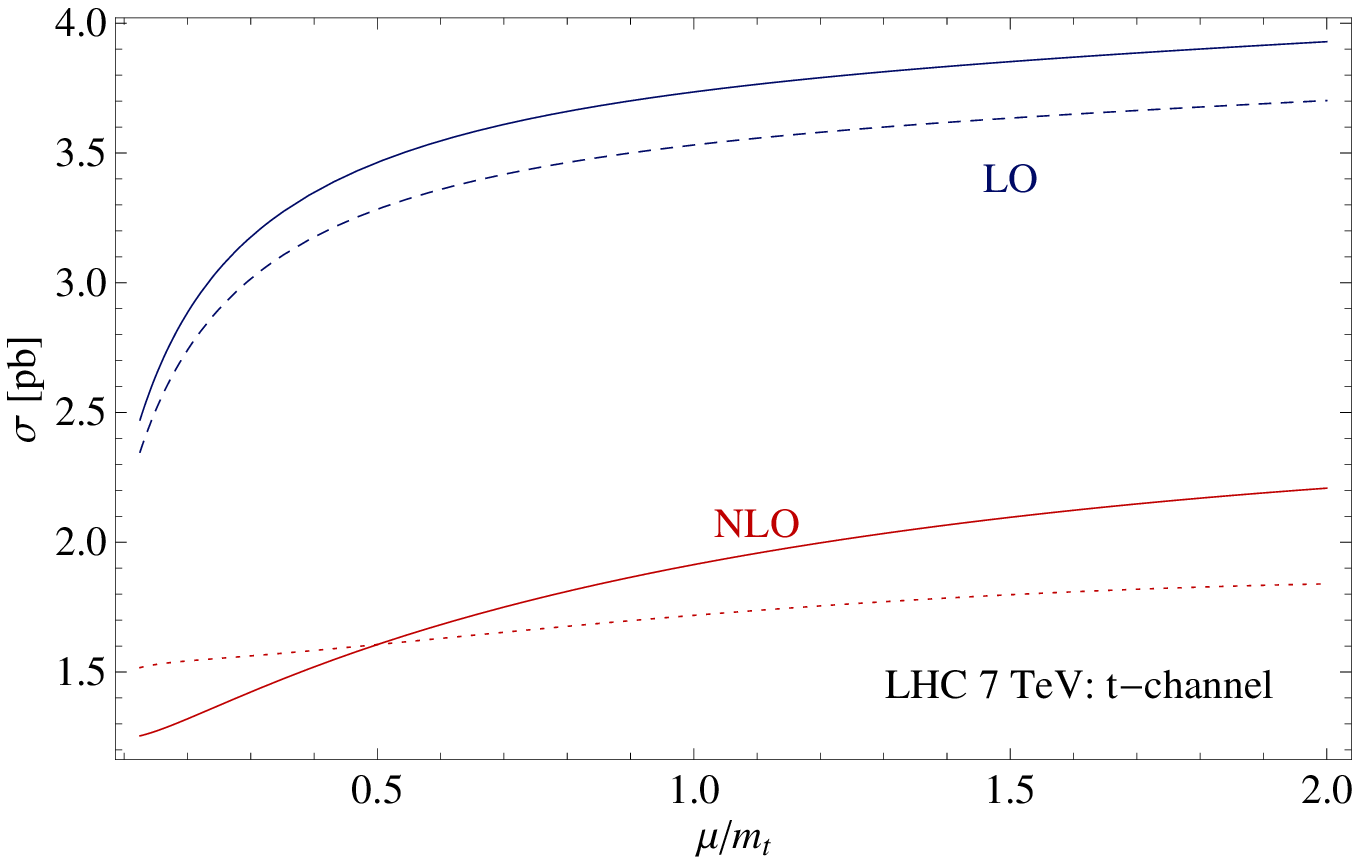}
\includegraphics[width=0.49 \linewidth]{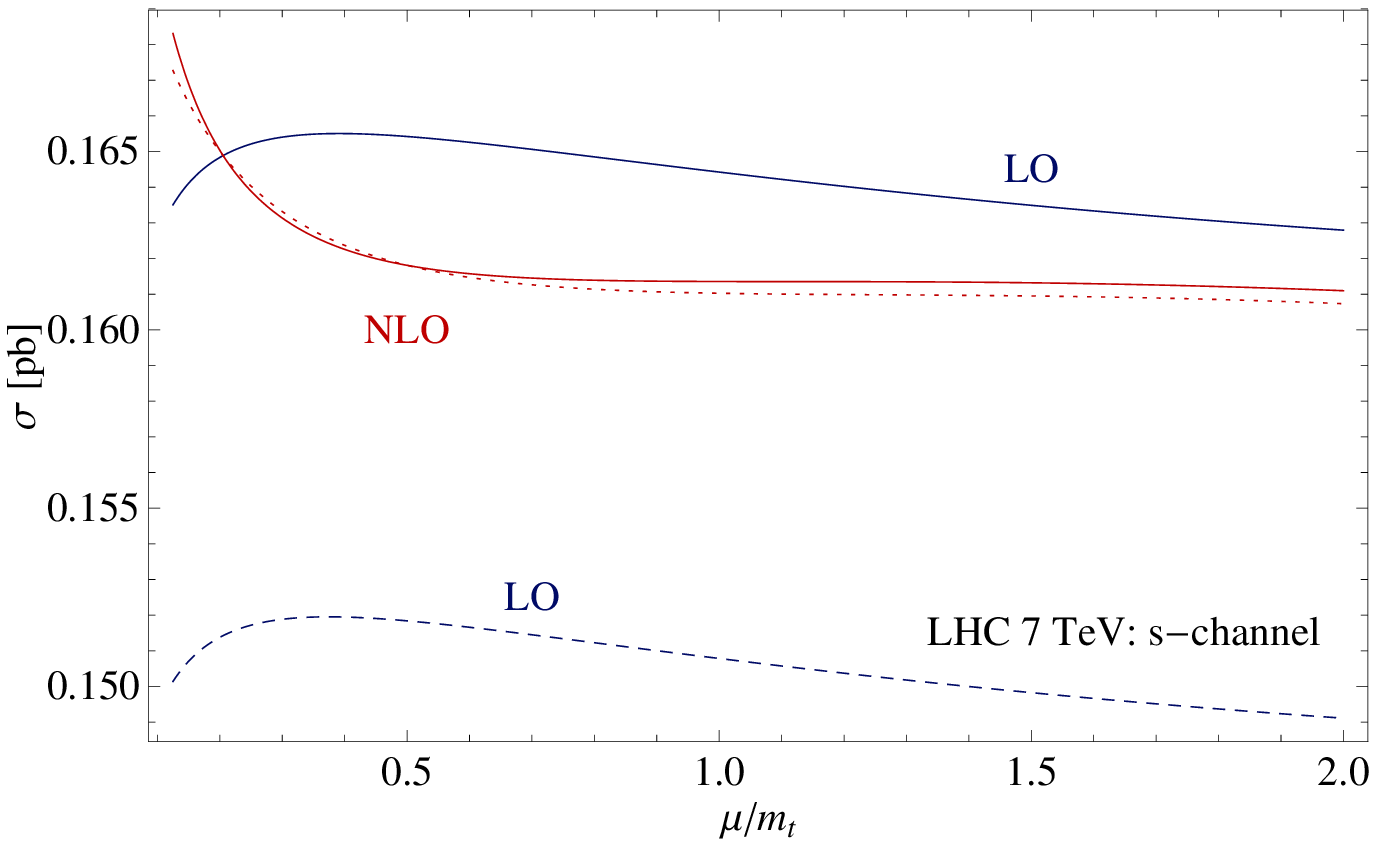}
\end{center}
\ccaption{}{Scale dependence of the total cross section for 
  $pp \rightarrow J_{b}J_{l} e^+\dirac{E}_T$ (left) and 
  $pp \rightarrow J_{b} J_{\bar{b}} e^+\dirac{E}_T$
  (right) at the 7~TeV LHC. The plot shows the LO cross section with LO
  (dashed blue) and NLO (solid blue) PDFs, and the NLO cross section
  with simultaneous variation of factorization and renormalization
  scale (solid red) and for fixed factorization scale (dashed
  red). \label{fig:scaleLHC}}
\end{figure}
\begin{figure}[t!]
\begin{center}
\includegraphics[width=0.7 \linewidth]{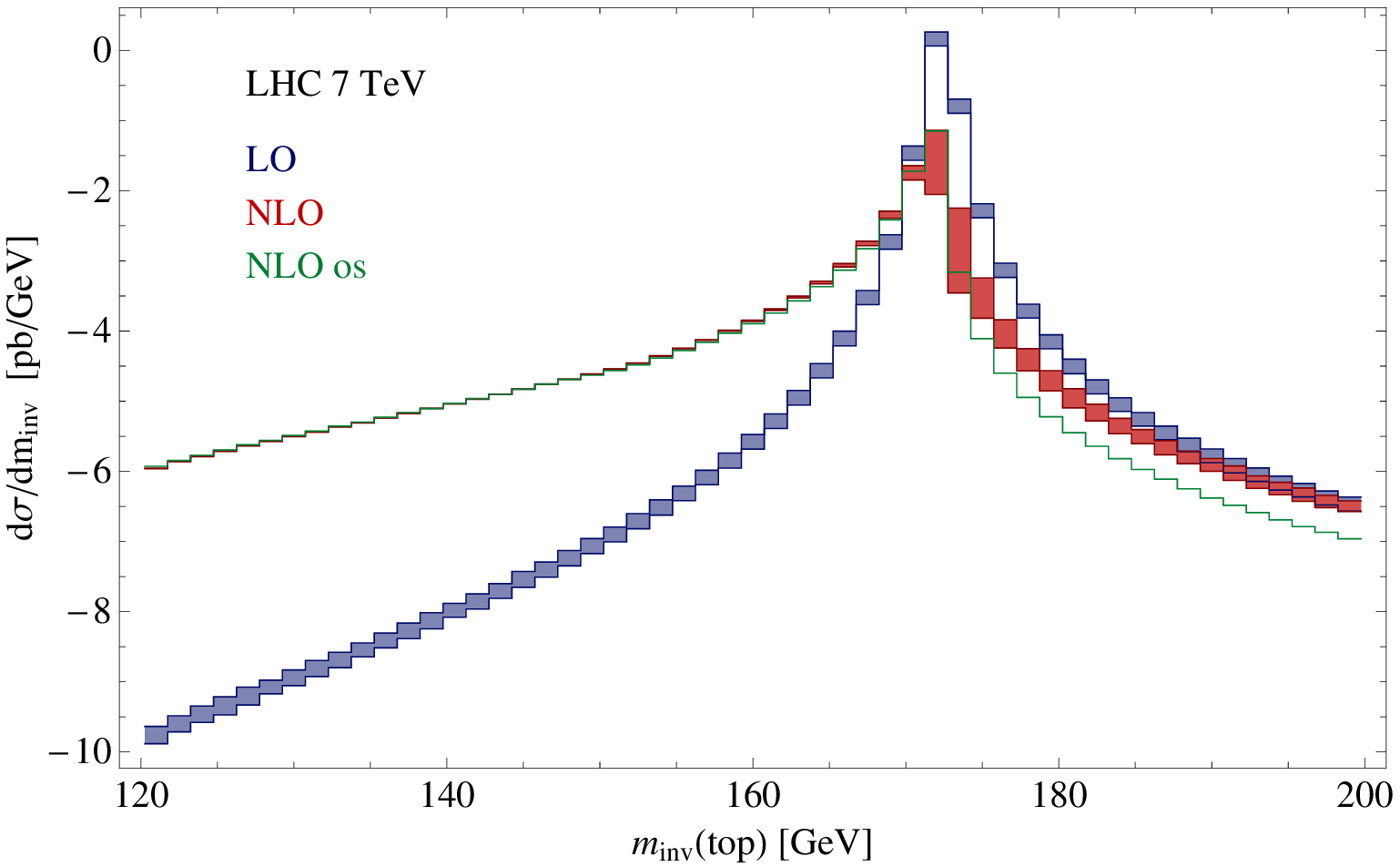}
\includegraphics[width=0.7 \linewidth]{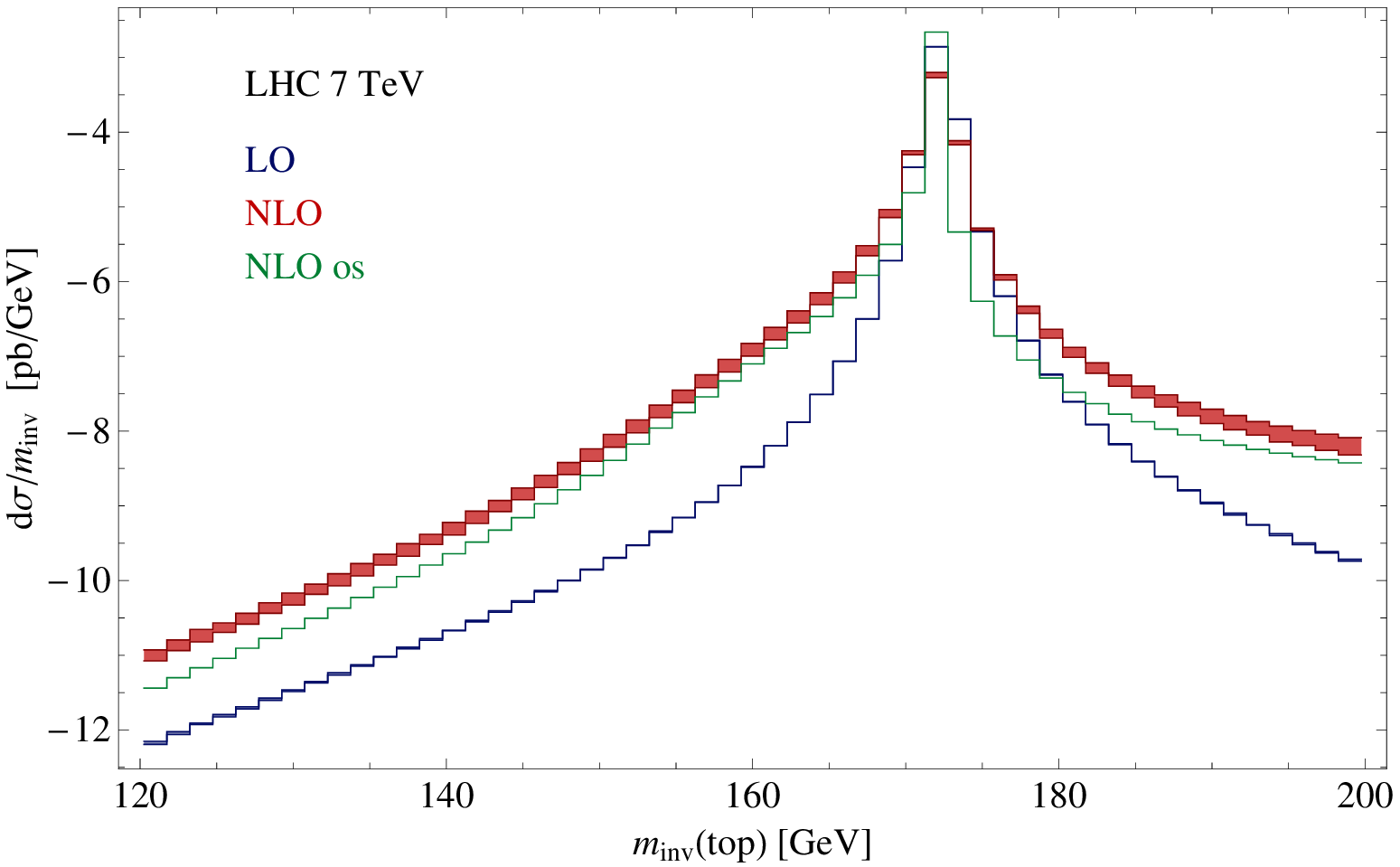} \ccaption{}{Top
  invariant-mass distributions for the process $pp \rightarrow
  J_{b}J_{l} e^+\dirac{E}_T$ (upper plot) and $pp \rightarrow J_{b}
  J_{\bar{b}} e^+\dirac{E}_T$ (lower plot) at the 7~TeV LHC. The blue
  band represents the LO ET result, the red band the NLO ET result,
  and the green curve the NLO spin-correlated NWA prediction. For the
  ET results the band width is obtained by varying the factorization
  and renormalization scales in the interval $m_t/4 \leq \mu_R = \mu_F
  \leq m_t$.
\label{fig:inv_mass_LHC}}
\end{center}
\end{figure}

Results for the total cross sections are given in Table~
\ref{tab:totalLHC}.  The total NLO corrections amount to $\sim -53 \%$
for the $t$-channel-like process, whereas for $p p \rightarrow J_{b}
J_{\bar{b}} e^+\dirac{E}_T+X$ they are very small ($\sim -2.5 \%$). As
for the Tevatron, the non-factorizable corrections are small ($\sim
1\%$), and confirm the na\"{i}ve expectation $\sim \Gamma_t/m_t$. Of
the total NLO cross section for $pp \rightarrow J_{b} J_{l}
e^+\dirac{E}_T+X$, only $0.5 \%$ arises from $s$-channel diagrams. On
the contrary, for $pp \rightarrow J_{b} J_{\bar{b}} e^+\dirac{E}_T+X$,
$t$-channel diagrams contribute about $59\%$ of the total NLO cross
section. This is a consequence of the much larger cross section of
$t$-channel single-top production compared to $s$-channel production
at the LHC. For the $t$-channel-like signal the scale dependence is only
mildly reduced at NLO. Again, this can be partly explained by the
additional renormalization scale dependence introduced at this order,
as is clear from the left plot in Figure~\ref{fig:scaleLHC}.  For the
$s$-channel-like signal, the scale dependence is increased at NLO. In
this case, this can be explained by our default choice for the scales,
$m_t/4 \leq \mu_R = \mu_F \leq m_t$, which is very close to the region
where the scale dependence of the NLO cross section is the strongest
and the scale dependence of the Born cross section the weakest. From
the right plot in Figure~\ref{fig:scaleLHC} it can be clearly seen
that above $\mu_R=\mu_F=0.75 \, m_t$ the scale dependence of the NLO
result is instead very mild, and much flatter than the LO result.

\begin{figure}[h!]
\begin{center}
\includegraphics[angle=-90,width=0.49 \linewidth]{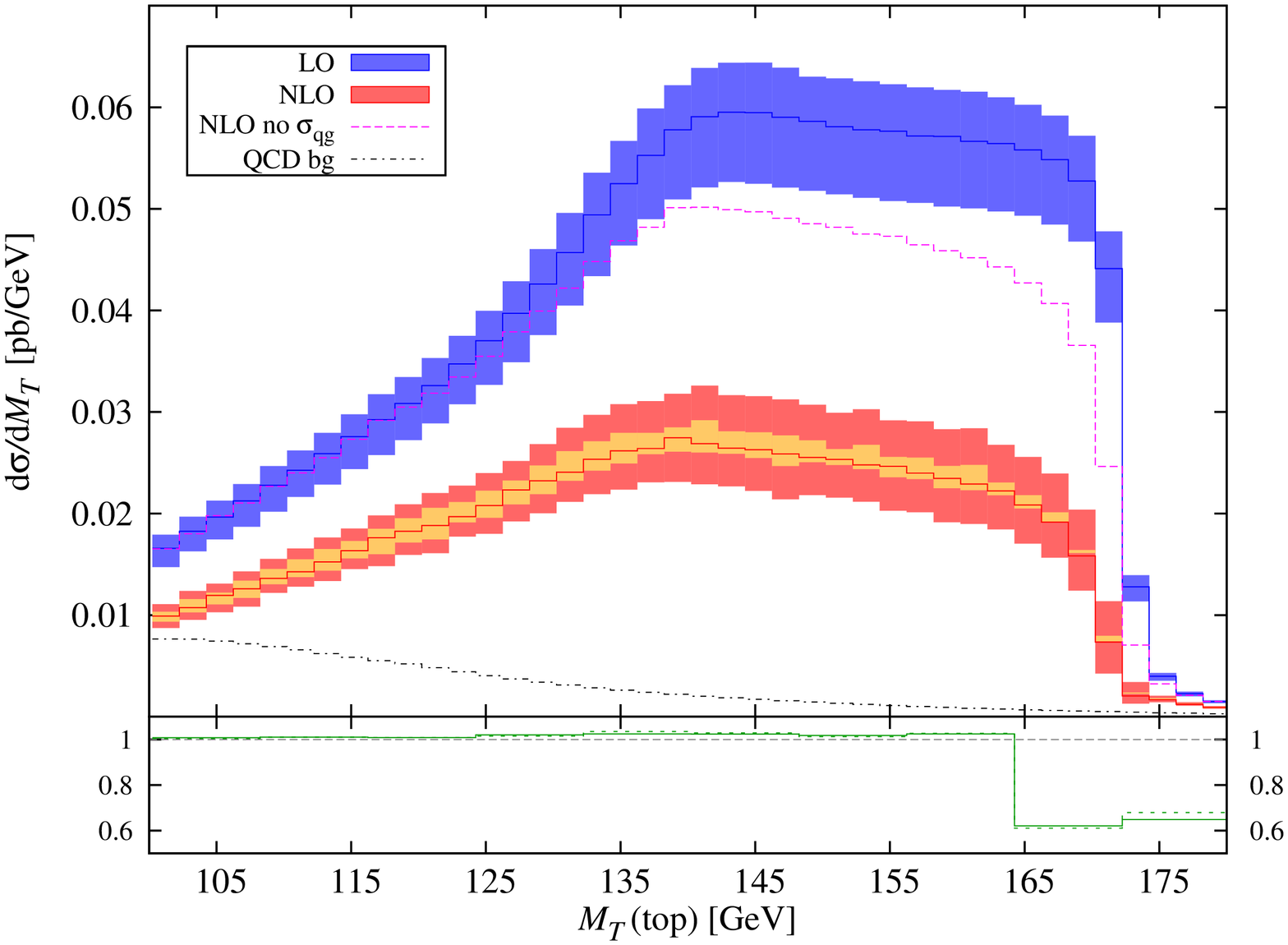}
\includegraphics[angle=-90,width=0.49 \linewidth]{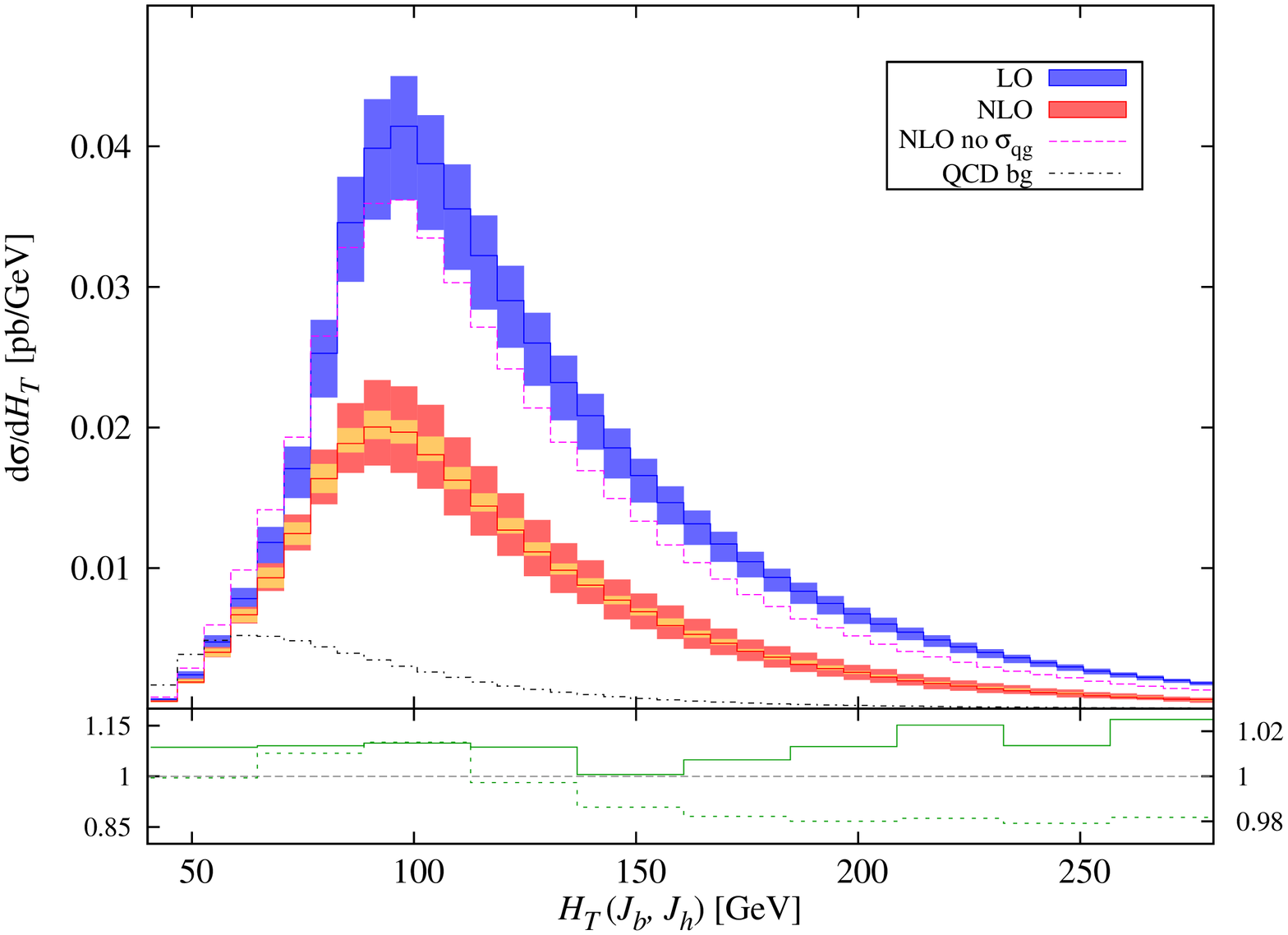}
\includegraphics[angle=-90,width=0.49 \linewidth]{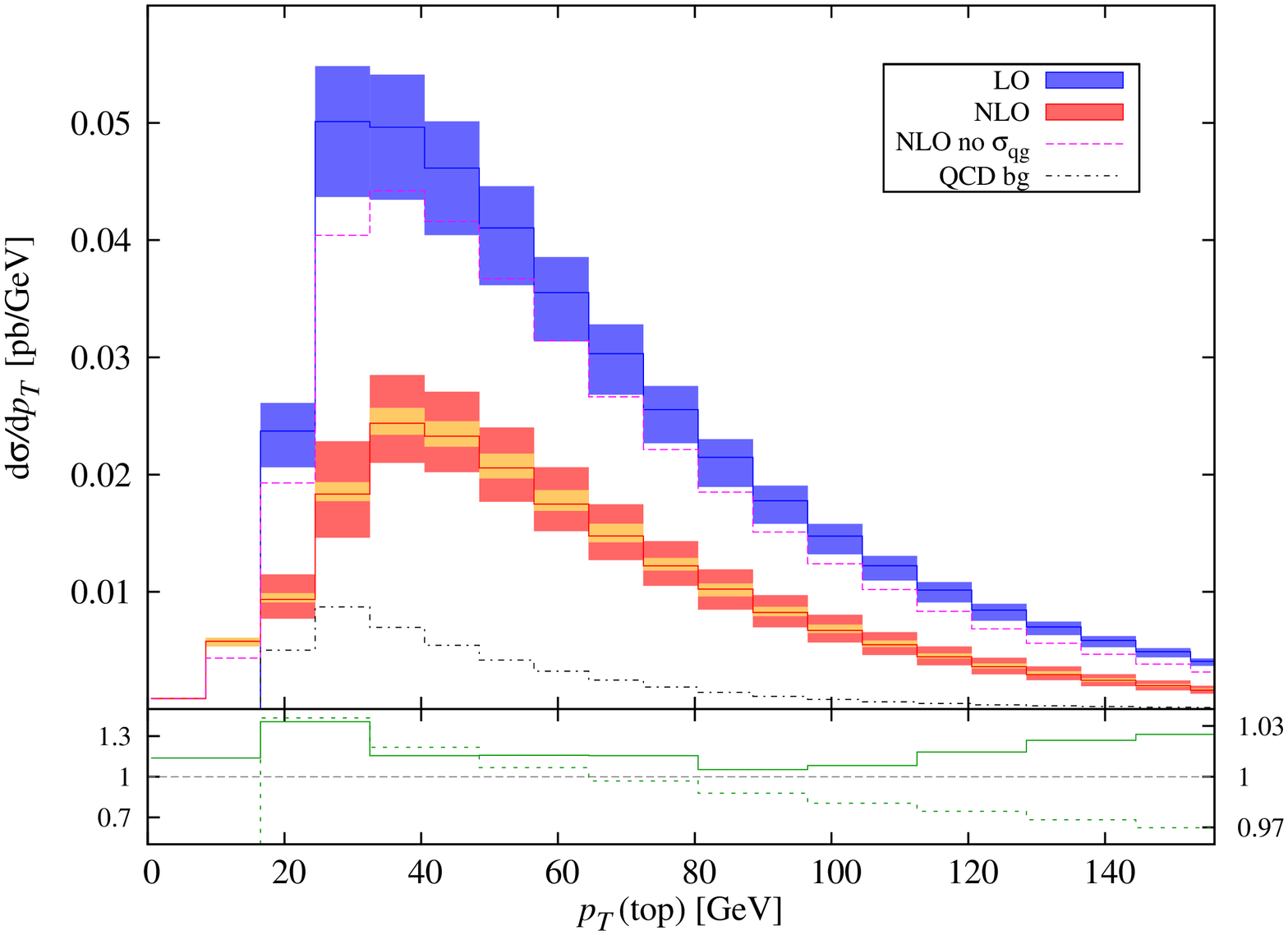}
\includegraphics[angle=-90,width=0.49 \linewidth]{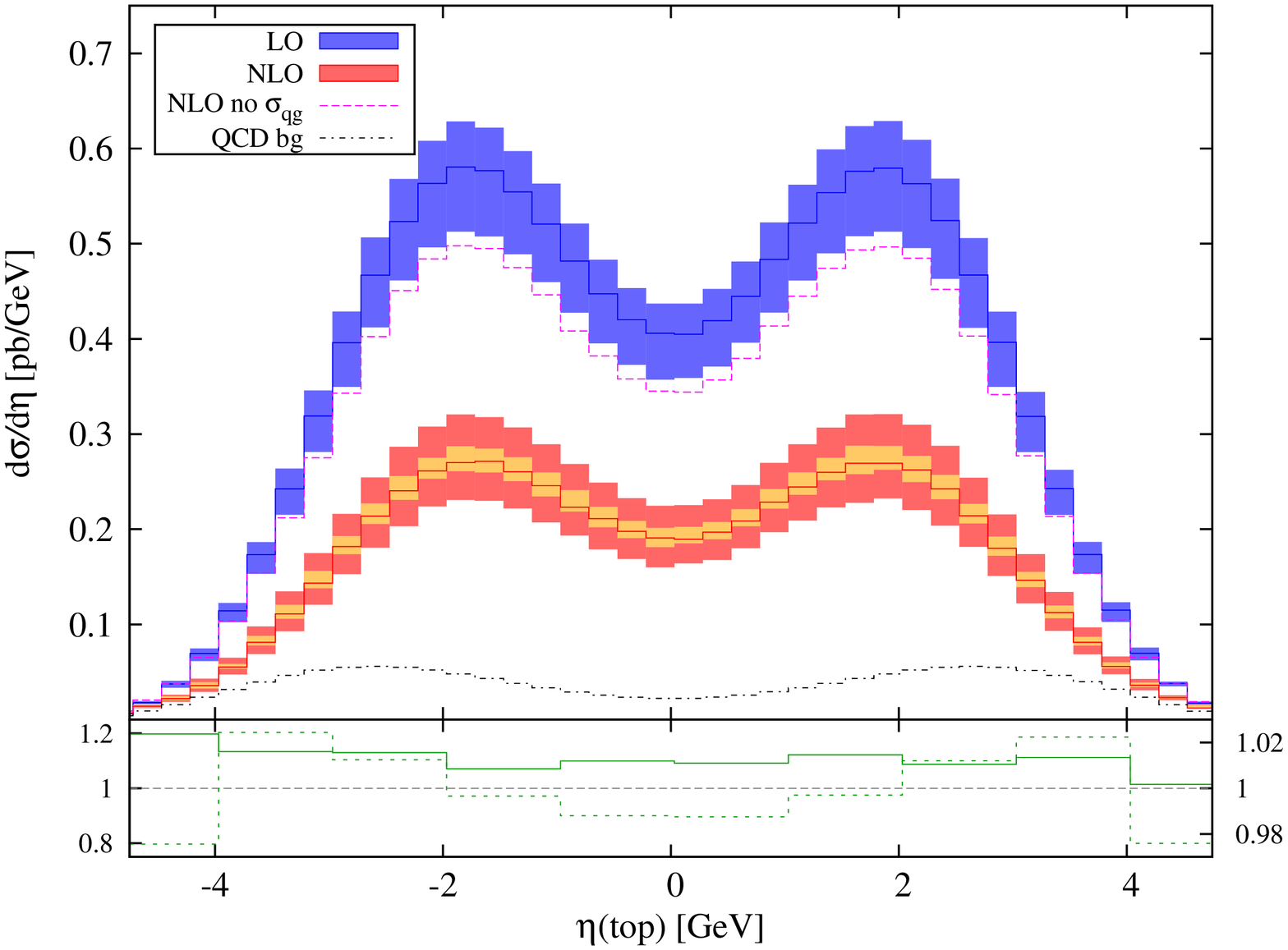}
\includegraphics[angle=-90,width=0.49 \linewidth]{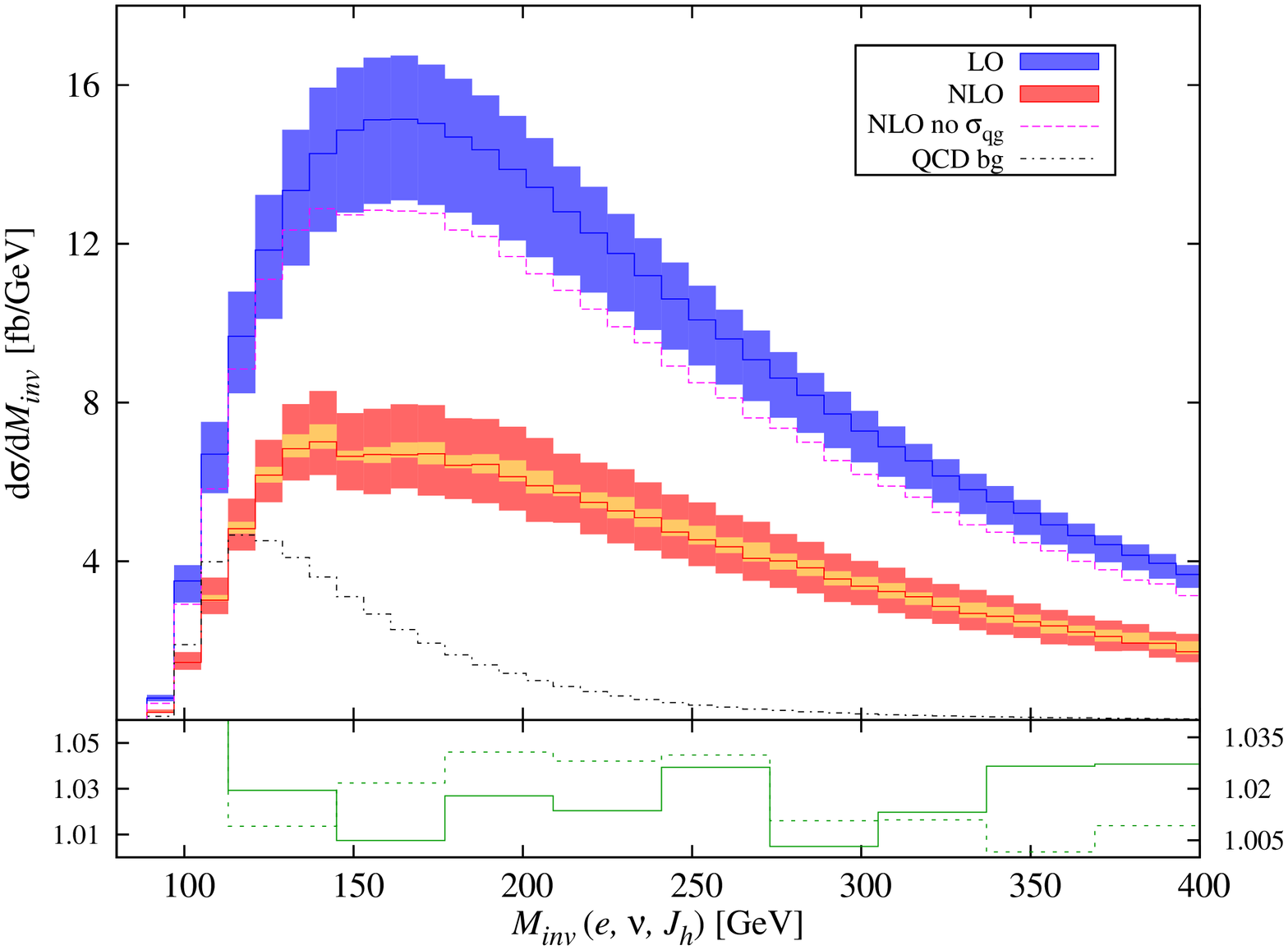}
\includegraphics[angle=-90,width=0.49 \linewidth]{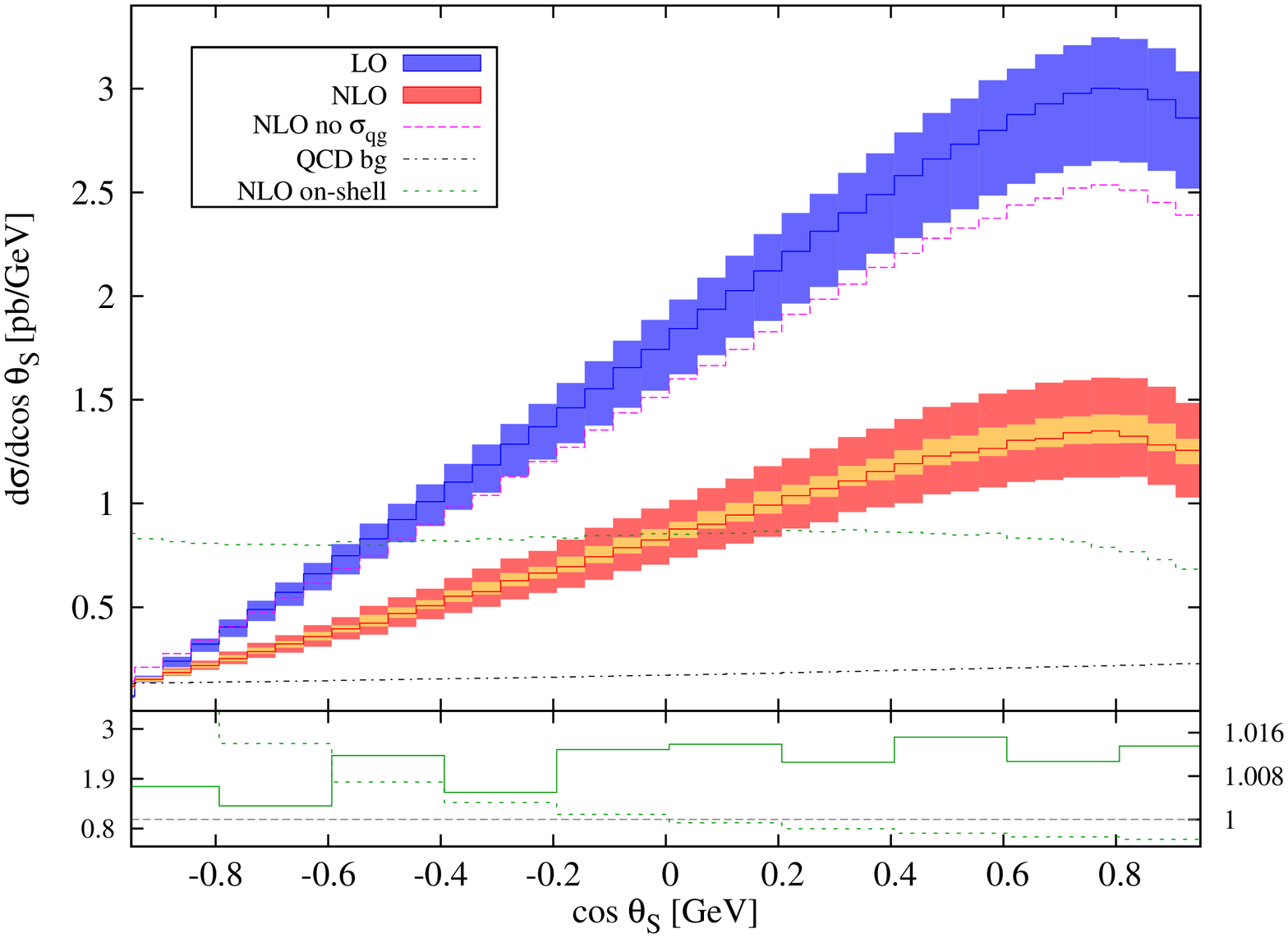}
\ccaption{}{Kinematical distributions for $p p \rightarrow J_{b} J_{l}
  e^+\dirac{E}_T+X$ for the 7~TeV LHC. Top: top-quark transverse mass
  (left) and hadronic transverse energy (right).  Centre: top-quark
  transverse momentum (left) and top-quark pseudorapidity (right).
  Bottom: $e^+ \nu J_h$ invariant mass (left) and $\cos\theta_S$
  (right). See the text for a precise definition of the observables
  and further explanations.
\label{fig:tchan_dis_LHC}}
\end{center}
\vspace{0.5 cm}
\end{figure}

\begin{figure}[h!]
\begin{center}
\includegraphics[angle=-90,width=0.49 \linewidth]{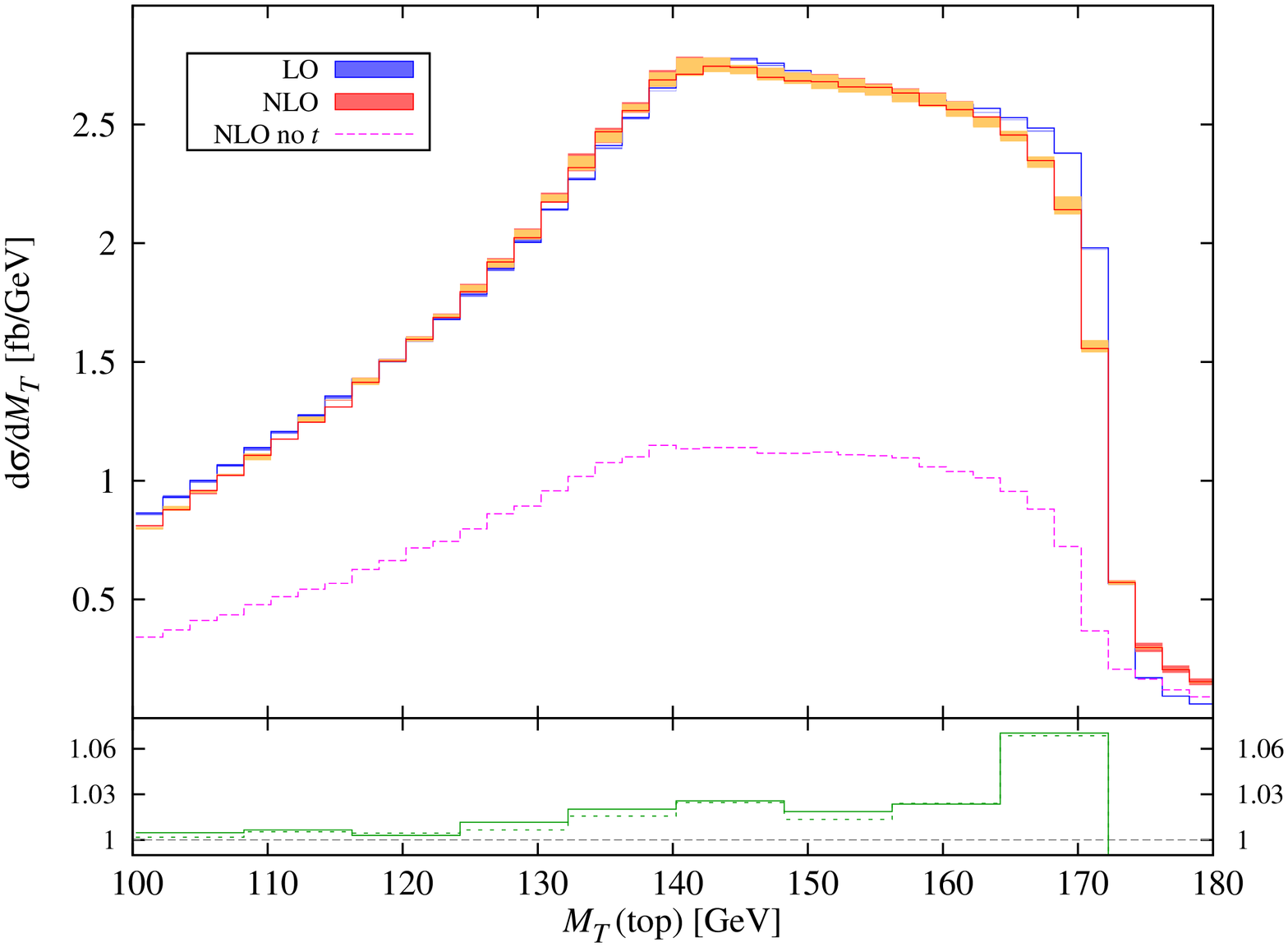}
\includegraphics[angle=-90,width=0.49 \linewidth]{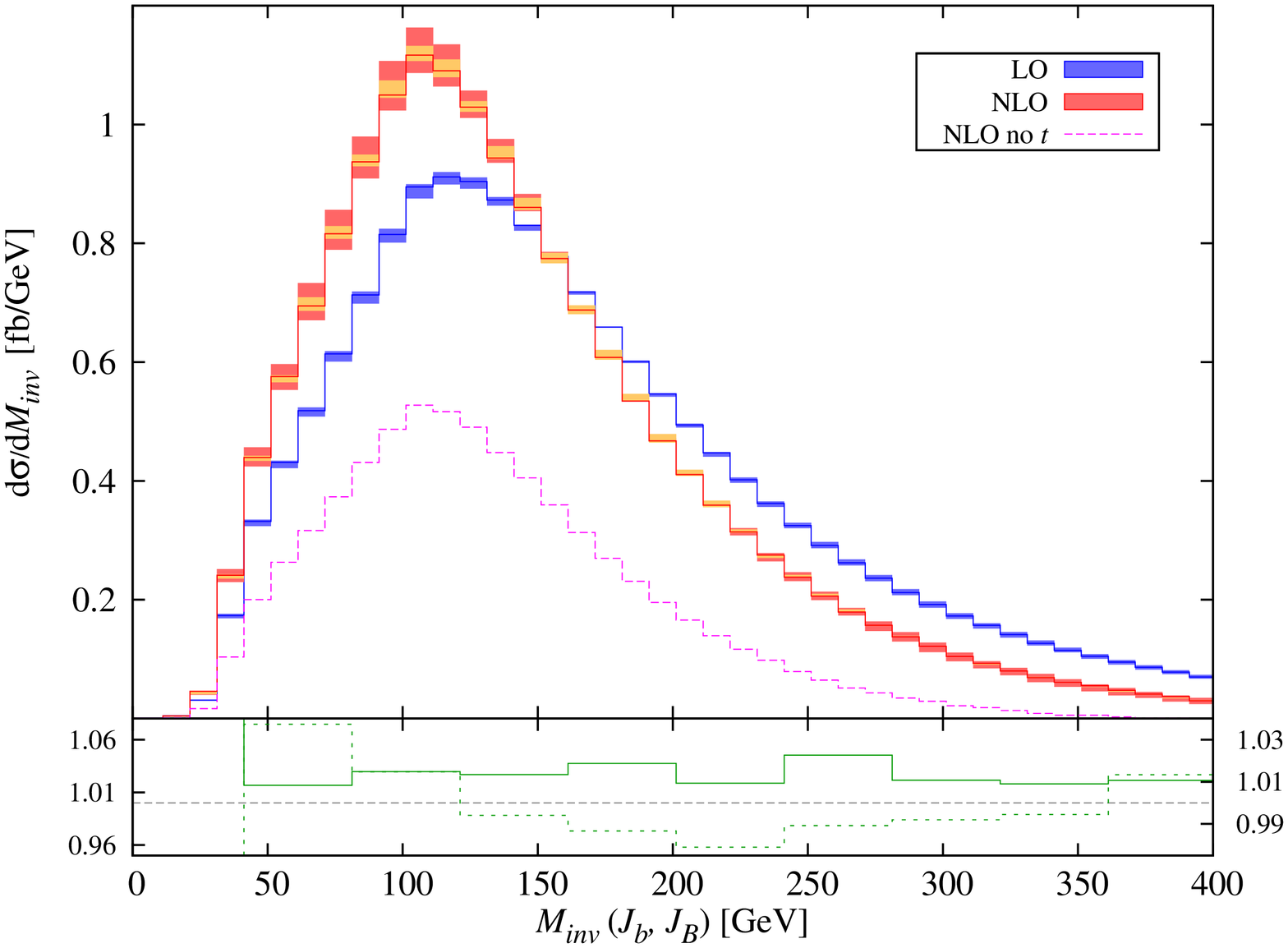}
\includegraphics[angle=-90,width=0.49 \linewidth]{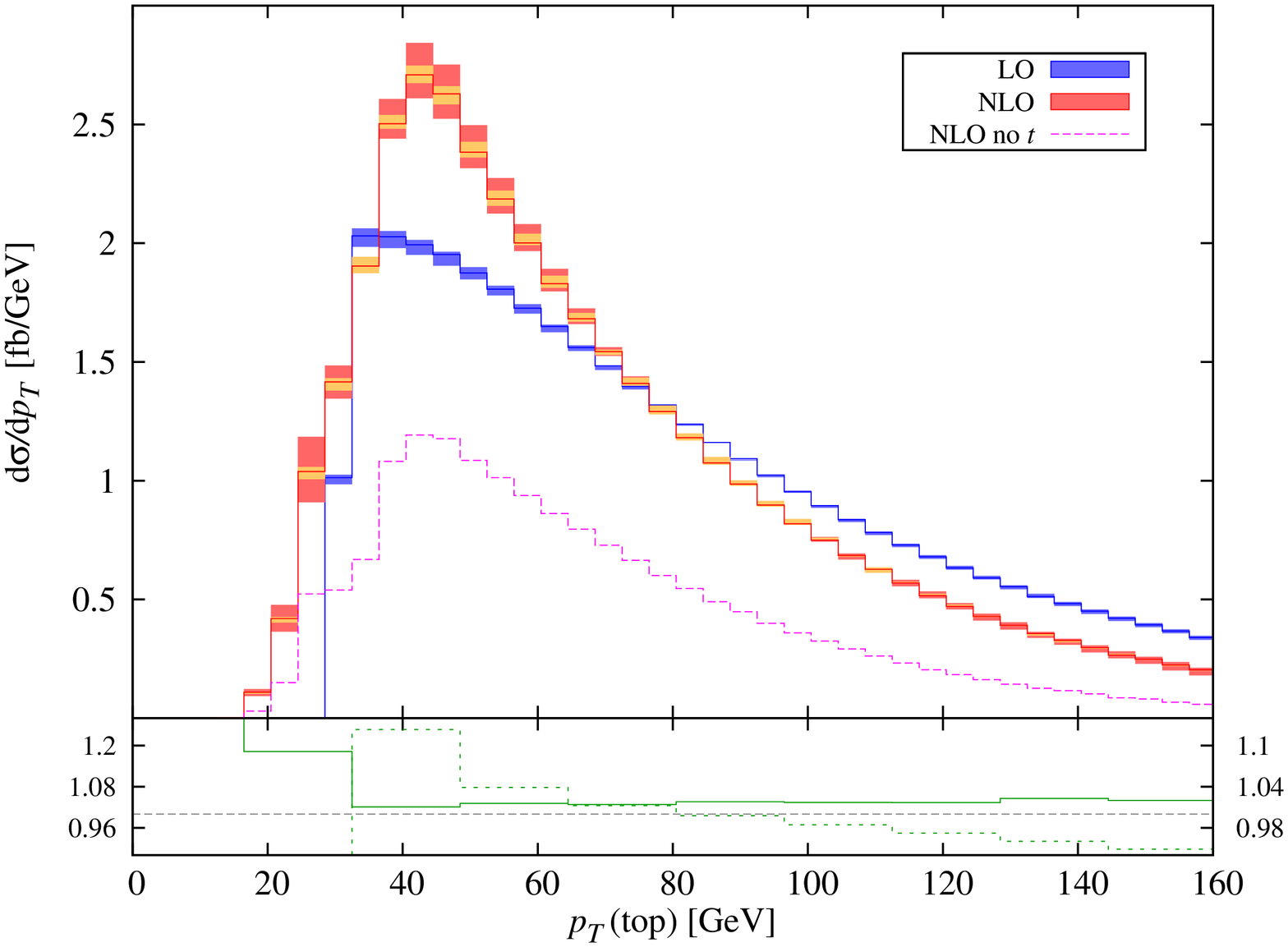}
\includegraphics[angle=-90,width=0.49 \linewidth]{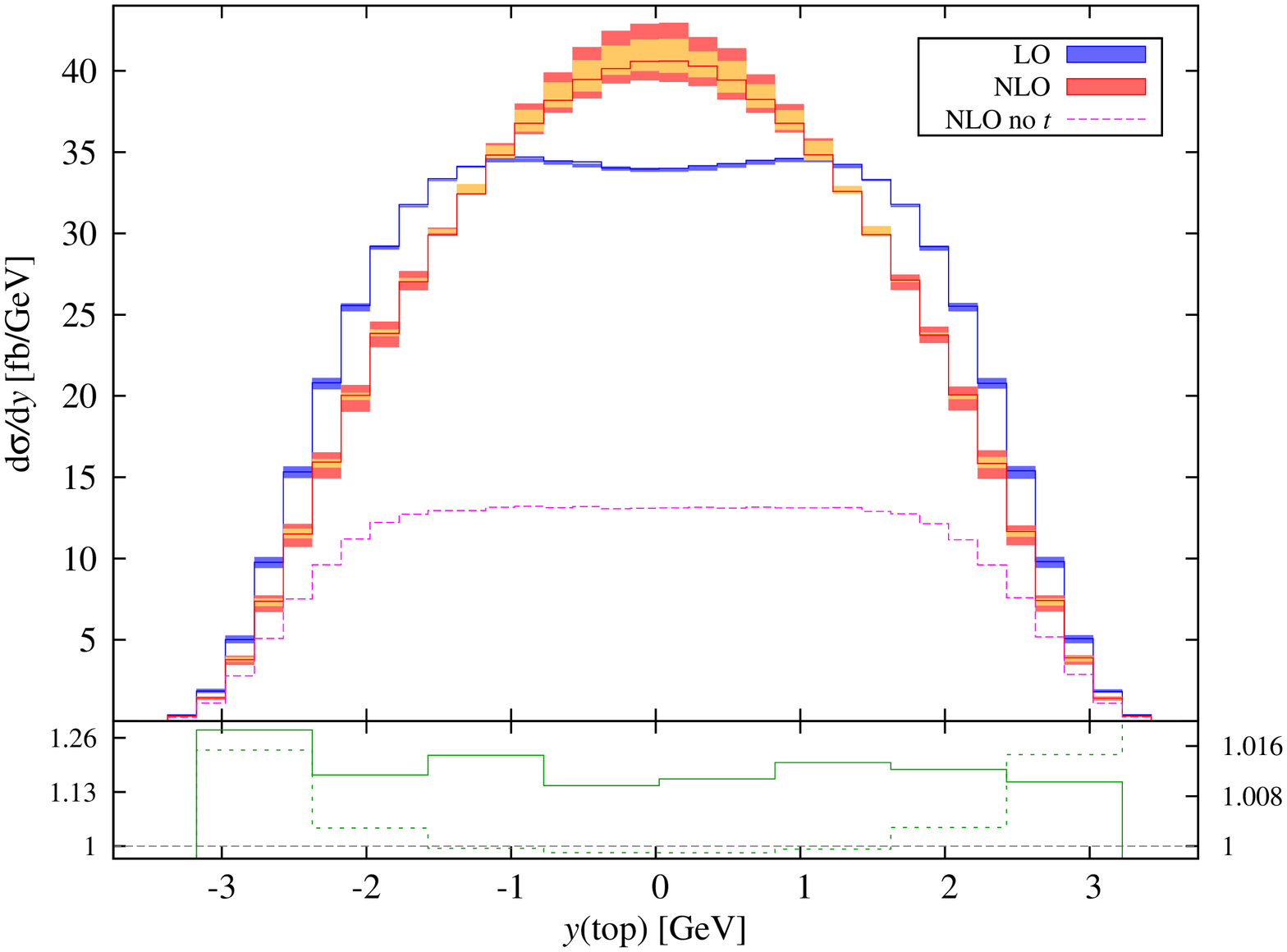}
\includegraphics[angle=-90,width=0.49 \linewidth]{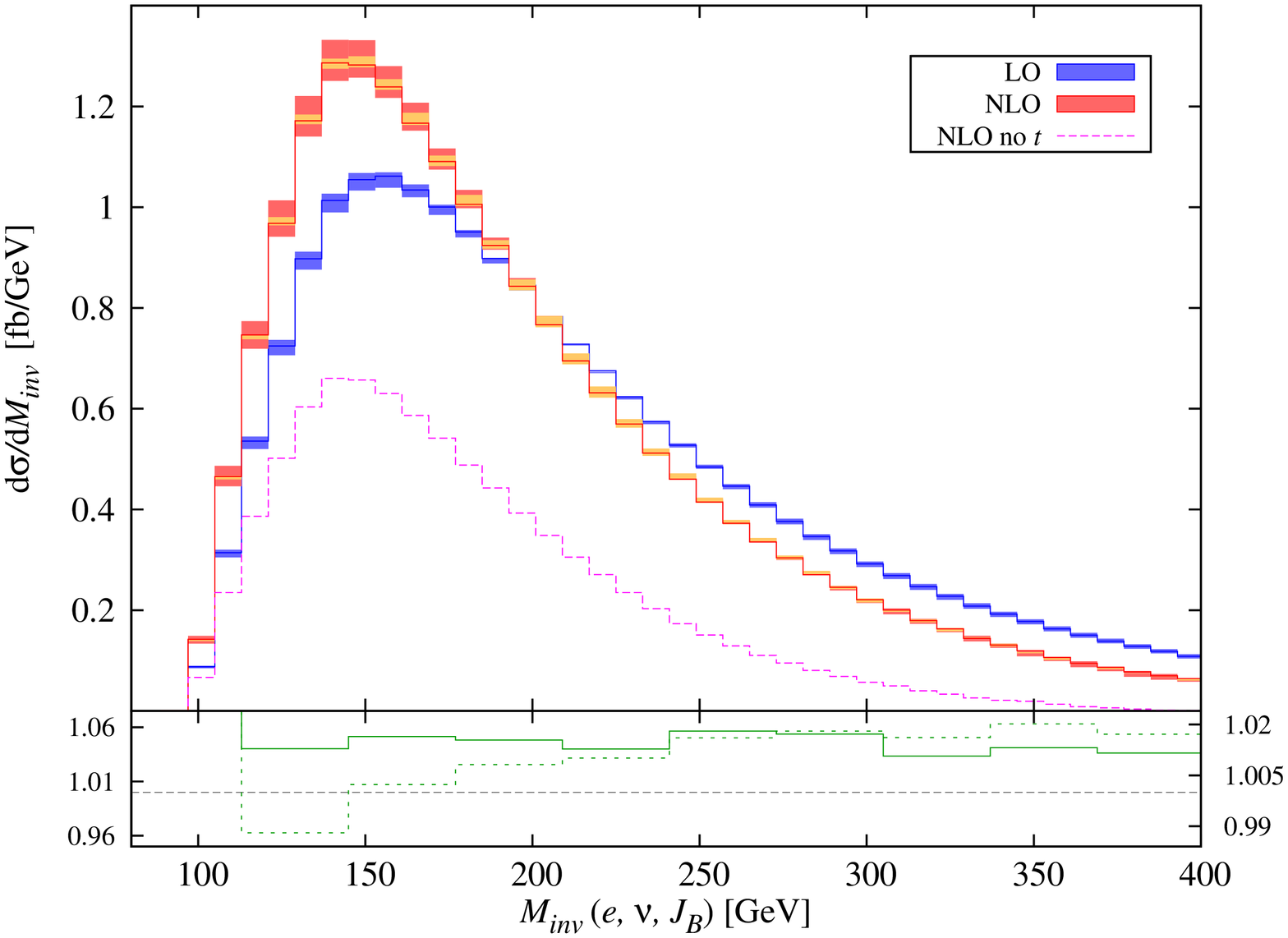}
\includegraphics[angle=-90,width=0.49 \linewidth]{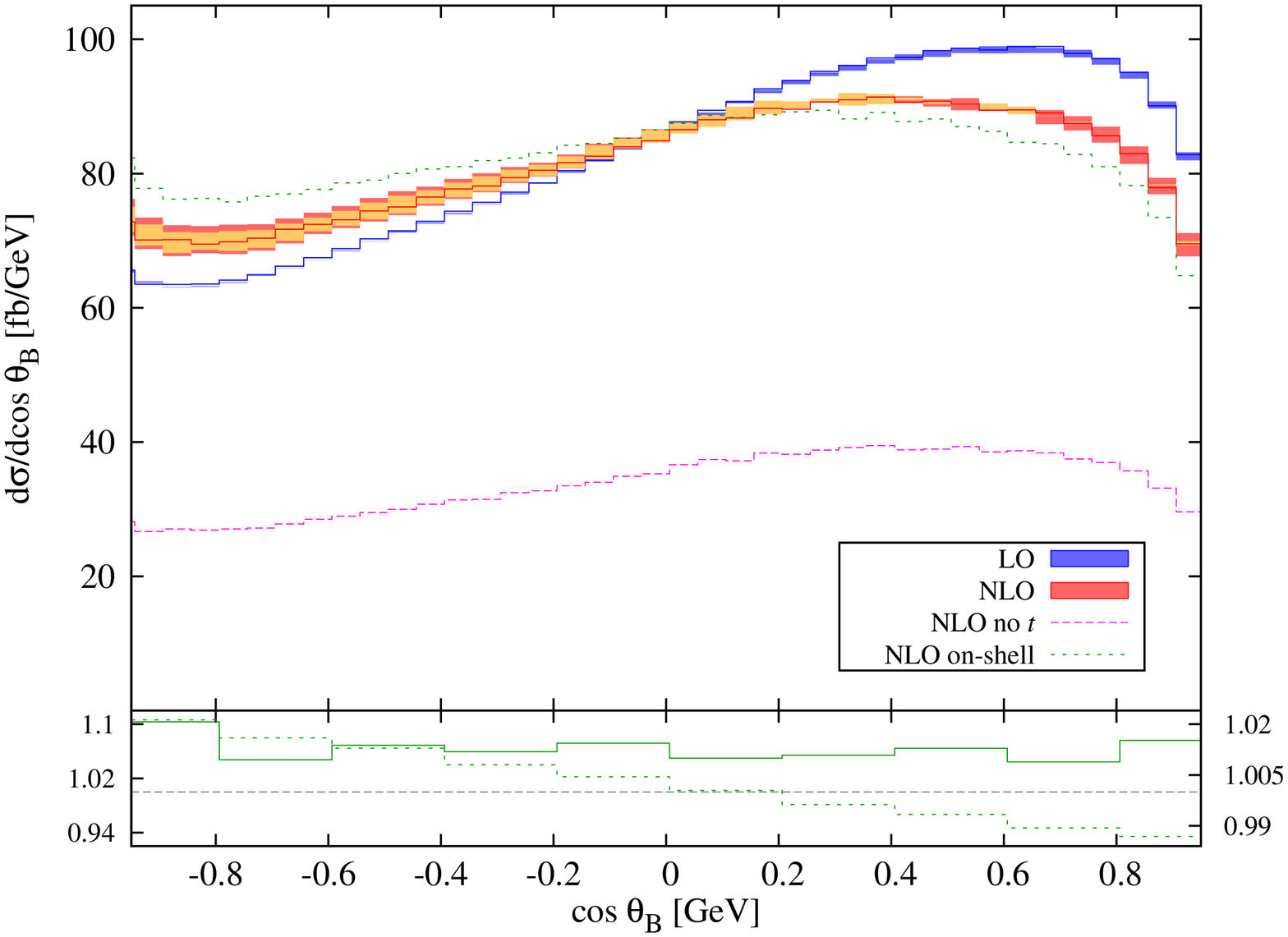}
\ccaption{}{Kinematical distributions for $p p \rightarrow J_{b}
  J_{\bar{b}} e^+\dirac{E}_T+X$ for the 7~TeV LHC. Top: top-quark
  transverse mass (left) and $J_b J_{\bar{b}}$ invariant mass (right).
  Centre: top-quark transverse momentum (left) and top-quark rapidity (right).
  Bottom: $e^+ \nu_e J_{\bar{b}}$ invariant mass (left) and
  $\cos\theta_B$ (right). See the text for a precise definition of the
  observables and further explanations.
\label{fig:schan_dis_LHC}}
\vspace{0.5 cm}
\end{center}
\end{figure}

Figure~\ref{fig:inv_mass_LHC} shows plots for the top-quark invariant-mass 
distribution for the two processes.  The blue band was obtained
varying the scale in the LO cross section in the interval $m_t/4 \leq
\mu_R=\mu_F \leq m_t$, and the red band is the analogue result for the NLO
cross section. The green curve is the NLO prediction in the
narrow-width approximation. As for the Tevatron, by comparing the two
NLO predictions one can see that the non-factorizable corrections are
sizeable (especially for the $s$-channel-like process) and negative at
the peak, while they turn positive above $m_{\text{inv}} \sim m_t$.
We again point out that the change in sign of the non-factorizable
contributions explains the small off-shell effects observed in the
total cross section. 

We now come to the discussion of kinematical distributions. In
Figure~\ref{fig:tchan_dis_LHC} we plot a selection of observables for
the process $p p \rightarrow J_{b} J_{l} e^+\dirac{E}_T+X$. The
observables chosen are the same ones used for the Tevatron analysis,
i.e. $M_T(t)$, $H_T(J_b,J_h)$, $p_T(t)$, $\eta(t)$,
$M_{\text{inv}}(e,\nu,J_h)$ and $\cos\theta_S$. As in the previous
section, the blue and red solid lines are the LO and NLO
effective-theory prediction for the central value of the scales, while
the blue and red bands are obtained by varying the scales
simultaneously in the interval $m_t/4 \leq \mu_R=\mu_F \leq m_t$.  The
orange band corresponds to variation of the factorization scale at
fixed renormalization scale.  The dashed magenta line is the NLO
result without the quark-gluon-initiated processes, and the black
dot-dashed line is the contribution of tree-level QCD background
diagrams. The lower plots show the ratio of the narrow-width
implementation with full spin correlations (solid green line, right
scale) and without (dashed green line, left scale) to the resonant
result at NLO.

As for the Tevatron, the total NLO corrections are large and negative,
up to $\sim 50-60\%$ in the central bins, and they mainly arise from
the gluon-initiated partonic channels. Non-factorizable corrections
are small, generally less than $\sim 5 \%$, except close to particular
kinematic thresholds. Spin-correlation effects are also small, and
comparable in size to non-factorizable corrections.  An exception to
this are observables depending upon angles, e.g. $\cos\theta_S$,
where the difference between the two NWA implementations is very
large, and where an exact treatment of spin-correlation effects is
clearly necessary. Also note that at the LHC the contribution of the QCD
background is smaller, compared to the signal, than for the Tevatron.
The NLO corrections found here are generally much larger than the ones
presented in Ref.~\cite{Schwienhorst:2010je}. Reasons for these large 
differences will be discussed in Section~\ref{sec:comparison}.

Kinematical distributions for $p p \rightarrow J_{b} J_{\bar{b}}
e^+\dirac{E}_T+X$ are given in Figure~\ref{fig:schan_dis_LHC}.  The
observables presented are $M_T(t)$, $M_{\text{inv}}(J_b,J_{\bar{b}})$,
$p_T(t)$, $y(t)$, $M_{\text{inv}}(e,\nu,J_{\bar{b}})$ and $\cos\theta_B$,
as for the Tevatron. In this case, the dashed magenta line represents
the NLO contribution from $s$-channel diagrams only.  We note that the
behaviour of the NLO corrections for $pp \rightarrow J_{b} J_{\bar{b}}
e^+\dirac{E}_T$ at the LHC is qualitatively very different from what is
observed at the Tevatron. This is a consequence of the large
contributions of the $t$-channel diagrams. The NLO corrections arising
from $s$-channel diagrams only are large and negative, whereas the
$t$-channel contributions are large and positive. As a result the
total correction to the transverse-mass distribution is very
small. For $M_{\text{inv}}(J_b,J_{\bar{b}})$, $p_T(t)$, $y(t)$ and
$M_{\text{inv}}(e,\nu,J_h)$ the inclusion of the $t$-channel
contributions leads to a distortion of the shape of the distribution,
with a sizeable positive total correction ($\sim20 \%$) in the peak
region, and negative corrections in the tail regions. This is
particularly dramatic for the rapidity distribution, where the plateaux
of the LO and NLO $s$-channel result is completely erased by the
$t$-channel corrections. Once more, non-factorizable corrections are
found to be small in most of the kinematical ranges considered
here. The same is true for spin-correlation effects, except for the
distribution for $\cos\theta_B$, as already observed for the Tevatron.
           
\subsection{Effects of a non-diagonal CKM matrix}
\label{sec:CKM}

The calculations presented in this paper so far have assumed that the
CKM matrix is diagonal. However, if we consider processes proportional 
to the off-diagonal elements of the matrix we can observe some interesting
results. In particular, the shape of the top-quark rapidity distribution 
is highly dependent on the flavour of the initial-state partons. This effect 
has been discussed, at tree-level, in Ref.~\cite{AguilarSaavedra:2010wf}.

\begin{figure}[t!]
\begin{center}
\includegraphics[width=0.7 \linewidth]{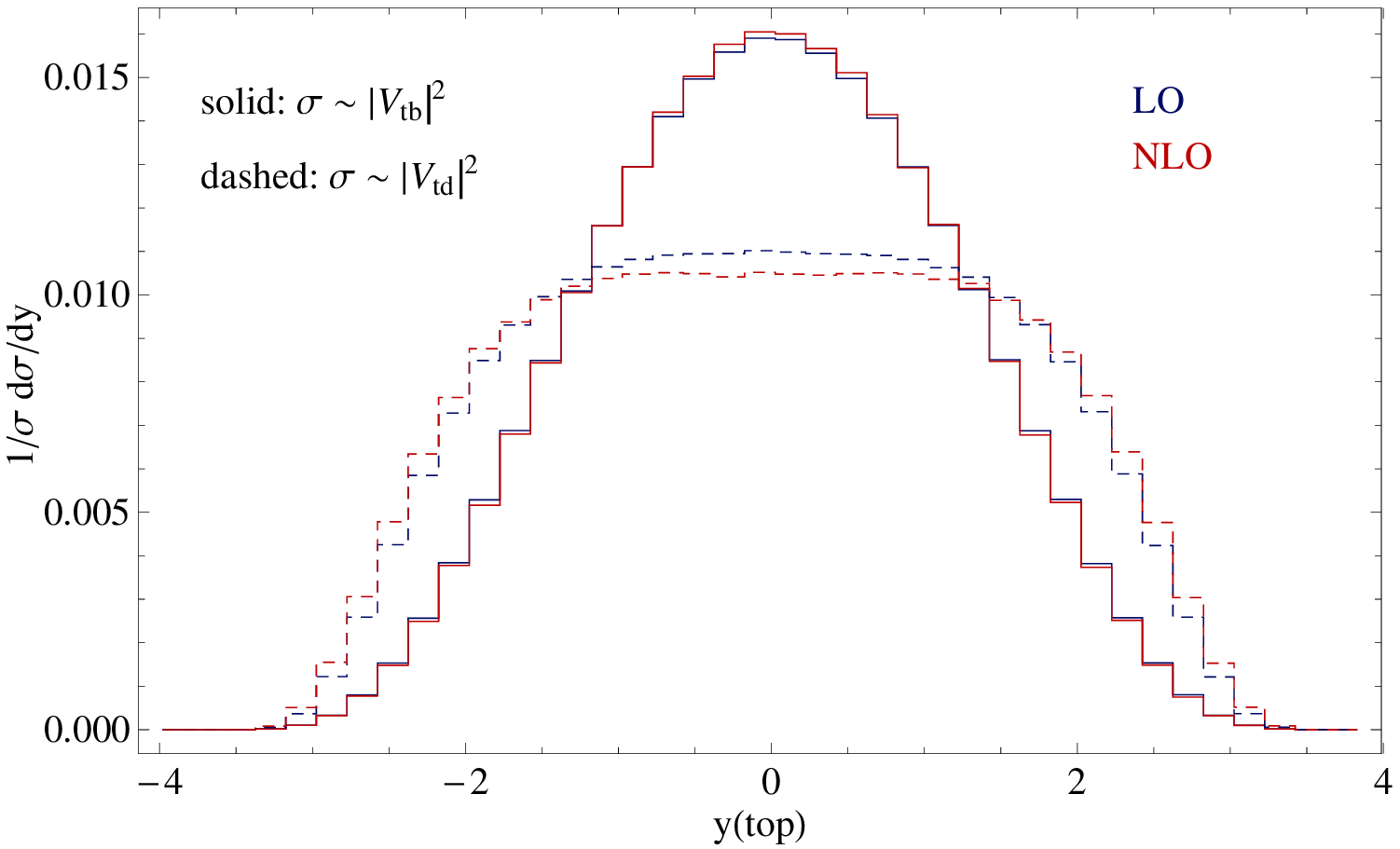}
\end{center}
\ccaption{}{Normalized top-quark rapidity distributions for $t$-channel
production at the LHC. Leading order results are shown in blue and 
next-to-leading order results in red. Solid and dashed lines represent 
contributions proportional to $|V_{tb}|^2$ and $|V_{td}|^2$, respectively. 
\label{fig:CKM_Rapidity}}
\end{figure}

The results shown in Figure~\ref{fig:CKM_Rapidity}, for $t$-channel
production at the LHC, show good agreement to those found in
Ref.~\cite{AguilarSaavedra:2010wf} at leading order. The events in
which the initial-state parton is a bottom quark are more central than
those in which the initial-state parton is a down quark.  The size of
the next-to-leading order effects means their impact on the overall
shape of the distributions is negligible. The tree-level differences
still dominate.  However, it is worth noting that for the
contributions proportional to $|V_{td}|^2$, the inclusion of
next-to-leading order effects seems to broaden the distribution,
further accentuating the differences in event centrality between
bottom-quark and down-quark initiated events.

As the main focus of this paper are the NLO corrections to single-top
production processes and we have seen that their impact on the shapes
of these distributions is negligible in comparison to the large differences
observed at leading order, further discussion of the effects of a
non-diagonal CKM matrix will not be made in this paper.

\subsection{Comparison to earlier results}
\label{sec:comparison}

As mentioned in Section~\ref{sec:LHC}, the size of the NLO corrections
presented here are, in general, much larger than those presented in,
for example, Ref.~\cite{Schwienhorst:2010je}. This is a consequence of
the different kinematical cuts chosen and, in particular, of the
strong veto imposed on extra $b$-tagged jets. As a validity check, we
performed our calculations again for $t$-channel production at the
LHC, this time using a different set of kinematical cuts and no veto
on extra $b$-tagged jets. Also, in this section we have evaluated the
LO results using LO parton distributions. A list of the cuts used is
given in Table~\ref{tab:no_veto}.

\begin{table}
  \begin{center}
  \begin{tabular}{lr}

  \hline
  \hline
  \multicolumn{2}{c}{$p \bar{p} \rightarrow J_{b} J_{l}
    e^+\dirac{E}_T+X$} \\[2pt] 
  \hline
  \hline
  $p_T(J_b) > 20$~GeV & $\dirac{E_T} > 20$~GeV \\[2pt]
  $p_T(\text{hardest} \, J_l) > 20$~GeV & $\eta(J_b)< 2.5$ \\[2pt]
  $p_T(e)> 20$~GeV & $\eta(\text{hardest}\, J_l)< 2.0$ \\[2pt] 
  $120 < m_{{\rm inv}} < 200$~GeV & $\eta(e)< 2.5$  \\[2pt]
  \hline
  \hline

  \end{tabular}
  \end{center}
  \ccaption{}{Kinematical cuts used for comparison to earlier
    results. \label{tab:no_veto}}
\end{table}

\begin{figure}[t!]
\begin{center}
\includegraphics[angle=-90,width=0.49 \linewidth]{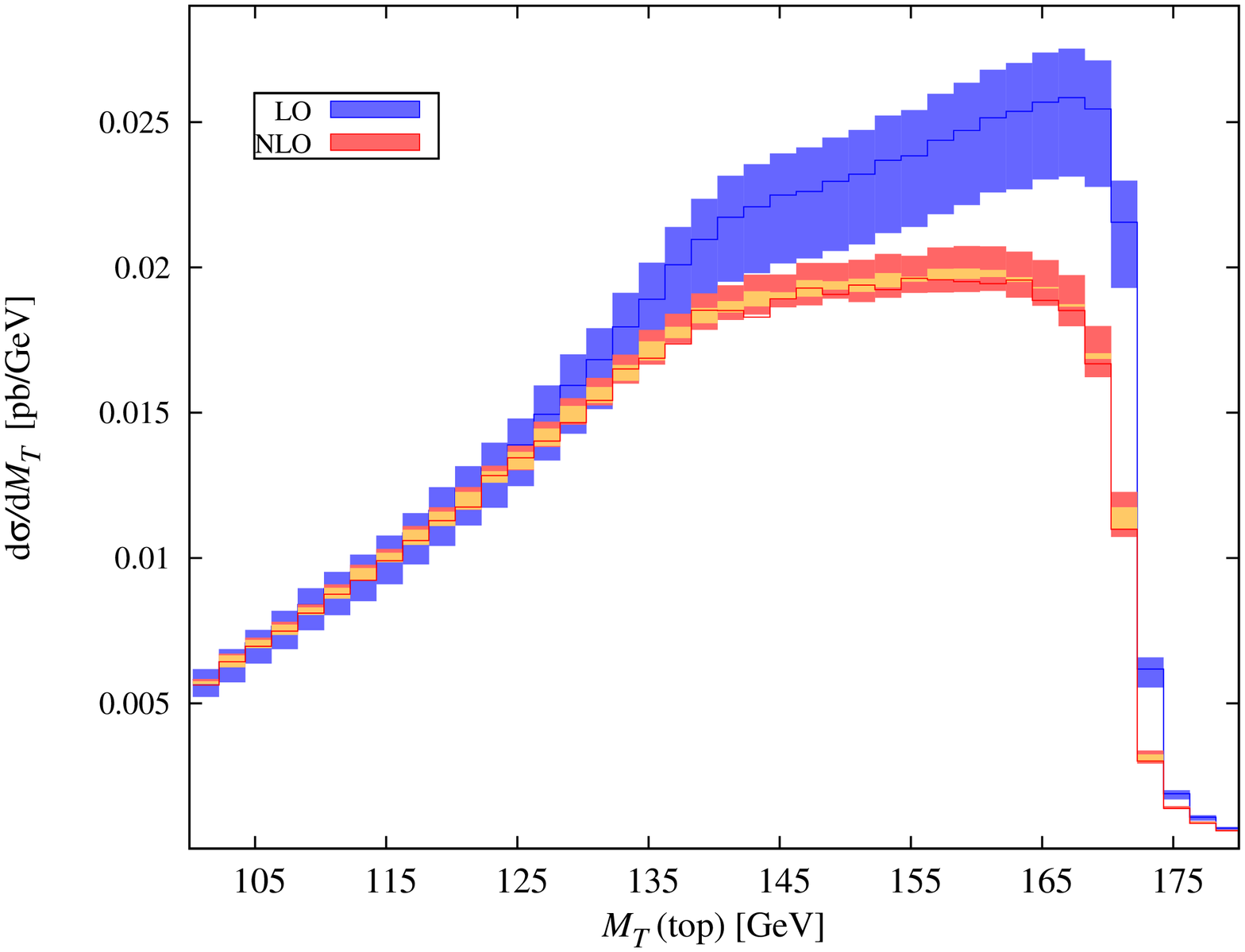}
\includegraphics[angle=-90,width=0.49 \linewidth]{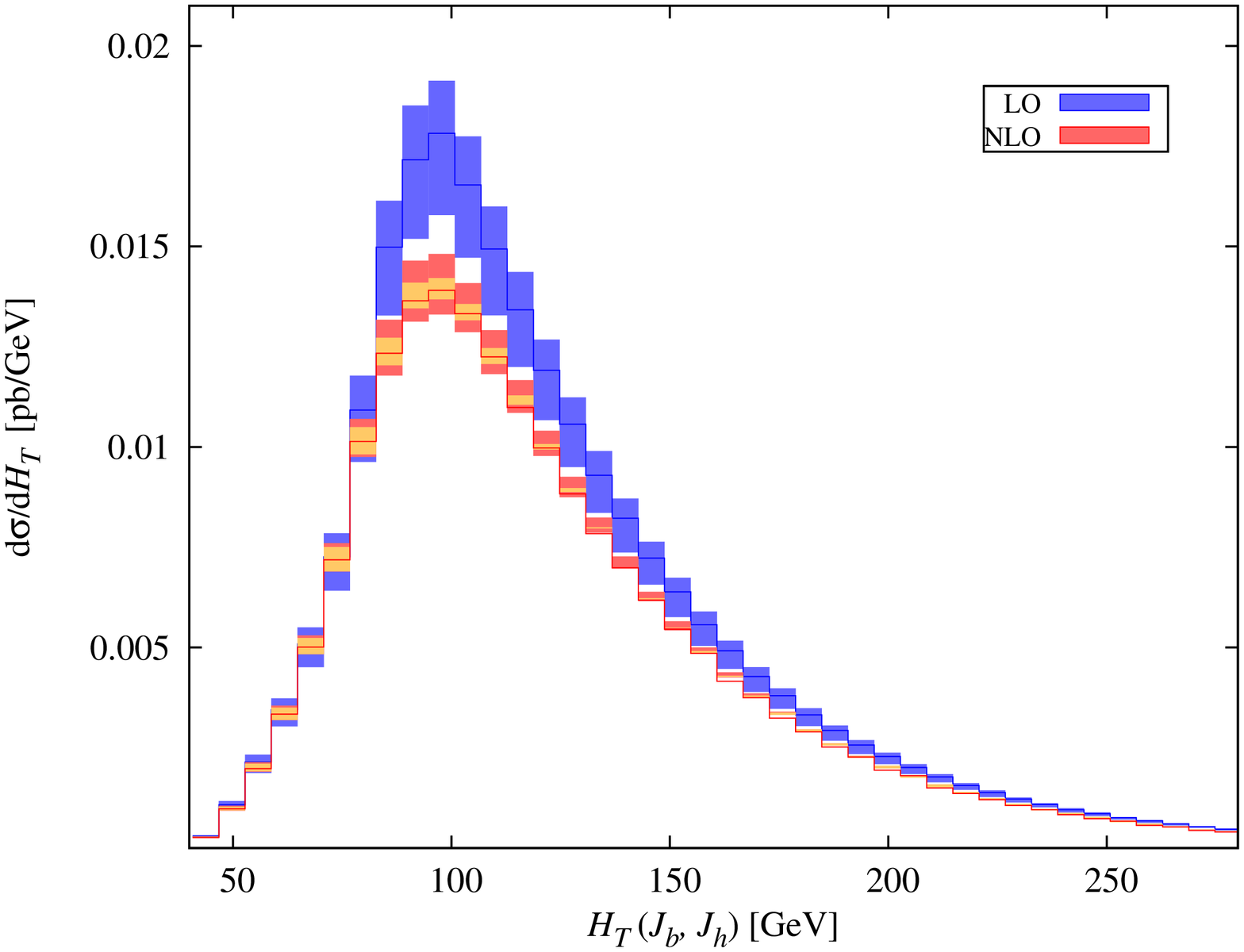}
\includegraphics[angle=-90,width=0.49 \linewidth]{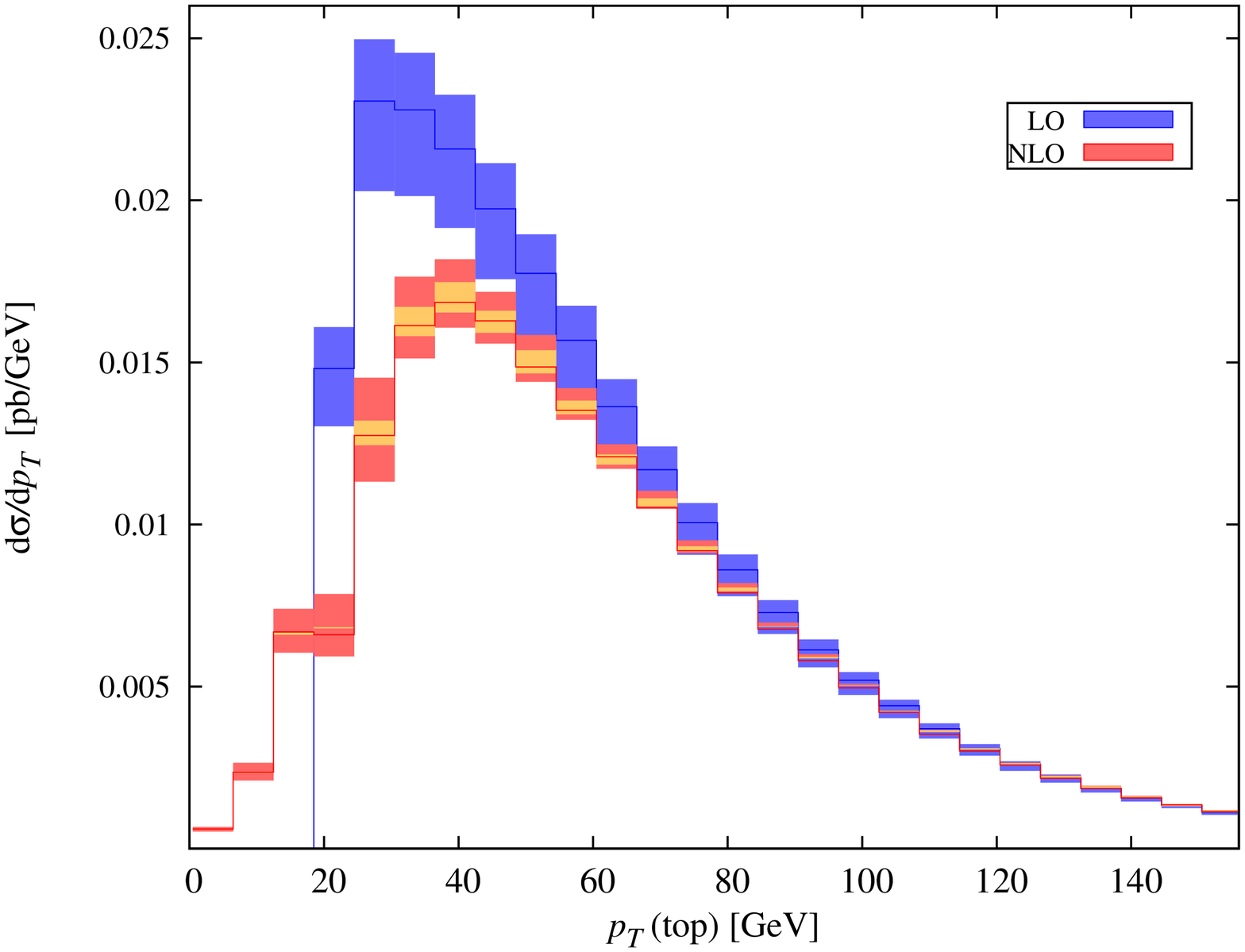}
\includegraphics[angle=-90,width=0.49 \linewidth]{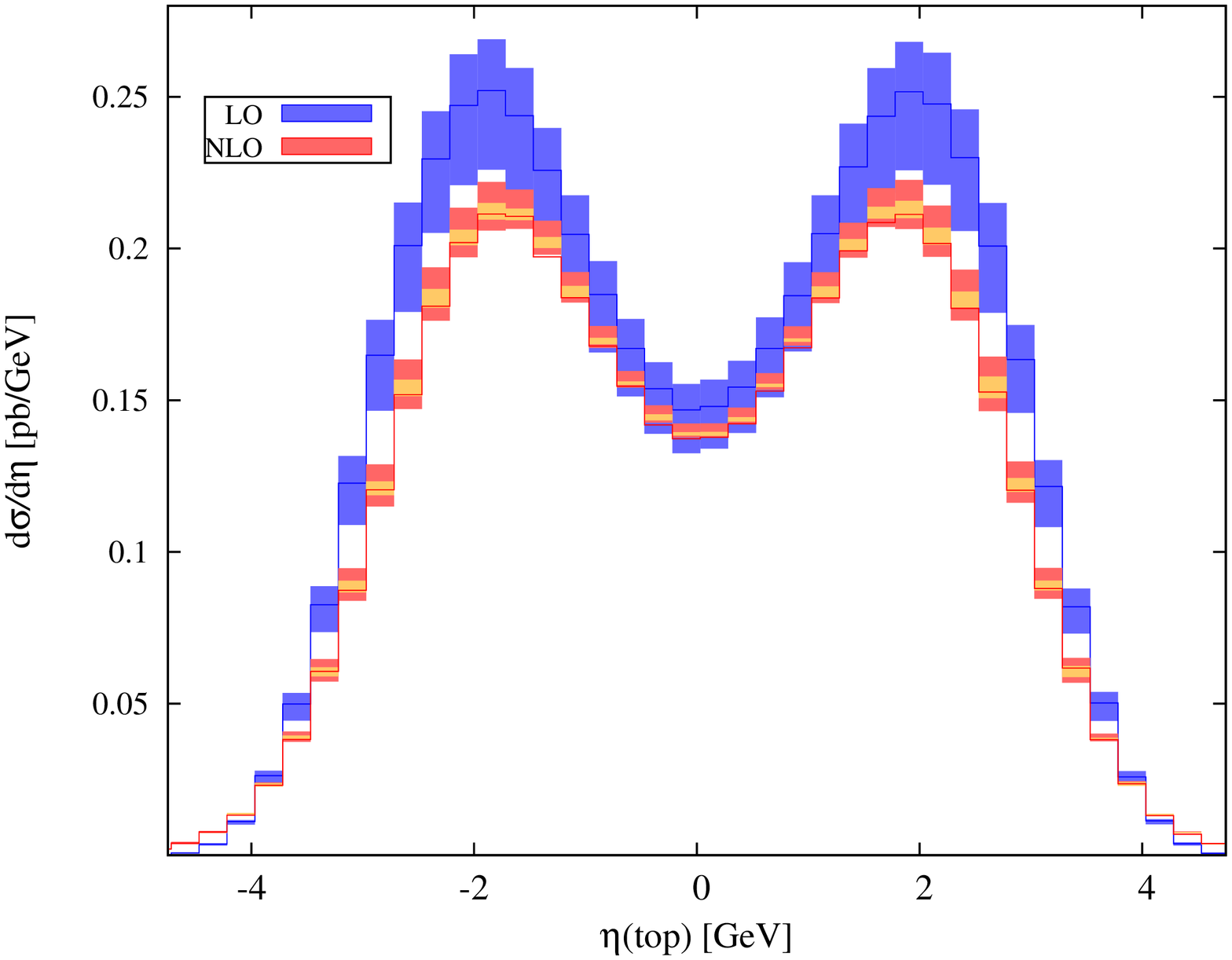}
\includegraphics[angle=-90,width=0.49 \linewidth]{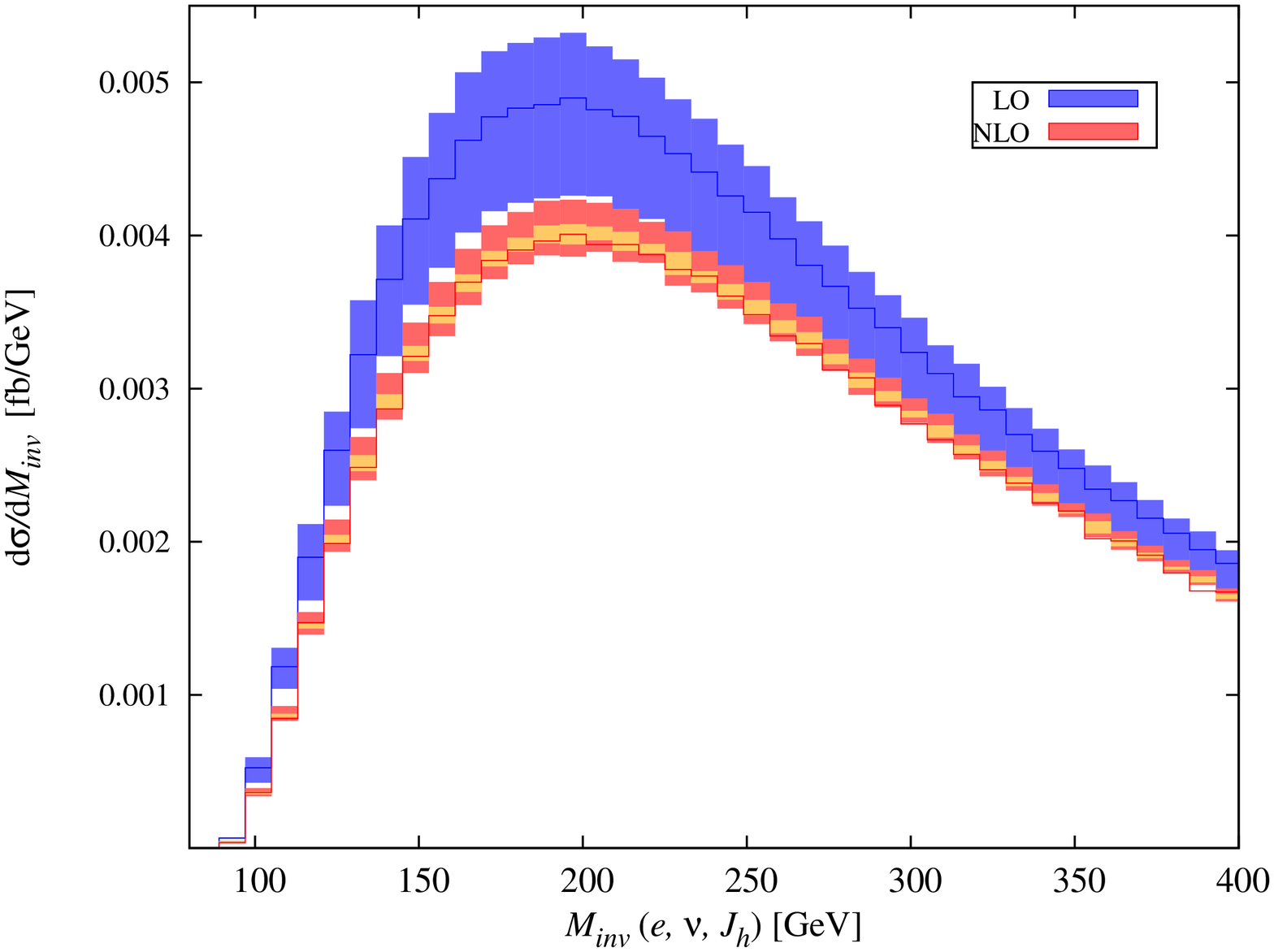}
\includegraphics[angle=-90,width=0.49 \linewidth]{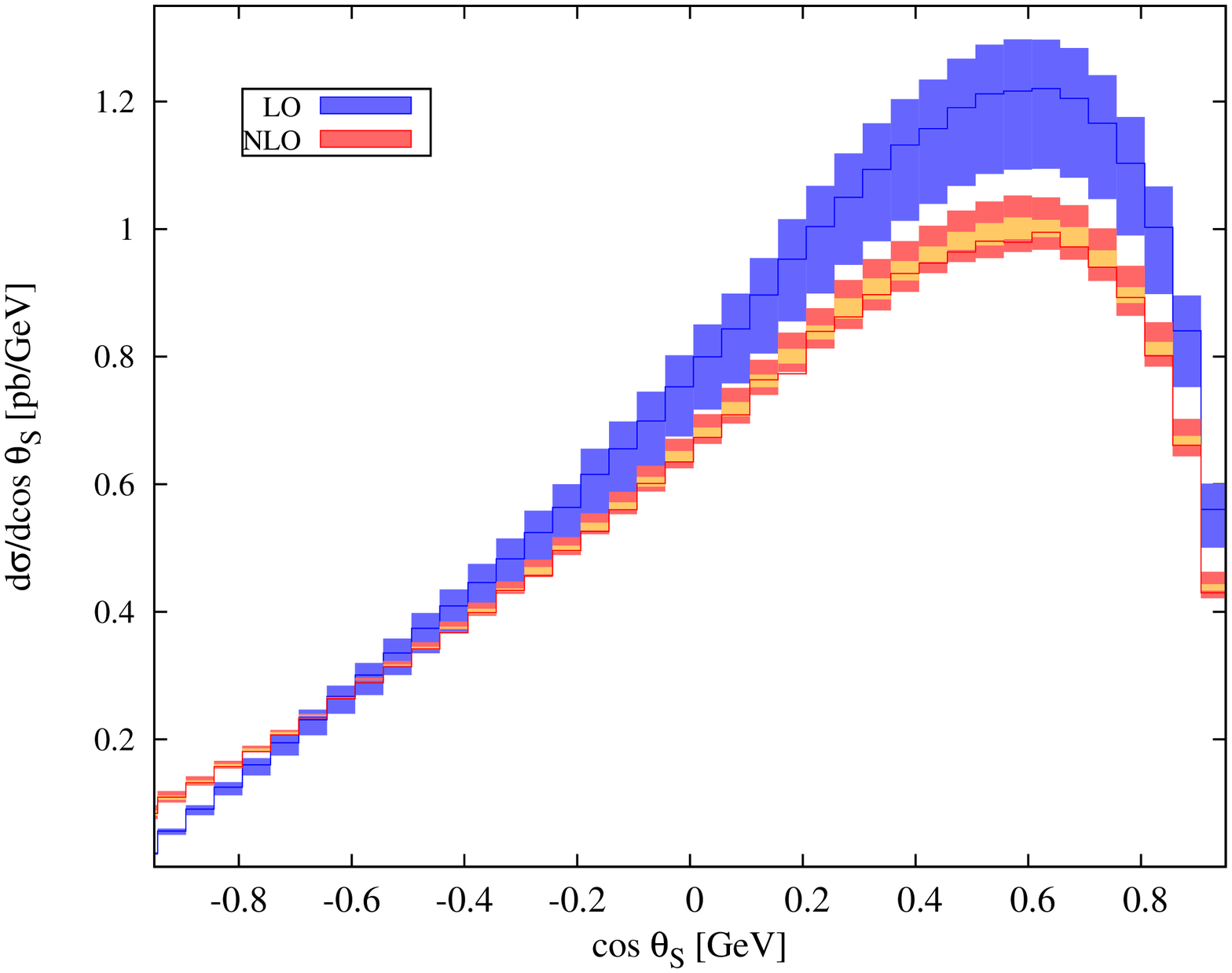}
\ccaption{}{Kinematical distributions for $p p \rightarrow J_{b} J_{l}
  e^+\dirac{E}_T+X$ for the 7~TeV LHC with the cuts shown in
  Table~\ref{tab:no_veto} applied. Top: top-quark transverse mass (left) and
  hadronic transverse energy (right).  Centre: top-quark transverse
  momentum (left) and top-quark pseudorapidity (right).  Bottom: $e^+
  \nu J_h$ invariant mass (left) and $\cos\theta_S$ (right). See the
  text for a precise definition of the observables and further
  explanations.
\label{fig:no_B_cut_LHC}}
\vspace{0.5 cm}
\end{center}
\end{figure}

Our results, calculated with these less stringent cuts, are shown in 
Figure~\ref{fig:no_B_cut_LHC}. Although a different set of kinematical 
cuts were used, it is clear to see by comparison to the results shown in
Figure~\ref{fig:tchan_dis_LHC}, that removing the veto on the extra
$b$-tagged jets leads to a large reduction in the size of the NLO
contributions. The size of these corrections are also in much better 
agreement with those found in Ref.~\cite{Schwienhorst:2010je}. 

The major factor contributing to the strong correlation between the
veto on extra $b$-tagged jets and the size of the NLO corrections are
the diagrams with initial-state $q$ and $g$ partons which result in a 
final state containing both a $b$ and a $\bar{b}$ quark. When using our
original cuts and vetoes (diagrams in Figure~\ref{fig:tchan_dis_LHC})
the NLO contribution to the total cross section due to these
diagrams amounted to a negative correction of $\sim 40\%$ of the
leading order value. This can clearly be seen in the diagrams by
comparing the full NLO result (red lines) to the NLO result with the
$qg$ contribution removed (dashed magenta lines). When we move to our
less stringent cuts, however, the NLO contribution to the total cross
section due to the $qg$ diagrams is reduced to a negative correction
of $\sim 10\%$ of the leading order value. These diagrams still have a
large impact on the size of the NLO corrections but their importance
is substantially reduced in comparison to the case where we employed
a strong veto.

\newpage

\section{Conclusion}\label{sec:conclusion}

We have applied a general method to include non-factorizable
corrections for processes at hadron colliders to single-top production
in the $t$- and $s$-channels. Within this effective-theory inspired
method, non-factorizable corrections are nothing but corrections due to
soft modes. The corrections to the production and decay parts of the
process are reproduced by hard effects. This method allows us to include
off-shell effects with a minimal amount of computation. Of course,
spin-correlation effects are also fully included.

As is well known, spin-correlation effects can be very large for
particular kinematic distributions. On the other hand, off-shell
effects are generally found to be small. For the total cross section
and most distributions they amount to a correction of the order of
1--3\%. However, they can be substantially larger for observables that
are not inclusive enough in the invariant mass of the decay products
of the top quark. Apart from the invariant mass distribution itself,
this is manifest at kinematic edges of certain distributions. 

Thus, even though off-shell effects are small, they can have an impact
on precise measurements of the top-quark mass. Furthermore, we stress
that we have evaluated the soft (non-factorizable) corrections with
the same large scale, $\mu_h \sim m_t$, as the hard corrections.
Na\"{i}vely, changing the scale of the soft corrections from $\mu_h$ to
$\mu_s \sim \Gamma_t$ would lead to a substantial increase in their
importance. However, a more careful study is required for a consistent
resummation of $\log\mu_s/\mu_h$. We are confident that these, and
related questions, are best addressed within an effective-theory
framework.

\subsection*{Acknowledgements}

This work has been supported in part by the European Commission
through the Initial Training Network PITN-GA-2010-264564
(LHCPhenoNet). P.F. acknowledges support by the `Stichting voor
Fundamenteel Onderzoek der Materie (FOM)'.  F.G. was supported, in
part, by the EU contract No. MRTN-CT-2006-035482, “FLAVIAnet” and by
the grant “Borse di ricerca in collaborazione internazionale” by
Regione Puglia, Italy. P.M. is supported by an STFC
studentship. P.F. and F.G. thank the IPPP, Durham for their kind
hospitality during part of this work.

\end{document}